\documentclass[12pt]{article}
\usepackage[margin=1in]{geometry}
\usepackage{lineno}
\usepackage[colorlinks=true,citecolor=blue,pdfcreator={},pdfproducer={},linkcolor=blue]{hyperref}
\modulolinenumbers[1]
\usepackage{amsmath}
\usepackage{amsfonts}
\usepackage{bm}
\usepackage{graphicx}
\allowdisplaybreaks
\usepackage{amsbsy}
\usepackage{amssymb}
\usepackage[ruled]{algorithm2e}
\usepackage{natbib}
\usepackage[export]{adjustbox}

\usepackage{chngcntr}

\usepackage{subfig}

\usepackage[ruled]{algorithm2e}

\usepackage{indentfirst}
\usepackage{enumitem}
\setlist{nolistsep}

\newcommand*\patchAmsMathEnvironmentForLineno[1]{%
\expandafter\let\csname old#1\expandafter\endcsname\csname #1\endcsname
\expandafter\let\csname oldend#1\expandafter\endcsname\csname end#1\endcsname
\renewenvironment{#1}%
{\linenomath\csname old#1\endcsname}%
{\csname oldend#1\endcsname\endlinenomath}}%
\newcommand*\patchBothAmsMathEnvironmentsForLineno[1]{%
\patchAmsMathEnvironmentForLineno{#1}%
\patchAmsMathEnvironmentForLineno{#1*}}%
\AtBeginDocument{%
\patchBothAmsMathEnvironmentsForLineno{equation}%
\patchBothAmsMathEnvironmentsForLineno{align}%
\patchBothAmsMathEnvironmentsForLineno{flalign}%
\patchBothAmsMathEnvironmentsForLineno{alignat}%
\patchBothAmsMathEnvironmentsForLineno{gather}%
\patchBothAmsMathEnvironmentsForLineno{multline}%
}

\usepackage[nomarkers]{endfloat}

\usepackage{xspace}

\usepackage{authblk}
\usepackage{times}

\title{\textbf{A hybrid eikonal solver for accurate first-arrival 
traveltime computation in anisotropic media with strong contrasts}}
\author[,1]{Kai Gao\thanks{Corresponding Author; Email: kaigao@lanl.gov}}
\author[1]{Lianjie Huang}
\affil[1]{Los Alamos National Laboratory, Geophysics Group, Los Alamos, 
NM
87545, USA}

\date{}
\providecommand{\keywords}[1]{\textbf{\textbf{Keywords:}} #1}

\begin{document}
\maketitle

\begin{abstract}
First-arrival traveltime computation is crucial for many applications 
such as traveltime tomography, Kirchhoff migration, etc. There exist two 
major issues in conventional eikonal solvers: the source singularity 
issue and insufficient numerical accuracy in complex media.  Some 
existing eikonal solvers also exhibit the stability issue in media with 
strong contrasts in medium properties.
We develop a stable and accurate hybrid eikonal solver for 2D and 3D 
transversely isotropic media with a tilted symmetry axis (TTI, or tilted 
transversely isotropic media). Our new eikonal solver combines the 
traveltime field factorization technique, the third-order Lax-Friedrichs 
update scheme, and a new method for computing the base traveltime field.
The solver has the following three advantages. First, there is no need to 
assign exact traveltime values in the near-source region, and the 
computed traveltime field near the source location is accurate even for 
TTI media with strong anisotropy. Second, the computed traveltime field 
is high-order accurate in space. Third, the solver is numerically stable 
for 2D and 3D TTI media with strong anisotropy, complex structures, and 
strong contrasts in medium properties.
We verify the stability and accuracy of our hybrid eikonal solver using 
several 2D and 3D TTI medium examples. The results show that our solver 
is stable and accurate in 2D and 3D complex TTI media, producing 
first-arrival traveltime fields that are consistent with full-wavefield 
solutions. 

\end{abstract}


\section{Introduction}

Traveltime computation is important for many applications, including 
underwater acoustics \cite[]{Martinelli_2012}, geometrical optics 
\cite[]{Qian_Leung_2006}, quantum mechanics \cite[]{Jin_etal_2005}, 
geophysics, etc. Many geophysical applications, such as Kirchhoff 
migration \cite[]{Samuel_etal_1994,Buske_1999} and first-arrival 
traveltime tomography \cite[]{Lin_etal_2009,Taillandier_etal_2009} for 
reconstructing subsurface structures and medium properties, rely on 
accurate and efficient traveltime computation. 

There exist roughly two categories of numerical methods for traveltime 
computation: ray-based methods and eikonal-equation-based methods. Ray 
approaches are based on ray equations approximated from wave equations.  
A ray-tracing system is then solved in the framework of either one-point 
initial value problem or two-point boundary value problem using different 
techniques 
\cite[]{Pereyra_etal_1980,Grechka_McMechan_1996,Sadeghi_etal_1999,Melendez_etal_2015}.  
These approaches are generally very efficient for sparse source and 
receivers, but the computational costs may increase dramatically as the 
number of source-receiver pairs increases. In addition, ray tracing 
cannot trace rays to cover the entire model, and cannot handle complex 
media with strong contrasts. A more flexible ray tracing approach is the 
so-called the wavefront construction (WFC) method 
\cite[]{Vinje_etal_1993,Lambare_etal_1996,Gibson_etal_2005,Chambers_Kendall_2008}, 
which computes traveltime and amplitude fields in a more intuitive manner   
by dynamically inserting new rays where necessary. WFC generally requires 
more programming efforts and is more computationally demanding compared 
with conventional ray-based approaches. 

\cite{Vidale_1988} first developed an eikonal-equation-based approach to 
generating first-arrival traveltimes in heterogeneous media. His method 
is also known as the expanding-box method. Eikonal-equation-based 
approaches have since gained fast development and wide applications.  
Currently, the most widely used approaches include expanding wavefront 
methods \cite[]{Podvin_Lecomte_1991,Qin_etal_1992}, fast marching methods 
\cite[]{Sethian_Popovici_1999,Rawlinson_Sambridge_2004,Zhang_etal_2006}, 
fast sweeping methods 
\cite[]{Tsai_etal_2003,Zhao_2004,Kao_etal_2005,Fomel_etal_2009,Luo_Qian_2011,Waheed_etal_2015}, 
etc. Eikonal solvers on triangular or unstructured mesh 
\cite[e.g.,][]{Qian_etal_2007,Bouteiller_etal_2019} can handle complex 
interfaces and domain boundaries. Various high-order and non-oscillatory 
numerical schemes 
\cite[]{Kim_Cook_1999,Kim_2002,Luo_Qian_2011,Luo_etal_2012} can improve 
the numerical accuracy and stability of eikonal solvers in complex media 
with strong contrasts. Different eikonal solvers have various 
computational complexities and numerical accuracy. A comparison on 
several popular eikonal solvers can be found in \cite{Gomez_etal_2019}.  
One of the most distinct advantage of eikonal solvers compared with 
ray-based methods is that the output of an eikonal solver is 
a first-arrival traveltime field in the entire computational domain, as 
opposed to ray tracing methods that compute only traveltimes on ray 
paths. In addition, eikonal solvers generally allow models to be 
arbitrarily heterogeneous and complex, whereas ray-based methods usually 
require simple or smooth media.

Finite-difference eikonal solvers were first developed for isotropic 
media. There is an intensive need for eikonal solvers in anisotropic 
media. Eikonal equations for various kinds of anisotropic media are often 
significantly more complex than that in isotropic media.  Most of the 
eikonal solvers in isotropic media require substantial modifications for 
anisotropic media if possible. In anisotropic media, the phase and group 
velocity directions are generally not the same \cite[]{Carcione_2015}. 
Therefore, eikonal solvers for anisotropic media require sophisticated 
numerical schemes for updating traveltime fields.  \cite{Eaton_1993} 
developed a high-order expanding-wavefront method on a hexagonal mesh to 
compute qP-, qSV- or qSH-wave traveltime fields for transversely 
isotropic (TI) media. \cite{Qian_Symes_2002} developed a paraxial eikonal 
equation system to compute the qP-wave traveltime field in TI media.  
\cite{Wang_etal_2006} developed an unconditionally-stable expanding 
wavefront method for the eikonal equation in TI media.  Their method 
explicitly tracks group velocity propagation directions to ensure correct 
causal stencils.  \cite{Waheed_etal_2015_jcp} developed a method based on 
perturbation expansion to solve the TTI eikonal equation.  There exist 
several other methods for solving eikonal equations in anisotropic media 
by assuming weak, elliptical anisotropy \cite[]{Ettrich_Gajewski_1998}, 
by perturbing from elliptical reference medium 
\cite[]{Soukina_etal_2003}, or with only low-order numerical accuracy 
\cite[]{Lecomte_1993}. 

The source singularity is a major problem in various eikonal solvers.  
Conventional numerical schemes for solving eikonal equations are based on 
local plane wave assumption for traveltime field update, and therefore 
cannot accurately handle large curvatures of the traveltime field around 
the source point. The numerical error in the near-source region can 
eventually deteriorate the overall numerical accuracy in the entire 
computational domain. The source singularity issue is generally solved 
with the traveltime factorization method. The total traveltime field is 
factorized into an addition or multiplication of a base traveltime and an 
additional or multiplicative traveltime field, i.e., $t = t_0 + \tau$ or 
$t = t_0\times \tau$, where $t$ is the total traveltime field, $t_0$ is 
the base traveltime field, and $\tau$ is the additional or multiplicative 
traveltime field \cite[]{Luo_Qian_2011,Luo_Qian_2012,Luo_etal_2012}. The 
base traveltime field is solved in a homogeneous isotropic or 
elliptically anisotropic medium, while the traveltime field $\tau$ is 
computed in the heterogeneous part of the medium. A notable approach 
based on the traveltime factorization is an iterative scheme developed by 
\cite{Waheed_etal_2015} and \cite{Waheed_Alkhalifah_2017} to solve the 
eikonal equation in TTI media.  Their method first decomposes the 
left-hand side of the eikonal equation into a tilted elliptically 
anisotropic term and an additional term, and moves the additional term to 
the right-hand side of the eikonal equation.  During each iteration, 
their method updates the right-hand side term to approximate the true TTI 
eikonal equation.

We develop a hybrid numerical scheme based on both the monotonic Godunov 
scheme and the high-order weighted essentially non-oscillatory (WENO) scheme to solve the eikonal equation in 2D 
and 3D anisotropic media. Our hybrid eikonal solver has three advantages.  
First, our solver is free of source singularity issue, and there is no 
need to assign traveltime around the source by using multiplicative 
factorization of the traveltime field. Second, the computed traveltime 
field of our solver is high-order accurate in space by using the 
third-order Lax-Friedrichs WENO scheme. Third, our solver is numerically 
stable for 2D and 3D TTI media with strong anisotropy, highly complex 
structures, and strong medium property contrasts, by using the weighted 
non-oscillatory scheme.  Our eikonal solver employs both the conventional 
first-order Godunov scheme and the third-order Lax-Friedrichs scheme to 
achieve numerical stability and high-order accuracy. Therefore, we call 
our solver a hybrid approach. To our knowledge, our hybrid eikonal solver 
is the first fast-sweeping-based method to date that simultaneously holds 
these three advantages.

Our paper is organized as follows. In the Methodology section, we 
describe the three computational steps of our anisotropic eikonal solver, 
including the first-order Godunov locking-sweeping step, the base 
traveltime field computation step in the arbitrary TTI medium, and the 
third-order Lax-Friedrichs fast sweeping step. We then use several 
numerical examples to verify the stability and accuracy of our method. In 
the Conclusions section, we summarize the most important features of our 
new anisotropic eikonal solver. 

\section{Methodology}

We derive the formulation for our hybrid eikonal solver in 2D anisotropic 
media without loss of generality. We give the 3D formulation in Appendix 
A. 

We adopt the following eikonal equation in TTI media \cite[]{Waheed_etal_2015}:
\begin{equation}
v_x^2 \left(\frac{\partial t}{\partial \hat{x}}\right)^2 + v_z^2 \left(\frac{\partial t}{\partial \hat{z}}\right)^2 \left[1-2(\varepsilon - \delta) v_z^2 \left(\frac{\partial t}{\partial \hat{x}}\right)^2 \right] = 1, \label{eq:aniso_eikonal_2d}
\end{equation}
where $v_x (\mathbf{x}) = V_p(\mathbf{x}) 
\sqrt{1+2\varepsilon(\mathbf{x})}$ is the qP-wave velocity along the 
$x$-axis, $v_z(\mathbf{x}) = V_p(\mathbf{x})$ is the qP-wave velocity 
along the $z$-axis (i.e., the anisotropy symmetry axis), $\varepsilon 
= \varepsilon(\mathbf{x})$ and $\delta = \delta(\mathbf{x})$ are Thomsen 
parameters describing anisotropy properties of a VTI medium.  
Equation~\eqref{eq:aniso_eikonal_2d} is written in the rotated 
coordinates $\hat{x}-\hat{z}$, and the spatial derivatives are 
combinations of the spatial derivatives in the unrotated coordinates 
$x-z$:
\begin{align}
\frac{\partial t}{\partial \hat{x}} &= a_x \frac{\partial t}{\partial x}+ c_x \frac{\partial t}{\partial z}, \\
\frac{\partial t}{\partial \hat{z}} &= a_z \frac{\partial t}{\partial x}+ c_z \frac{\partial t}{\partial z},
\end{align}
with the coordinate transformation matrix $\mathbf{R}$ written as
\begin{equation}
\mathbf{R} = \begin{bmatrix}
a_x & c_x \\
a_z & c_z 
\end{bmatrix} = \begin{bmatrix}
\cos\theta & \sin\theta \\
-\sin\theta & \cos\theta
\end{bmatrix},
\end{equation}
where $\theta = \theta(\mathbf{x})$ is the tilt angle of a VTI medium's 
symmetry axis (i.e., the counterclockwise rotation angle of the symmetry 
axis with respect to the $y$-axis). 

For notation clarity and derivation convenience, we further define 
\begin{align}
\alpha_x &= v_x a_x = v_x \cos \theta, \\
\gamma_x &= v_x c_x = v_x \sin \theta, \\
\alpha_z &= v_z a_z = -v_z\sin \theta, \\
\gamma_z &= v_z c_z = v_z \cos \theta,
\end{align}
and 
\begin{equation}
\xi = \frac{2(\varepsilon - \delta)}{1+2\varepsilon},
\end{equation}
leading to
\begin{equation}
\left(\frac{\partial t}{\partial \hat{x}}\right)^2 + \left(\frac{\partial t}{\partial \hat{z}}\right)^2 \left[1-\xi \left(\frac{\partial t}{\partial \hat{x}}\right)^2 \right] = 1, \label{eq:aniso_eikonal_2d_simple}
\end{equation}
with 
\begin{align}
\frac{\partial t}{\partial \hat{x}} &= \alpha_x \frac{\partial t}{\partial x}+ \gamma_x \frac{\partial t}{\partial z}, \\
\frac{\partial t}{\partial \hat{z}} &= \alpha_z \frac{\partial 
t}{\partial x}+ \gamma_z \frac{\partial t}{\partial z}.
\end{align}
Generally,  $v_x \neq v_z$ and $\theta\neq 0$ in TTI media. 

\subsection{Step I: First-order Godunov fast locking-sweeping}

The first step of our hybrid eikonal solver is to solve 
equation~\eqref{eq:aniso_eikonal_2d_simple} using an iterative 
first-order fast sweeping method. The iterative approach is based on 
rewriting equation~\eqref{eq:aniso_eikonal_2d_simple} in the form of
\begin{equation}
\left[1 - \frac{1}{2}\xi \left(\frac{\partial t}{\partial \hat{z}}\right)^2\right]\left(\frac{\partial t}{\partial \hat{x}}\right)^2 + \left[1 - \frac{1}{2}\xi \left(\frac{\partial t}{\partial \hat{x}}\right)^2\right]\left(\frac{\partial t}{\partial \hat{z}}\right)^2  = 1. \label{eq:aniso_eikonal_2d_iterative}
\end{equation}
By setting two coefficients
\begin{align}
c_x = 1-\frac{1}{2}\xi\left(\frac{\partial t}{\partial \hat{z}}\right)^2, \label{eq:cx}\\
c_z = 1-\frac{1}{2}\xi\left(\frac{\partial t}{\partial \hat{x}}\right)^2, \label{eq:cz}
\end{align}
and absorbing them into $\partial t/\partial \hat{x}$ and $\partial t/ \partial \hat{z}$, we have
\begin{equation}
\left(\frac{\partial t}{\partial \tilde{x}}\right)^2 +\left(\frac{\partial t}{\partial \tilde{z}}\right)^2  = 1,  \label{eq:aniso_eikonal_2d_iterative2} 
\end{equation}
with 
\begin{align}
\frac{\partial t}{\partial \tilde{x}}& = \sqrt{|c_x|}\alpha_x \frac{\partial t}{\partial x}+ \sqrt{|c_x|}\gamma_x \frac{\partial t}{\partial z}, \\
\frac{\partial t}{\partial \tilde{z}}&= \sqrt{|c_z|}\alpha_z \frac{\partial t}{\partial x}+ \sqrt{|c_z|}\gamma_z \frac{\partial t}{\partial z}.
\end{align}
The left-hand side of equation~\eqref{eq:aniso_eikonal_2d_iterative2} is 
the eikonal correspondence in an elliptically transversely isotropic 
medium, and can be solved using the first-order Godunov scheme. 

We solve the eikonal equation~\eqref{eq:aniso_eikonal_2d_iterative2} 
using the following procedure. We first set $c_x (\mathbf{x}) 
= c_z(\mathbf{x}) = 1$, and solve 
equation~\eqref{eq:aniso_eikonal_2d_iterative2} using fast sweeping; Then 
we update $c_x(\mathbf{x})$ and $c_z(\mathbf{x})$ using the computed 
traveltime $t$ based on equations~\eqref{eq:cx} and \eqref{eq:cz}, and 
solve equation~\eqref{eq:aniso_eikonal_2d_iterative2} again. The reason 
for such a reformatting is that a Godunov scheme for 
equation~\eqref{eq:aniso_eikonal_2d_simple} can be fairly difficult to 
derive, and possibly leads to high computational complexity for local 
solvers. Instead, a first-order Godunov scheme for the left-hand side of 
equation~\eqref{eq:aniso_eikonal_2d_iterative} (or 
equation~\eqref{eq:aniso_eikonal_2d_iterative2}) is usually simple to 
derive \cite[e.g.,][]{Tsai_etal_2003}. 

We use a locking-sweeping procedure to reduce computational costs. At 
each fast sweeping iteration, we simply lock the points where 
$t(\mathbf{x})^{(m)} = t(\mathbf{x})^{(m-1)}$ where the superscript $(m)$ 
represents the $m$-th round of fast sweeping, and update traveltime 
values only at unlocked points. The criterion $t(\mathbf{x})^{(m)} 
= t(\mathbf{x})^{(m-1)}$ might not be as accurate as the one given in 
\cite{Bak_etal_2010} and \cite{Gomez_etal_2019} based on checking the 
changes of neighbor points of a certain spatial point, but is much more 
efficient to compute. In practice, we find that even for complex media, 
this criterion results in accurate traveltime fields. The 
locking-sweeping procedure can significantly reduce the computational 
costs for the first step. 

\subsection{Step II: Base traveltime field computation}

In Steps II and III, we express the traveltime field $t = t(\mathbf{x})$ in heterogeneous TTI media using a multiplicative factorization \cite[]{Luo_Qian_2012,Waheed_Alkhalifah_2017} as
\begin{equation}
t(\mathbf{x}) = t_0 (\mathbf{x}) \tau(\mathbf{x}),
\end{equation}
where $t_0(\mathbf{x})$ is the base traveltime field in the homogeneous 
TTI media, and $\tau(\mathbf{x})$ is the multiplicative traveltime field 
that accounts for heterogeneities of the model.

This multiplicative factorization results in
\begin{align}
\frac{\partial t}{\partial \hat{x}} &= \alpha_x \left(t_0 \frac{\partial \tau}{\partial x} + \frac{\partial t_0}{\partial x} \tau\right)+ \gamma_x  \left(t_0 \frac{\partial \tau}{\partial z} + \frac{\partial t_0}{\partial z} \tau\right), \label{eq:facx}\\
\frac{\partial t}{\partial \hat{z}} &= \alpha_z \left(t_0 \frac{\partial \tau}{\partial x} + \frac{\partial t_0}{\partial x} \tau\right)+ \gamma_z  \left(t_0 \frac{\partial \tau}{\partial z} + \frac{\partial t_0}{\partial z} \tau\right), \label{eq:facz}
\end{align}
which transform equation~\eqref{eq:aniso_eikonal_2d_simple} into 
a factorized eikonal equation for TTI media. 

For the factorized eikonal equation, the base traveltime field $t_0$ and 
its first-order spatial derivatives, say, $t_{0x} = \partial t_0/\partial 
x$ and $t_{0z} = \partial t_0/\partial z$, are computed analytically, and 
are fixed during fast-sweeping iterations 
\cite[]{Luo_Qian_2012,Waheed_Alkhalifah_2017}. The base traveltime field 
is essential to avoid the source singularity issue in 
eikonal-equation-based traveltime computation without specifying exact 
traveltime values at the points near the point source location. 

Existing techniques for solving the factorized isotropic or anisotropic 
eikonal equation assume that the background media is either isotropic 
where $\varepsilon = \delta = 0$ 
\cite[]{Fomel_etal_2009,Luo_Qian_2011,Luo_etal_2012} or elliptically 
anisotropic where $\varepsilon = \delta \neq 0$ 
\cite[]{Luo_Qian_2011,Luo_Qian_2012,Waheed_Alkhalifah_2017}, because 
there exist closed-form expressions for computing the traveltime and its 
spatial derivatives in the case of $\varepsilon = \delta$. 

However, the requirement of $\varepsilon=\delta$ also limits the 
application of multiplicative factorization in complex or strong 
anisotropic media. For instance, to apply multiplicative factorization to 
the case where $\varepsilon \neq \delta$, \cite{Waheed_Alkhalifah_2017} 
had to use an iterative scheme to update the right-hand side of the 
eikonal equation, and update the analytic $t_0$, $t_{0x}$ and $t_{0z}$ 
after several iterations. Even with such an iterative scheme, in each 
iteration, the background anisotropic medium is assumed to be 
elliptically anisotropic.  Therefore, the base traveltime field $t_0$ 
never truly approximates that in the anisotropic media where 
$\varepsilon\neq \delta$. For strongly anisotropic media, the group 
velocity curve can significantly deviate from an ellipse. The resulting 
base traveltime field might significantly differ from the true solution, 
eventually leading to suboptimal traveltime solutions, even though it 
helps avoid the source singularity issue. 

In contrast to existing techniques where $t_0$ is computed for an 
elliptically anisotropic medium, we develop a semi-analytic approach to 
directly computing $t_0$ for anelliptically anisotropic medium where 
$\varepsilon \neq \delta$. Therefore, the background anisotropic medium 
in our method is the non-degenerated TTI medium at the reference point. 
This is the most important difference between our hybrid method and 
existing factorized eikonal solvers. 

In the following derivations, to distinguish the rotation angle $\theta$ 
of the TTI symmetry axis in the following description, we use $\vartheta$ 
to represent the phase velocity angle in a VTI medium, which measures the 
deviation angle from the vertical symmetry axis of a VTI medium. The 
phase velocity of the qP-wave in a VTI medium can be written as 
\cite[]{Tsvankin_2012}
\begin{equation}
v_{\text{phase}}(\vartheta) = V_p \sqrt{1 + \varepsilon \sin^2 \vartheta - \frac{f}{2}\left[1- \sqrt{\left(1+ \frac{2\varepsilon\sin^2\vartheta}{f}\right)^2 - \frac{2(\varepsilon-\delta)\sin^2 2\vartheta}{f} }\right]}, 
\end{equation}
where $f = 1-V_s^2/V_p^2$, and $V_p$ and $V_s$ are qP- and qS-wave velocities along the symmetry axis, respectively. By setting $V_s = 0$, we have
\begin{equation}
v_{\text{phase}}(\vartheta) = V_p \sqrt{\frac{1}{2} + \varepsilon \sin^2 \vartheta + \sqrt{\left(1+ 2\varepsilon\sin^2\vartheta\right)^2 - 2(\varepsilon-\delta)\sin^2 2\vartheta }}. \label{eq:phase_vel}
\end{equation}
Note that there is no approximation in equation~\eqref{eq:phase_vel}, and 
therefore it is accurate even for strong anisotropy. We also have the 
group velocity angle $\psi$ in terms of the phase velocity angle 
$\vartheta$ as \cite[]{Tsvankin_2012}
\begin{equation}
\psi(\vartheta) = \arctan\dfrac{\tan\vartheta  + \dfrac{1}{v_{\text{phase}}(\vartheta)} \dfrac{d v_{\text{phase}}(\vartheta)}{d\vartheta}}{1 - \dfrac{\tan\vartheta}{v_{\text{phase}}(\vartheta)}\dfrac{d v_{\text{phase}}(\vartheta)}{d\vartheta}}, \label{eq:group_angle}
\end{equation}
and the magnitude of the group velocity in terms of the phase velocity 
angle $\vartheta$ as \begin{equation}
v_{\text{group}}(\vartheta) = v_{\text{phase}}(\vartheta) \sqrt{1 + \frac{1}{v_{\text{phase}}(\vartheta)} \frac{d v_{\text{phase}}(\vartheta)}{d\vartheta} }, \label{eq:group_vel}
\end{equation}
where based on equation~\eqref{eq:phase_vel}, we have
\begin{equation}
\frac{d v_{\text{phase}}(\vartheta)}{d\vartheta} = \frac{V_p^2  \sin 
2\vartheta}{2v_{\text{phase}}(\vartheta)} \left[\frac{ \varepsilon  
\left(1+2 \varepsilon  \sin ^2 \vartheta\right)-2 (\varepsilon -\delta 
) \cos 2\vartheta }{\sqrt{\left(1+ 2 \varepsilon  \sin ^2\vartheta 
\right)^2-2 (\varepsilon -\delta ) \sin ^2 2\vartheta}}+ \varepsilon 
\right].
\end{equation}

For an arbitrary spatial point $\mathbf{x} = (x,z)$ in the computational domain, the group velocity angle at this point can be evaluated as 
\begin{equation}
 \psi = \arctan\frac{x-x_0}{z - z_0}, \label{eq:group_angle_x}
\end{equation}
where $\mathbf{x}_0 = (x_0, z_0)$ is the position of the point source. 

Our goal is to compute the semi-analytic magnitude of the group velocity 
at this point, so that we can find the exact first-arrival traveltime at 
$\mathbf{x}$. This requires the determination of the corresponding phase 
velocity angle $\vartheta$ at $\mathbf{x}$, by which we can evaluate the 
magnitude of the group velocity at $\mathbf{x}$ using 
equation~\eqref{eq:group_vel}. Unfortunately, there is no closed-form 
expression to compute $\vartheta$ from $\psi$ based on 
equation~\eqref{eq:group_angle}, because equation~\eqref{eq:group_angle} 
is a complicated transcendental equation and is extremely difficult to 
solve analytically, if not impossible. 

We therefore adopt a numerical method to compute the magnitude of group 
velocity $v_{\text{group}}$ at $\mathbf{x}$. We first compute a series of 
group velocity values $v_{\text{group}}(\vartheta_1), 
v_{\text{group}}(\vartheta_2), \cdots, v_{\text{group}}(\vartheta_n)$, 
where $\vartheta_1, \vartheta_2, \cdots, \vartheta_n$ is an equal 
division of the phase angle range $[0, \frac{\pi}{2}]$. Meanwhile, we 
compute the corresponding group velocity angles $\psi(\vartheta_1), 
\psi(\vartheta_2), \cdots, \psi(\vartheta_n)$ based on these phase 
velocity angles. In any VTI medium, the group velocity angles 
$\psi(\vartheta_1), \psi(\vartheta_2), \cdots, \psi(\vartheta_n)$ range 
exactly from $[0, \frac{\pi}{2}]$, but are generally not equally 
distributed within this range. Therefore, we obtain a series of group 
angle-velocity pairs $\left\{\psi(\vartheta_i), 
v_{\text{group}}(\vartheta_i)\right\}$ with $i=1,2,\cdots, n$. We then 
use cubic spline interpolation to obtain the group velocity for the 
spatial location $\mathbf{x}$, which corresponds to a group velocity 
angle $\psi_{\mathbf{x}}$ based on equation~\eqref{eq:group_angle_x}. The 
interpolant function of this cubic spline interpolation is built from the 
group angle-velocity pairs $\left\{\psi(\vartheta_i), 
v_{\text{group}}(\vartheta_i)\right\}$.  In practical computations, we 
use a large $n$ to divide the range $[0, \pi/2]$, leading to high 
accuracy for cubic spline interpolation. We repeat the process until all 
the spatial points in the model are covered. 

Therefore, there is no need to analytically compute the phase velocity 
angle $\vartheta$ for the spatial point $\mathbf{x}$ in our numerical 
scheme. The group velocity value at any spatial point is obtained through 
an 1D interpolation process with smooth and continuous interpolants built 
from the analytic group angle-velocity pairs. As a result, the computed 
group velocities are practically of analytic accuracy. The computational 
cost associated with this part is small in the entire eikonal equation 
solving process. 

Because for any VTI medium, the phase or group velocity is symmetric with 
respect to both axes, it is sufficient to build a complete group velocity 
profile for the entire $2\pi$ range based on the computed group velocity 
values in $[0,\pi/2]$. In addition, because any TTI medium is simply 
a coordinate rotation result of some VTI medium, the group velocity 
values for the TTI medium can be easily computed using the scheme 
described above. Assume that the tilt angle of a TTI medium is $\theta$, 
then for a normalized spatial location $\mathbf{x} =(x-x_0,z-z_0)$ where 
$(x_0,z_0)$ is the source location, the corresponding directional vector 
in the unrotated coordinate is
\begin{equation}
\mathbf{x}' = \begin{bmatrix}
x' \\
z'
\end{bmatrix} = \begin{bmatrix}
\cos \theta & \sin \theta \\
-\sin \theta & \cos \theta 
\end{bmatrix}\begin{bmatrix}
x - x_0 \\
z - z_0,
\end{bmatrix}
\end{equation}
which indicates the group velocity angle corresponding to $\mathbf{x}$ should be 
\begin{equation}
\psi' = \arctan\frac{\left|x'\right|}{\left|z'\right|} \label{eq:gangle}.
\end{equation}
We take absolute value in equation~\eqref{eq:gangle} to ensure that the 
angle $\psi'$ falls in $[0, \pi/2]$. 

We then compute the group velocity value $v_{\text{group}}|_{\psi'}$ for $\mathbf{x}$ by the aforementioned interpolation procedure at $\psi'$, and the traveltime at $\mathbf{x}$ is
\begin{equation}
t_0 (\mathbf{x}) = \frac{\sqrt{(x-x_0)^2 + (z-z_0)^2}}{v_{\text{group}}|_{\psi'}}.
\end{equation}

Finally, we use a high-order centered finite-difference scheme to compute the spatial derivatives of the base traveltime field $t_{0x}$ and $t_{0z}$: 
\begin{align}
t_{0x}(i,j) & = \frac{1}{\Delta x} \sum_{l=1}^{M} c_l [t_0(i+l,j) - t_0(i-l,j)], \\
t_{0z}(i,j) & = \frac{1}{\Delta z} \sum_{l=1}^{M} c_l [t_0(i,j+l) - t_0(i,j-l)],
\end{align}
where $c_l$ are finite-difference coefficients, $M$ is the half length of 
the finite-difference operator \cite[]{Fornberg_1988}, and $\Delta x$ and 
$\Delta z$ are the grid sizes in the $x$- and $z$-directions, 
respectively. In our computation, we use $M=10$, and compute the 
associate coefficients $c_l$ using the procedure described in 
\citet[]{Fornberg_1988}. Again, these fields are practically of analytic 
accuracy because $t_0$ is practically of analytic accuracy and $M$ is 
large. 

\begin{figure}
	\centering
\subfloat[]{\includegraphics[width=0.48\textwidth]{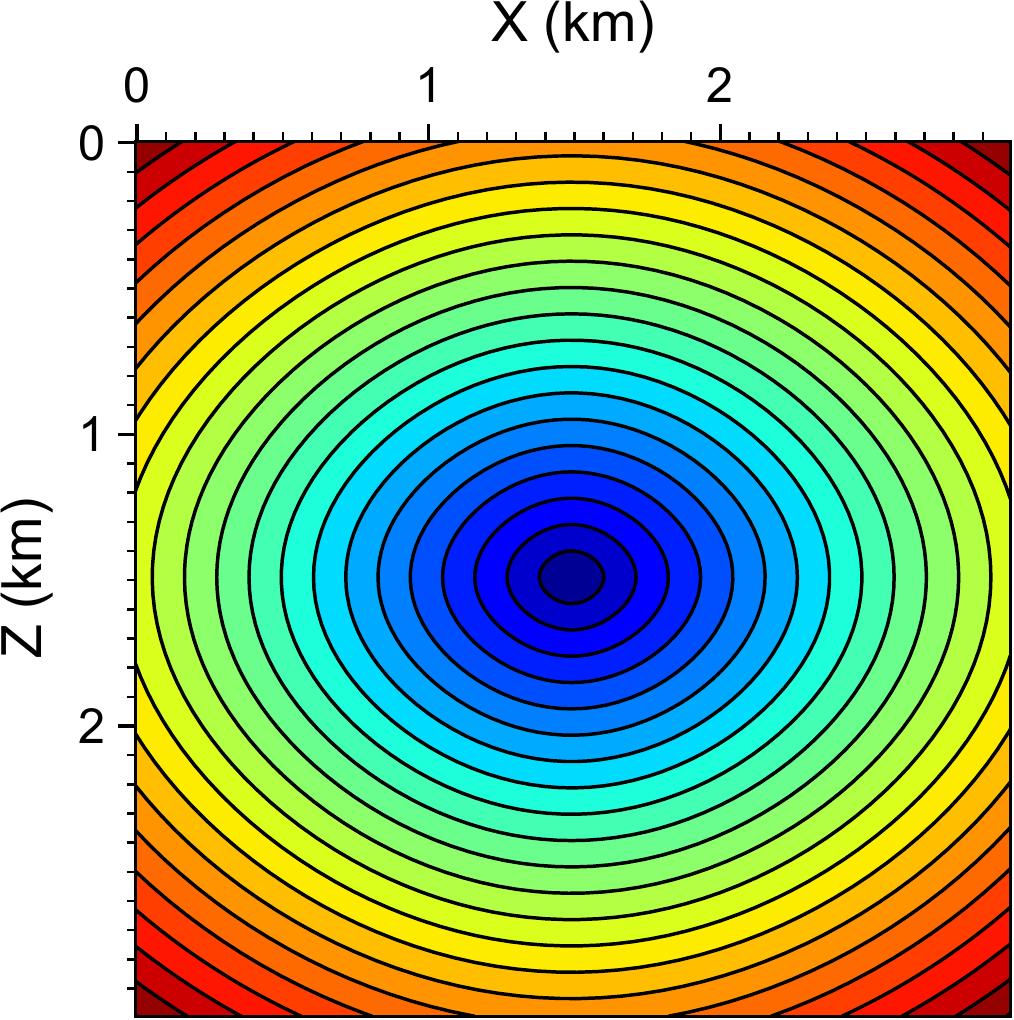}} \quad
\subfloat[]{\includegraphics[width=0.48\textwidth]{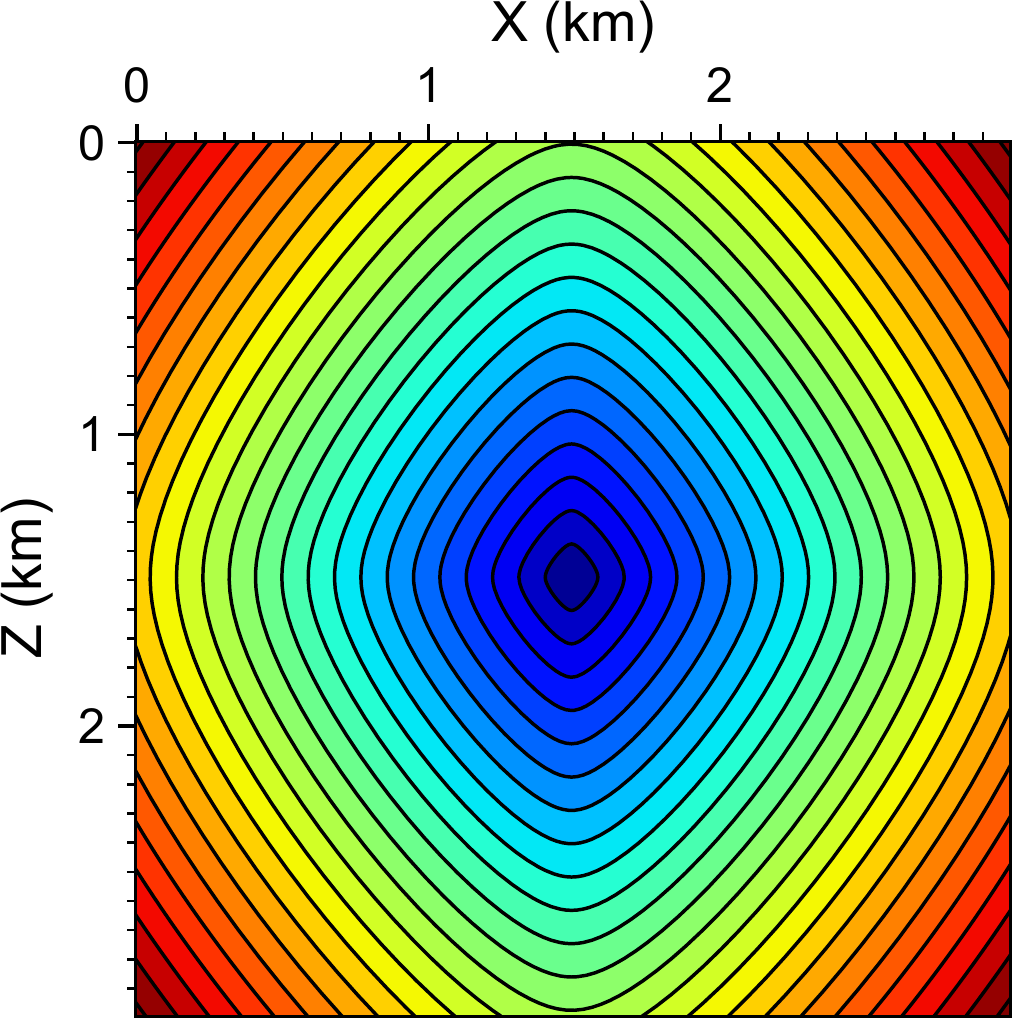}} \\ 
\subfloat[]{\includegraphics[width=0.48\textwidth]{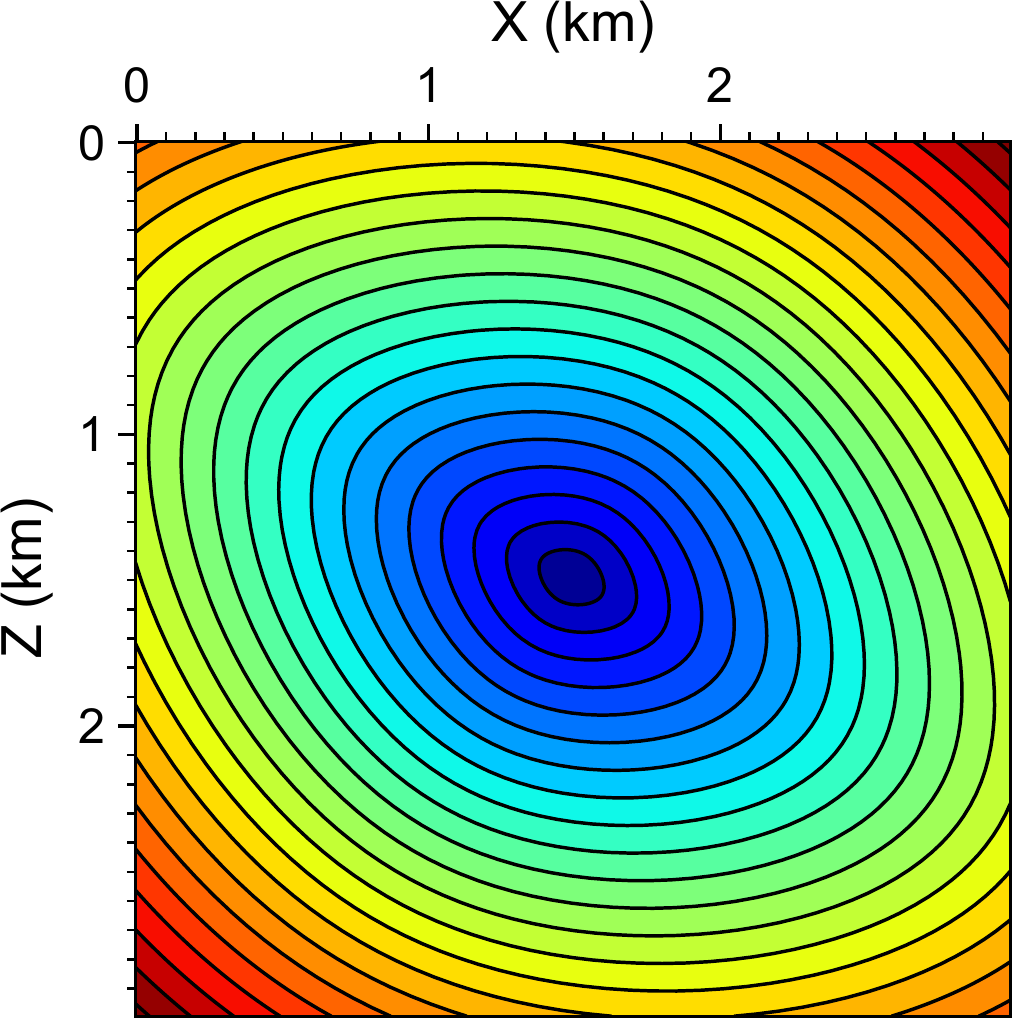}} \quad
\subfloat[]{\includegraphics[width=0.48\textwidth]{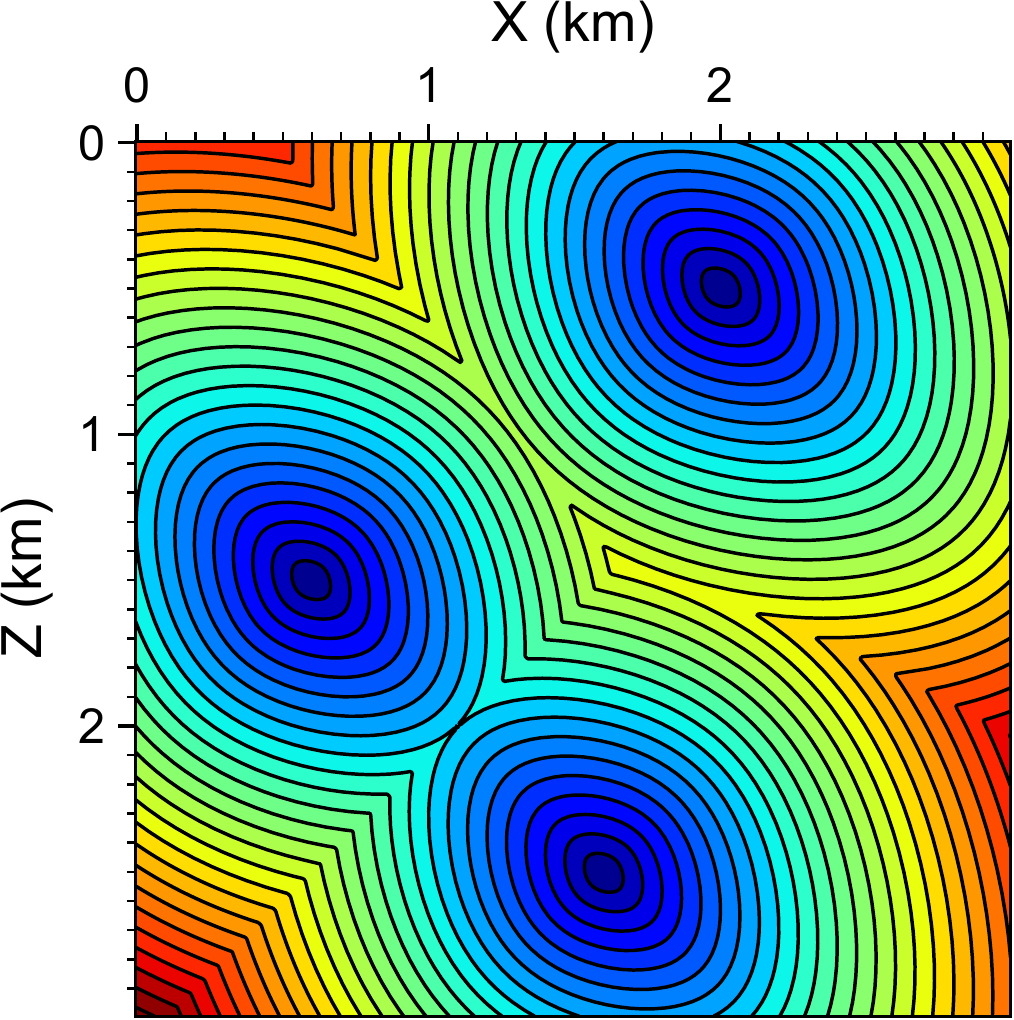}}
\caption{Base traveltime fields computed using the semi-analytic approach 
for (a) a VTI medium with $V_p = 2000$~m/s, $\varepsilon = 0.25$, 
$\delta=0.1$, $\theta = 0$, (b) a HTI medium with $V_p = 2000$~m/s, 
$\varepsilon = 0.3$, $\delta=-0.3$, $\theta = \pi/2$, (c) a TTI medium 
with $V_p = 2000$~m/s, $\varepsilon = 0.4$, $\delta=0.1$, $\theta 
= \pi/6$ for single point sources, and (d) $V_p = 2000$~m/s, $\varepsilon 
= 0.35$, $\delta=0.05$, $\theta = \pi/4$ for three random point sources.  
In all the panels, the interval between any two adjacent contours is 
0.03~s. The models are homogeneous in all the panels. }
\label{fig:base2d}
\end{figure}

\begin{figure}
	\centering
	\includegraphics[width=0.65\textwidth]{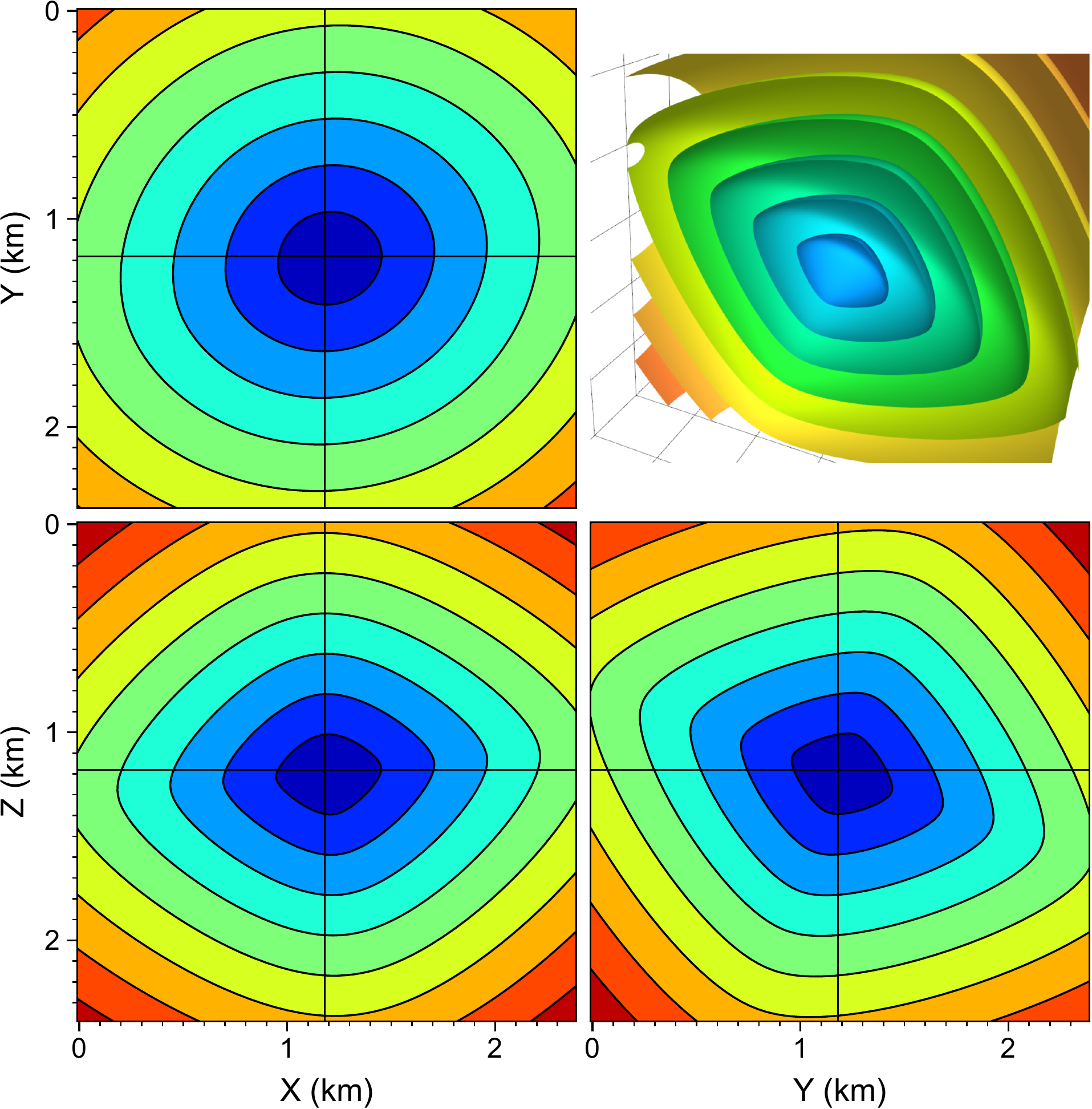}
        \caption{Base traveltime fields computed using the semi-analytic 
        approach for a 3D TTI medium with $V_p = 2000$~m/s, $\varepsilon 
        = 0.3$, $\delta=-0.3$, $\theta = \pi/9$ and $\phi=5\pi/12$ for 
        a point source. The interval between any two adjacent contours or 
     isosurfaces is 0.1~s.}
	\label{fig:base3d}
\end{figure}

In Figure~\ref{fig:base2d}, we show three examples for base traveltime 
computation in anelliptically anisotropic TI media, including a VTI 
medium (Figure~\ref{fig:base2d}a), a HTI medium 
(Figure~\ref{fig:base2d}b), and a TTI media (Figure~\ref{fig:base2d}c), 
all containing strong anisotropies. All group velocity curves 
significantly deviate from an ellipse. The results indicate that our 
semi-analytic approach can accurately compute the base traveltime fields 
in complex anisotropic media. The numerical scheme to compute the base 
traveltime field in arbitrary TTI media is also applicable to 3D scenario 
as shown in Figure~\ref{fig:base3d}. 

When a model contains multiple simultaneous sources, we need to compute 
the base traveltime fields for all the point sources, and compute 
a minimum base traveltime field by finding the minimal value from all the 
base traveltime field values at each point.  That is, \begin{equation}
t_0(\mathbf{x}) = \min \left\{t_0^{s_1}(\mathbf{x}), t_0^{s_2}(\mathbf{x}), \cdots, t_0^{s_n}(\mathbf{x})\right\},
\end{equation}
where $t_0^{s_i}$ represents the base traveltime field for the $i$-th 
point source. The medium properties at different point source locations 
can be different. Figure~\ref{fig:base2d}d shows a simple example of base 
traveltime field computation with multiple simultaneous point sources in 
a 2D TTI medium.  

The above procedure also implies that, in homogeneous media, we can 
obtain the semi-analytic solution to the eikonal equation regardless of 
TTI anisotropy type and the number of simultaneous point sources, without 
the need of the aforementioned Step I or the Step III described below. 

\subsection{Step III: Third-order Lax-Friedrichs fast sweeping}

For heterogeneous media, once the initial traveltime filed $t$, the base 
traveltime field $t_0$ and its derivatives are computed, we solve the 
factorized eikonal equation~\eqref{eq:aniso_eikonal_2d_simple} along with 
equations~\eqref{eq:facx} and \eqref{eq:facz} in TTI media using the 
Lax-Friedrichs scheme based on a third-order WENO discretization 
\cite[]{Zhang_etal_2006}. The initial multiplicative traveltime field 
$\tau$ is computed as $\tau(\mathbf{x}) = t(\mathbf{x})/t_0(\mathbf{x})$, 
with $\tau(x_0, z_0) \equiv 1$. The value of $\tau$ at location $(x_0, 
z_0)$ is kept unchanged during iterations. For multiple simultaneous 
point source applications, the values of $\tau$ at all the point source 
locations are kept unchanged during iterations. In contrast to existing 
high-order schemes where neighbor points of the source point should be 
assigned and kept unchanged during iterations, our algorithm requires 
only the value at the source point fixed during iterations thanks to the 
use of the multiplicative factorization. 

To facilitate our description, we define two functionals associated with the multiplicative time field $\tau$ and its spatial derivatives $\tau_x$ and $\tau_z$:
\begin{align}
F_x(\tau, \tau_x, \tau_z) &= \alpha_x \left(t_0 \tau_x + t_{0x} \tau \right) + \gamma_x \left(t_0 \tau_z + t_{0z} \tau\right), \\
F_z(\tau, \tau_x, \tau_z) &= \alpha_z \left(t_0 \tau_x + t_{0x} \tau\right) + \gamma_z \left(t_0 \tau_z + t_{0z} \tau\right). 
\end{align}
Then the Hamiltonian for equation~\eqref{eq:aniso_eikonal_2d_simple} can 
be written as
\begin{equation}
H(\tau, \tau_x, \tau_z) = F_x^2 (\tau, \tau_x, \tau_z) + F_z^2(\tau, \tau_x, \tau_z) - \xi F_x^2(\tau, \tau_x, \tau_z) F_z^2(t, \tau_x, \tau_z). \label{eq:hamiltonian}
\end{equation}

In the Lax-Friedrichs update scheme, it is necessary to compute a set of 
artificial viscosities. To improve numerical stability, we use the 
following artificial viscosities:
\begin{align}
\omega_x &= \max\limits_{\Omega} \left(\left|\frac{\partial H}{\partial \tau}\right| + \left|\frac{\partial H}{\partial \tau_x}\right|\right), \\
\omega_z &= \max\limits_{\Omega} \left(\left|\frac{\partial H}{\partial \tau}\right| + \left|\frac{\partial H}{\partial \tau_z}\right|\right), 
\end{align}
where $\Omega$ represents the entire computational domain. Based on 
equation~\eqref{eq:hamiltonian}, we have
\begin{align}
\frac{\partial H}{\partial \tau} & = 2\left[(1-\xi F_z^2) \frac{\partial F_x}{\partial \tau} F_x + (1-\xi F_x^2) \frac{\partial F_z}{\partial \tau} F_z\right] \nonumber \\
& = 2\left[(1-\xi F_z^2) (\alpha_x t_{0x} + \gamma_x t_{0z}) F_x + (1-\xi F_x^2) (\alpha_z t_{0x} + \gamma_z t_{0z}) F_z\right], \\
\frac{\partial H}{\partial \tau_x} & = 2\left[(1-\xi F_z^2) \frac{\partial F_x}{\partial \tau_x} F_x + (1-\xi F_x^2) \frac{\partial F_z}{\partial \tau_x} F_z\right] \nonumber \\
& = 2t_0\left[(1-\xi F_z^2) \alpha_x F_x + (1-\xi F_x^2) \alpha_z F_z\right], \\
\frac{\partial H}{\partial \tau_z} & =2\left[(1-\xi F_z^2) \frac{\partial F_x}{\partial \tau_z} F_x + (1-\xi F_x^2) \frac{\partial F_z}{\partial \tau_z} F_z\right] \nonumber \\
& = 2t_0\left[(1-\xi F_z^2) \gamma_x F_x + (1-\xi F_x^2) \gamma_z F_z\right]. 
\end{align}
In our algorithm, we compute the quantities $\left|\partial H/\partial 
t\right|$,  $\left|\partial H/\partial \tau_x\right|$ and $\left|\partial 
H/\partial \tau_z\right|$ at every spatial points of the model, and find 
$\omega_x$ and $\omega_z$ based on the maximum values of these 
quantities. Note that we do not adopt the artificial viscosities defined 
by \cite{Luo_Qian_2012} and \cite{Luo_etal_2012}, which may lead to 
unstable results in complex anisotropic media with large medium property 
contrasts.

We then obtain the following Lax-Friedrichs scheme to update the multiplicative traveltime field $\tau$ at the spatial grid point $(i,j)$:
\begin{align}
\tau_{i,j}^{(m+1)} = \dfrac{1 - H\left(\tau_{i,j}^{(m)}, \tau_{0x; i,j}^{(m), \dagger}, \tau_{0z; i,j}^{(m),\dagger}\right) + \omega_x \tau_{0x; i,j}^{(m), *} + \omega_z \tau_{0z; i,j}^{(m), *}}{\omega_x/\Delta x + \omega_z /\Delta z} + \tau_{i,j}^{(m)}, 
\end{align}
where the superscripts $(m)$ and $(m+1)$ represent the values of 
$\tau_{i,j}$ at the $m^{\text{th}}$ and $(m+1)^{\text{th}}$ sweeping 
iterations, respectively, and according to \cite{Zhang_etal_2006},
\begin{align}
\tau_{0x; i,j}^{\dagger} &  = \frac{1}{2} \left(\tau_{0x; i,j}^+ + \tau_{0x; i,j}^-\right), \\
\tau_{0x; i,j}^{*} &  = \frac{1}{2} \left(\tau_{0x; i,j}^+ - \tau_{0x; i,j}^-\right), 
\end{align}
with the third-order WENO discretizations 
\begin{align}
\tau_{0x; i,j}^+ & = (1-w_x^+) \frac{\tau_{i+1,j} - \tau_{i-1,j}}{2\Delta x} + w_x^+ \frac{-\tau_{i+2,j} + 4\tau_{i+1,j} - 3\tau_{i,j}}{2\Delta x}, \\
\tau_{0x; i,j}^- & = (1-w_x^-) \frac{\tau_{i+1,j} - \tau_{i-1,j}}{2\Delta x} - w_x^- \frac{-\tau_{i-2,j} + 4\tau_{i-1,j} - 3\tau_{i,j}}{2\Delta x}, \\
w_x^+ & = \left\{1+2\left[\frac{\epsilon + (\tau_{i+2,j} - 2\tau_{i+1,j} + \tau_{i,j})^2}{\epsilon + (\tau_{i+1,j} - 2\tau_{i,j} + \tau_{i-1,j})^2}\right]^2\right\}^{-1}, \\
w_x^- & = \left\{1+2\left[\frac{\epsilon + (\tau_{i-2,j} - 2\tau_{i-1,j} + \tau_{i,j})^2}{\epsilon + (\tau_{i+1,j} - 2\tau_{i,j} + \tau_{i-1,j})^2}\right]^2\right\}^{-1}, 
\end{align} 
where $\epsilon$ is a small number to avoid singularity. The expressions for $\tau_{0z; i,j}^{\dagger}$ and $\tau_{0z; i,j}^{*}$ can be analogously derived. 

Because we actually use the first-order traveltime field computed at Step 
I as the initial solution for Step III, Step III in our algorithm 
requires much fewer iterations to achieve accurate results compared with 
the LF-3 method that directly solves the eikonal equation from a rough or 
constant initial guess. 

\subsection{Workflow of our hybrid eikonal solver}
We summarize the workflow of our hybrid eikonal solver as follows:
\begin{enumerate}
\item Compute an initial, first-order accurate traveltime field 
   $t(\mathbf{x})$ using the first-order Godunov algorithm based on the 
   locking-sweeping procedure in the following order:
\begin{align}
& i = 1, \cdots, N_x, \quad j = 1, \cdots, N_z,  \label{eq:sweep1}\\
& i = N_x, \cdots, 1, \quad j = 1, \cdots, N_z, \\
& i = 1, \cdots, N_x, \quad j = N_z, \cdots, 1, \\
& i = N_x, \cdots, 1, \quad j = N_z, \cdots, 1, \label{eq:sweep4}
\end{align}
where $i$ and $j$ are indices of the finite-difference grids, and $N_x$ and $N_z$ are the number of grids in the model in the $x$- and $z$-directions, respectively. The order of fast-sweeping is not important. 

\item Compute the base traveltime field $t_0(\mathbf{x})$ and its spatial derivatives using the semi-analytic approach described in Step II. 

\item Compute the multiplicative traveltime field $\tau(\mathbf{x})$ 
   using the numerical scheme described in Step III. The sweeping follows 
   the order listed in equations~\eqref{eq:sweep1}-\eqref{eq:sweep4}.

\end{enumerate} 

\section{Numerical Results}

We use five numerical examples to verify the stability and accuracy of 
our hybrid eikonal solver for 2D and 3D TTI anisotropic media. We compare  
the results from four different methods:
\begin{enumerate}
\item Godunov: the first-order Godunov method based on the iterative 
   scheme developed in \cite{Waheed_etal_2015}.
\item Factorized Godunov: the first-order Godunov method based on the 
iterative scheme for the factorized eikonal equation developed in 
\cite{Waheed_Alkhalifah_2017}.
\item LF-3: the third-order direct Lax-Friedrichs method without 
traveltime factorization developed in \cite{Luo_Qian_2012}.
\item Hybrid: our hybrid eikonal solver in this paper.
\end{enumerate}

In our tests, we adapt and program all methods to solve the eikonal 
equation~\eqref{eq:aniso_eikonal_2d_simple}. The meaning of ``iterative'' 
in the first-two approaches is that we need to update the right-hand side 
of the degenerated TTI eikonal 
equation~\eqref{eq:aniso_eikonal_2d_iterative} during sweeping 
iterations. The meaning of ``factorized'' in the second approach is that 
we use the traveltime field factorization scheme. The meaning of 
``direct'' in the Lax-Friedrichs approach is that we directly discretize 
equation~\eqref{eq:aniso_eikonal_2d_simple} based on the Lax-Friedrichs 
update scheme and the third-order WENO finite-difference scheme, without 
any degeneration or right-hand-side iteration as in the first two 
approaches. 

In all the implementations, we do not assign exact values for the points 
around the source to study the efficacy of these methods in realistic 
computational tasks. In practical applications, the media around the 
source can be heterogeneous, where assigning exact traveltime values can 
be very difficult, if not impossible, particularly for heterogeneous, 
anelliptically anisotropic media. In all the implementations, we only fix 
the traveltime value at the source point (i.e., where $t=0$) over 
sweeping iterations. Traveltime field values at all other spatial points 
can change during iterations. 

Note that without accurate and fixed values around the point source, it 
can be very difficult to achieve convergence for LF-3. Therefore, we use 
the result from the Godunov method as the initial guess for LF-3.

\subsection{Homogeneous model}

Since we are able to obtain traveltime field with analytic accuracy 
solely using the algorithm in Step II, traveltime computation in 
homogeneous TTI media is trivial for our hybrid method. However, it can 
still be challenging for conventional eikonal solvers. In the first 
numerical example, we first compare the traveltime field computed using 
our method with those obtained using three conventional methods. 

\begin{figure}
	\centering
	\subfloat[]{\includegraphics[width=0.7\textwidth]{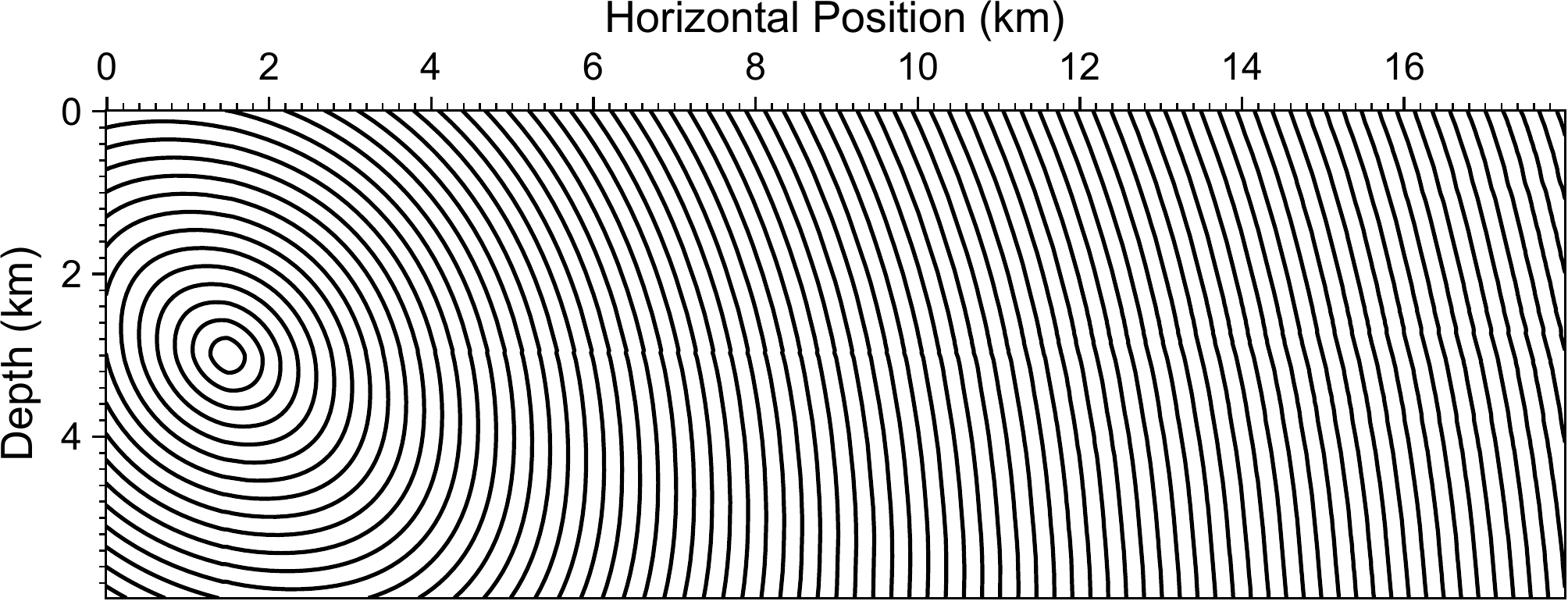}} \\
	\subfloat[]{\includegraphics[width=0.7\textwidth]{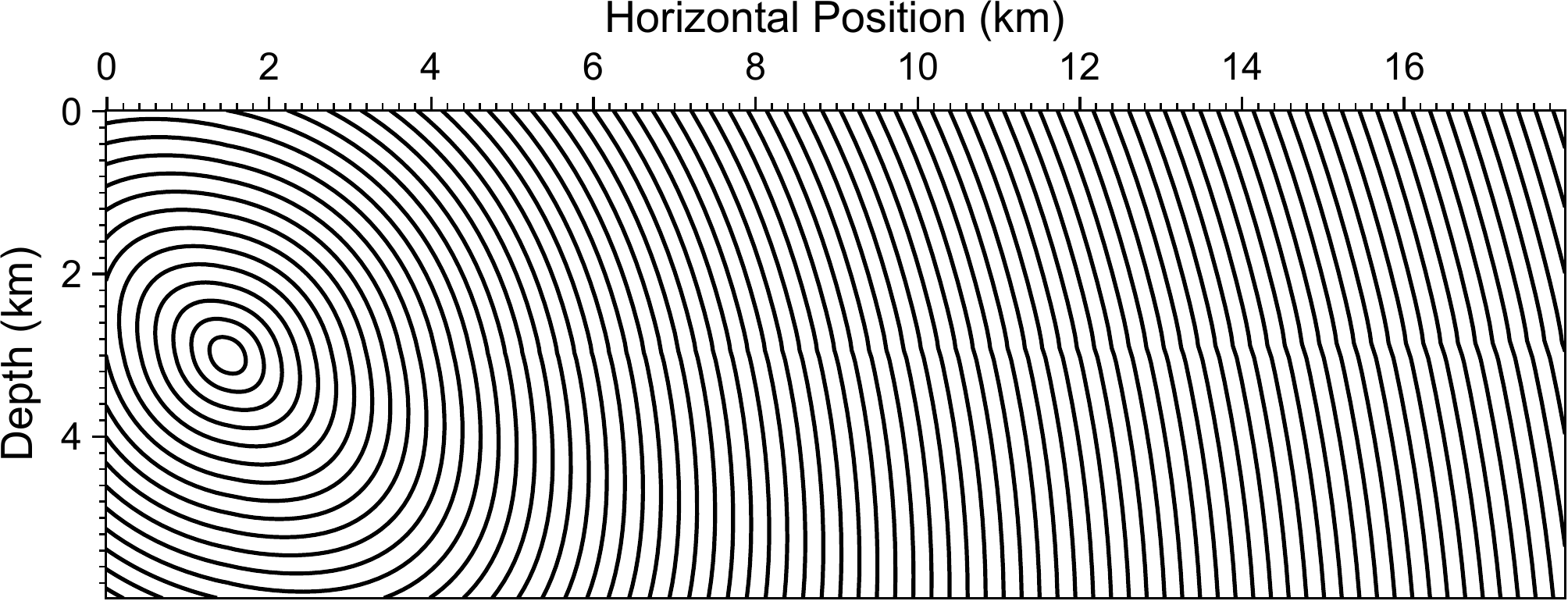}} \\
	\subfloat[]{\includegraphics[width=0.7\textwidth]{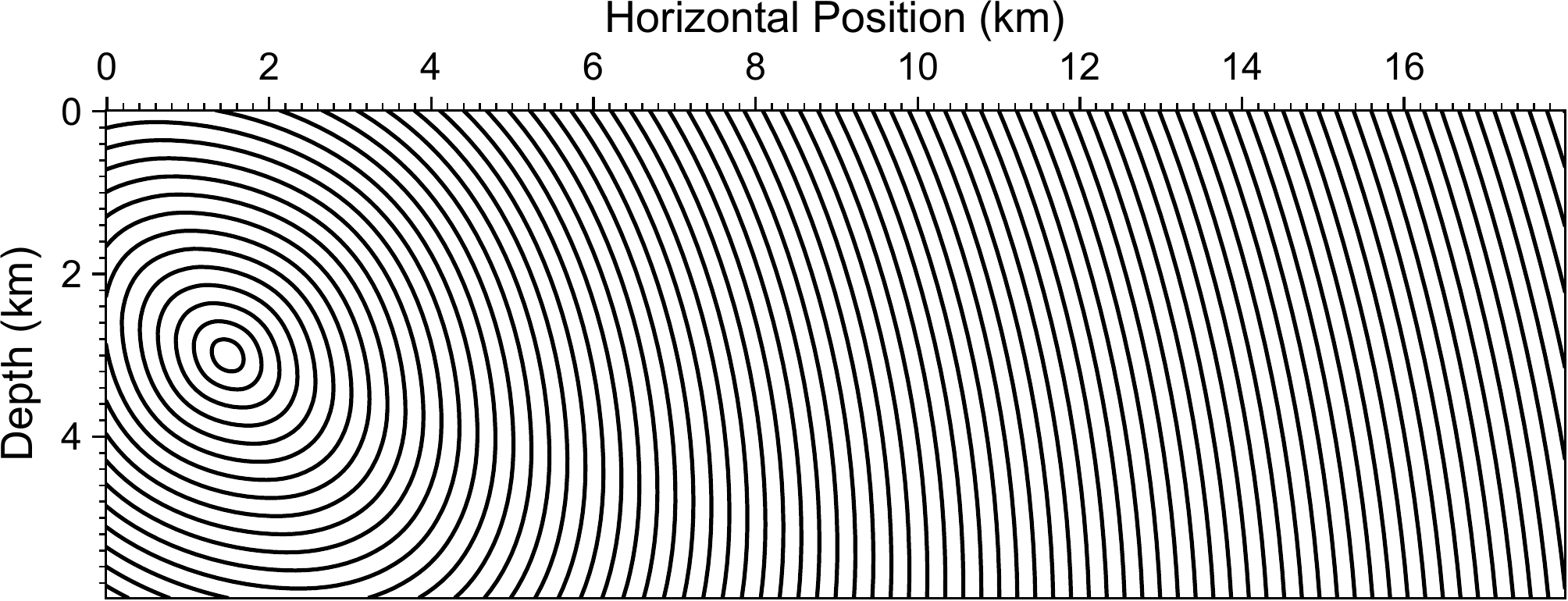}}\\
	\subfloat[]{\includegraphics[width=0.7\textwidth]{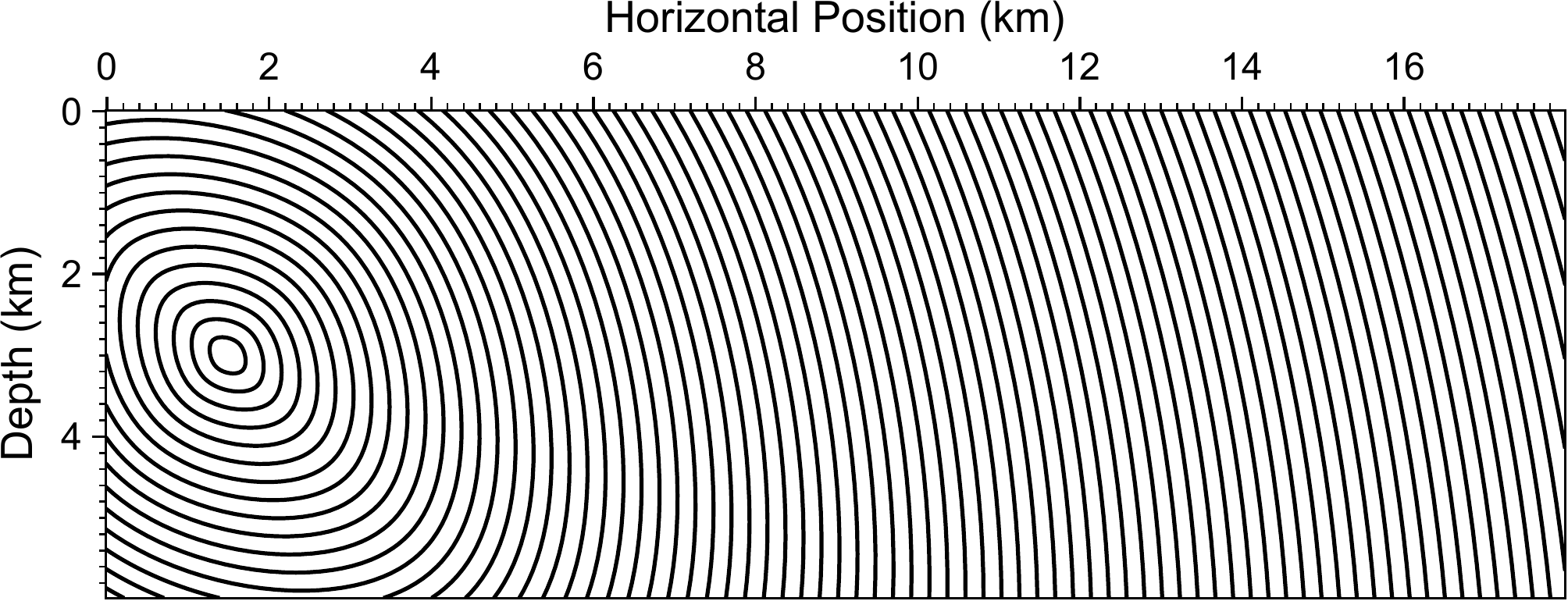}}
        \caption{Traveltime fields in the homogeneous TTI model computed 
        using (a) Godunov method, (b) factorized Godunov method, (c) LF-3 
     method, and (d) our hybrid method. The interval between any two 
  adjacent contours in all panels is 0.1~s. }
	\label{fig:tt_homo}
\end{figure}

The model parameters for a homogeneous TTI medium are $V_p = 2000$~m/s, 
Thomsen parameters $\varepsilon=0.25$ and $\delta=0.05$, and TTI symmetry 
axis tilt angle $\theta = \pi/4$.  Figures~\ref{fig:tt_homo}a-d show the 
traveltime fields in the homogeneous TTI model computed using Godunov, 
factorized Godunov, LF-3 and our hybrid method, respectively. The 
solution computed using our method is taken as the reference solution. 

\begin{figure}
	\centering
	\subfloat[]{\includegraphics[width=0.8\textwidth]{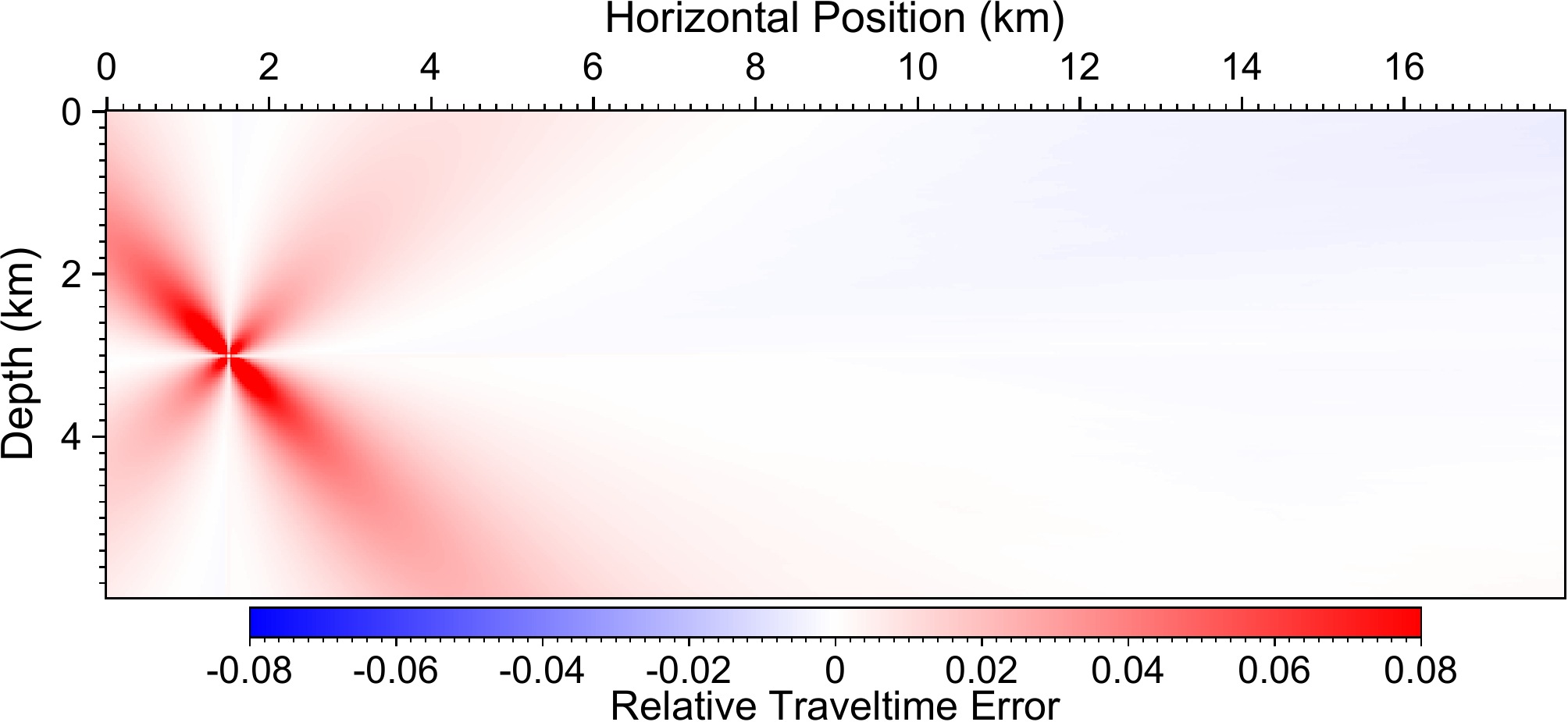}} \\
	\subfloat[]{\includegraphics[width=0.8\textwidth]{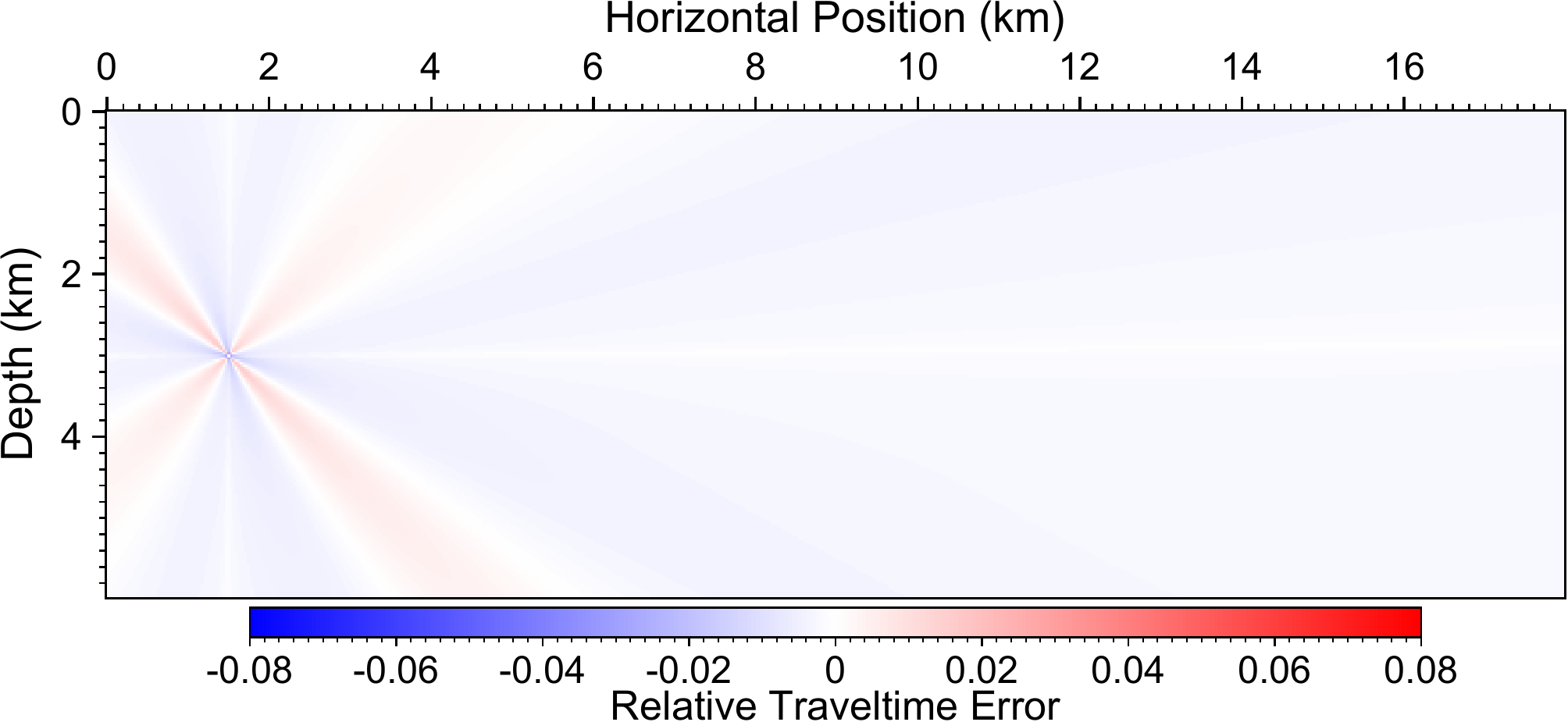}} \\
	\subfloat[]{\includegraphics[width=0.8\textwidth]{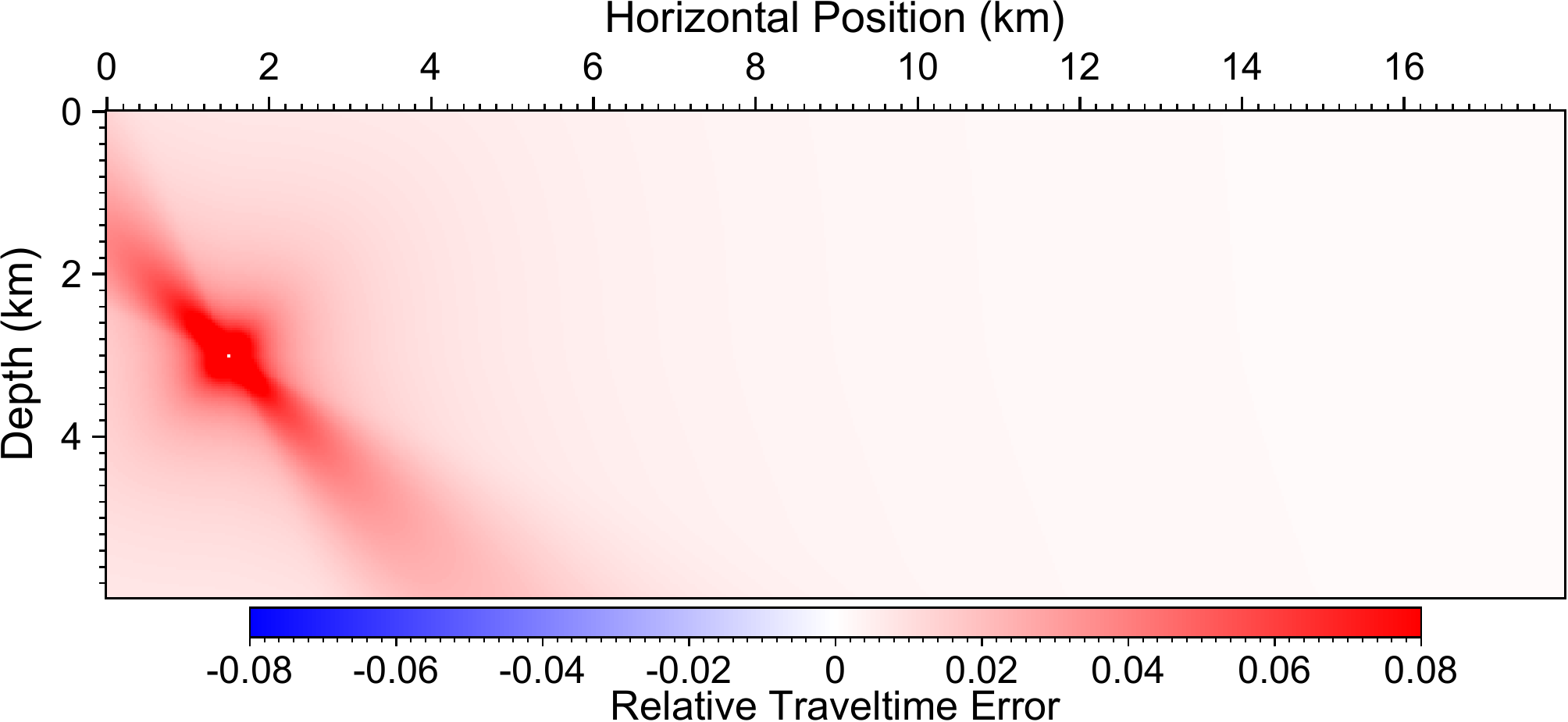}} 
	\caption{Difference between the traveltime field computed using our hybrid method (the semi-analytic solution) and the traveltime field computed using (a) Godunov method, (b) factorized Godunov method, and (c) LF-3 method.}
	\label{fig:tt_homo_diff}
\end{figure}

Visually the solutions from different methods are close to one another, 
except the near-source traveltime field contours in 
Figure~\ref{fig:tt_homo}a (Godunov method), which clearly deviate from 
the reference traveltime contours shown in Figure~\ref{fig:tt_homo}d (our 
hybrid method). We compute the differences between the traveltime fields 
obtained using the three conventional methods and the reference solution 
(Figures~\ref{fig:tt_homo_diff}a-c). Different conventional methods have 
different error levels. The result of the LF-3 method is the least 
accurate, partially because we do not assign exact values in the 
near-source region for this method. The two Godunov methods have higher 
accuracy compared with the LF-3 method. Nevertheless, obvious errors 
occur at the near-source region in both solutions, and the errors become 
larger with the increased distance away from the source position.  

\subsection{Gradient model}

In the second example, we study the convergence of our hybrid method 
using an isotropic constant gradient model. We choose this model because 
we can analytically compute the first-arrival traveltime in such 
a medium.

The model is 4~km in both spatial directions, with a constant gradient of 
velocity
\begin{equation}
	\frac{1}{s(\mathbf{x})} = \frac{1}{s_0} + \mathbf{G}_0 \cdot (\mathbf{x} - \mathbf{x}_0),
\end{equation}
where $s(\mathbf{x})$ is the spatially variant slowness, $s_0$ is the 
slowness at the source point $\mathbf{x}_0$, and $\mathbf{G}_0$ is the 
constant gradient of the model. The analytical first-arrival traveltime 
for this medium is \cite[]{Fomel_etal_2009}:
\begin{equation}
t(\mathbf{x}) = \frac{1}{|\mathbf{G}_0|} \mathrm{arccosh} \left(1 + \frac{1}{2}s(\mathbf{x})s_0|\mathbf{G}_0|^2|\mathbf{x} - \mathbf{x}_0|^2\right),  
\label{eq:tt_analytical}
\end{equation}
with
\begin{equation}
\mathrm{arccosh}(x) = \ln\left(x + \sqrt{x^2 - 1}\right).
\end{equation}

\begin{figure}
\centering
	\subfloat[]{\includegraphics[height=0.3\textheight]{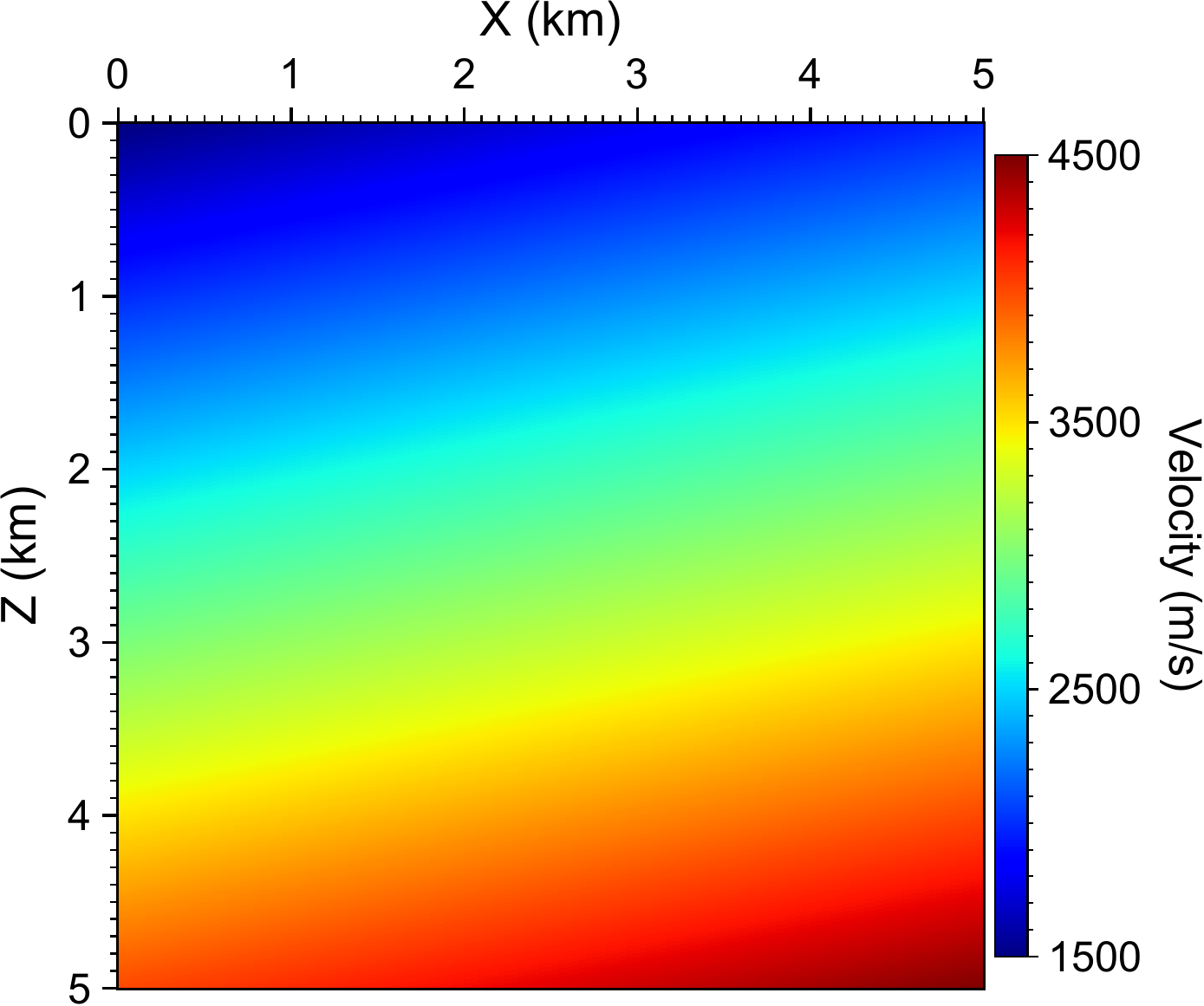}} \quad
\subfloat[]{\includegraphics[height=0.3\textheight]{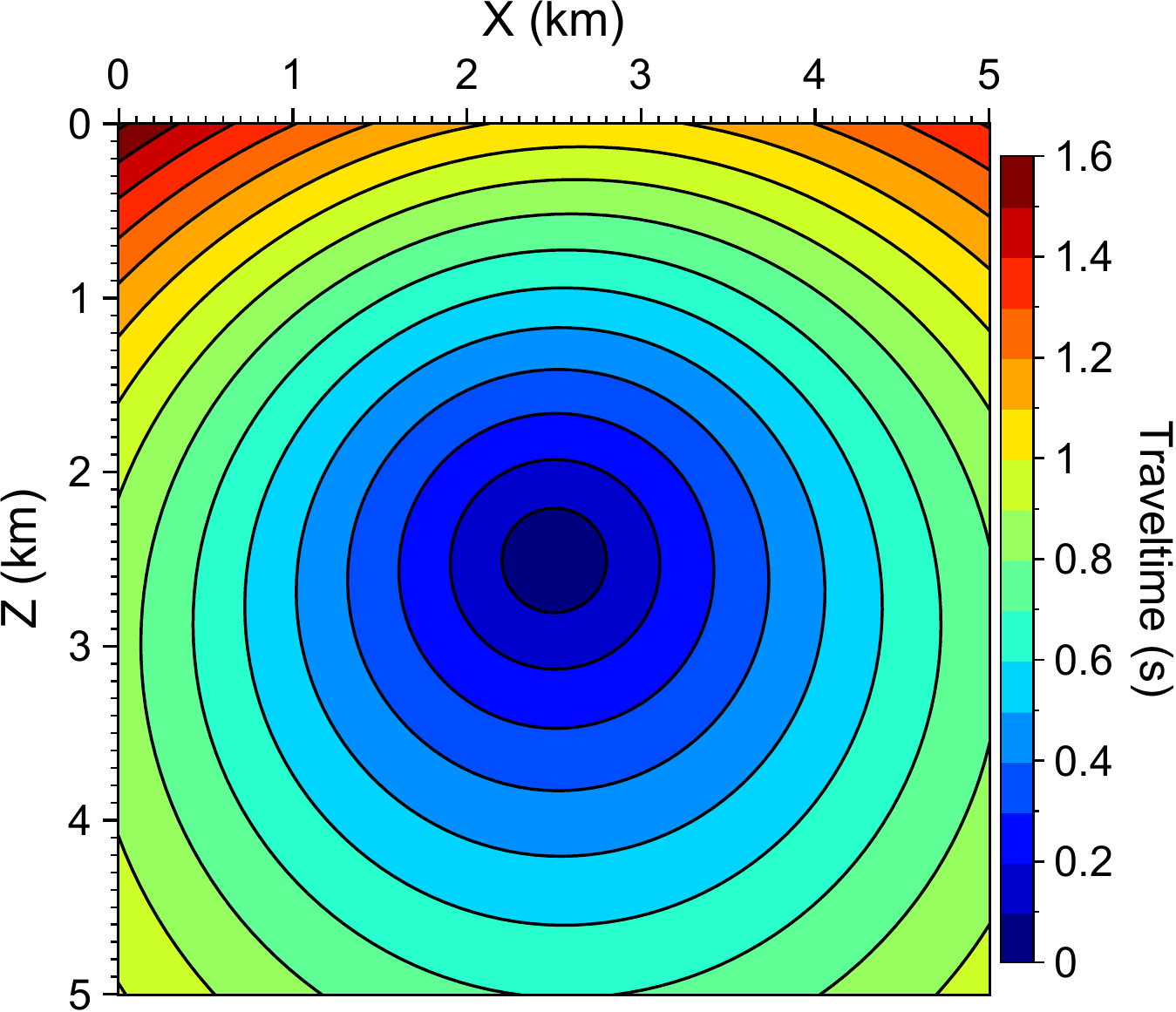}}
\caption{(a) A constant-gradient velocity model. (b) The analytical traveltime field computed using equation~\eqref{eq:tt_analytical}.} 
	\label{fig:analytical}
\end{figure}

We set a point source at $\mathbf{x}_0 = (2.5, 2.5)$~km in the model, 
with a background constant slowness $s_0 = 1/3000$~s/m. The constant 
gradient is $\mathbf{G}_0 = (G_x, G_z) = (0.1, 0.5)$~s$^{-1}$. The 
velocity in the model varies from 1500~m/s to 4500~m/s as shown in 
Figure~\ref{fig:analytical}a. Figure~\ref{fig:analytical}b depicts the 
corresponding analytical traveltime field computed using 
equation~\eqref{eq:tt_analytical}.

\begin{figure}
\centering 
\includegraphics[width=0.6\textwidth]{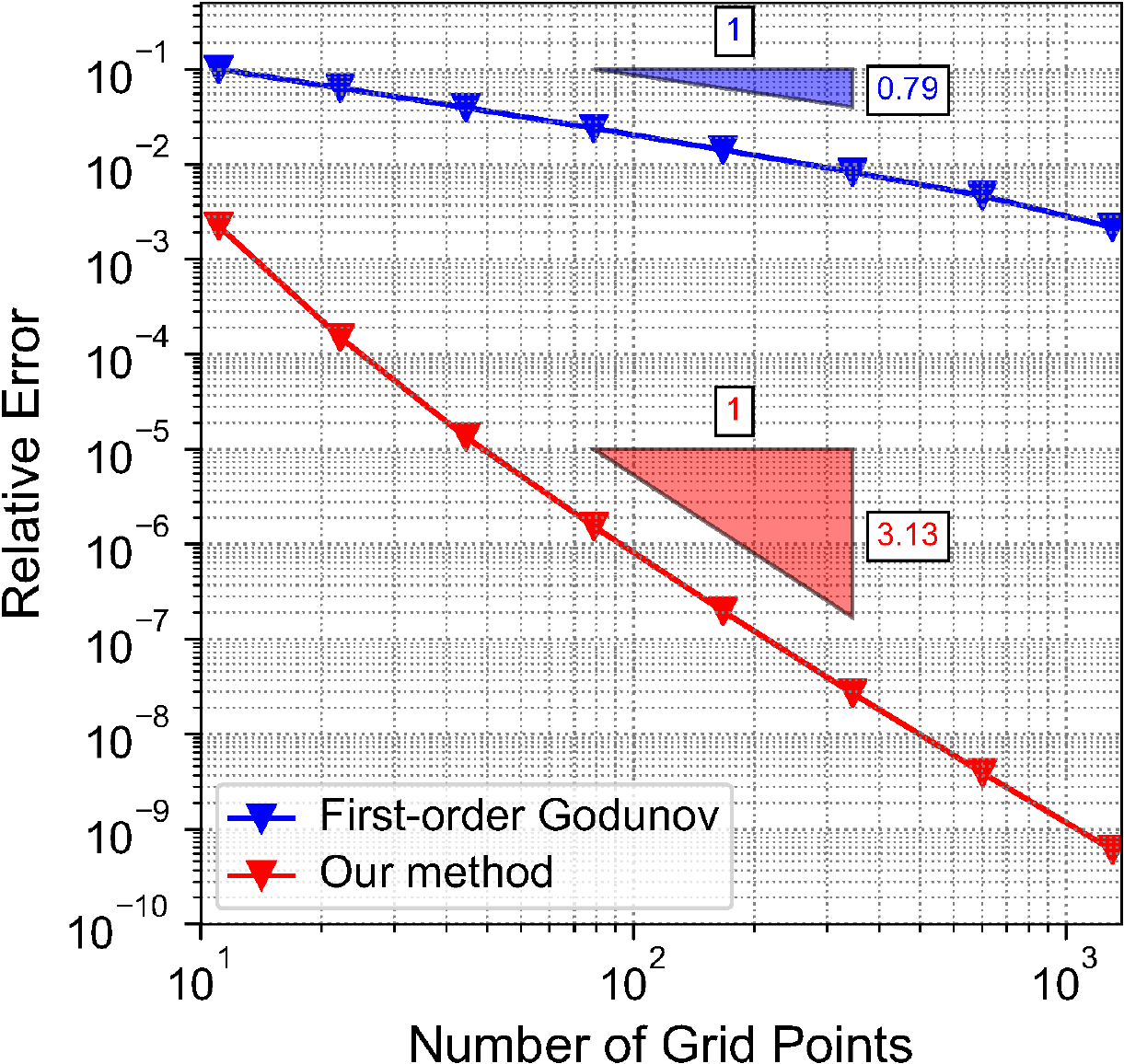}
\caption{Comparison between the convergence curves for the Godunov method 
(blue) and our hybrid method (red). }
\label{fig:error_analytical}
\end{figure}

We compute the traveltime fields with different model grid spacing using 
the Godunov and our hybrid method in this isotropic model. We compare the 
relative $L_2$-norm misfit between the numerical solutions and the 
analytical solution in Figure~\ref{fig:error_analytical}, demonstrating 
that our method is more accurate than the Godunov method.  The 
convergence order of our method is approximately 3.13 while that of the 
Godunov method is approximately 0.79. Even at the largest grid size where 
the number of grids along each direction is 10, our hybrid method is 
almost two orders of magnitude more accurate than the Godunov method. 

\subsection{Ball model}

\begin{figure}
	\centering
	\subfloat[]{\includegraphics[width=0.48\textwidth]{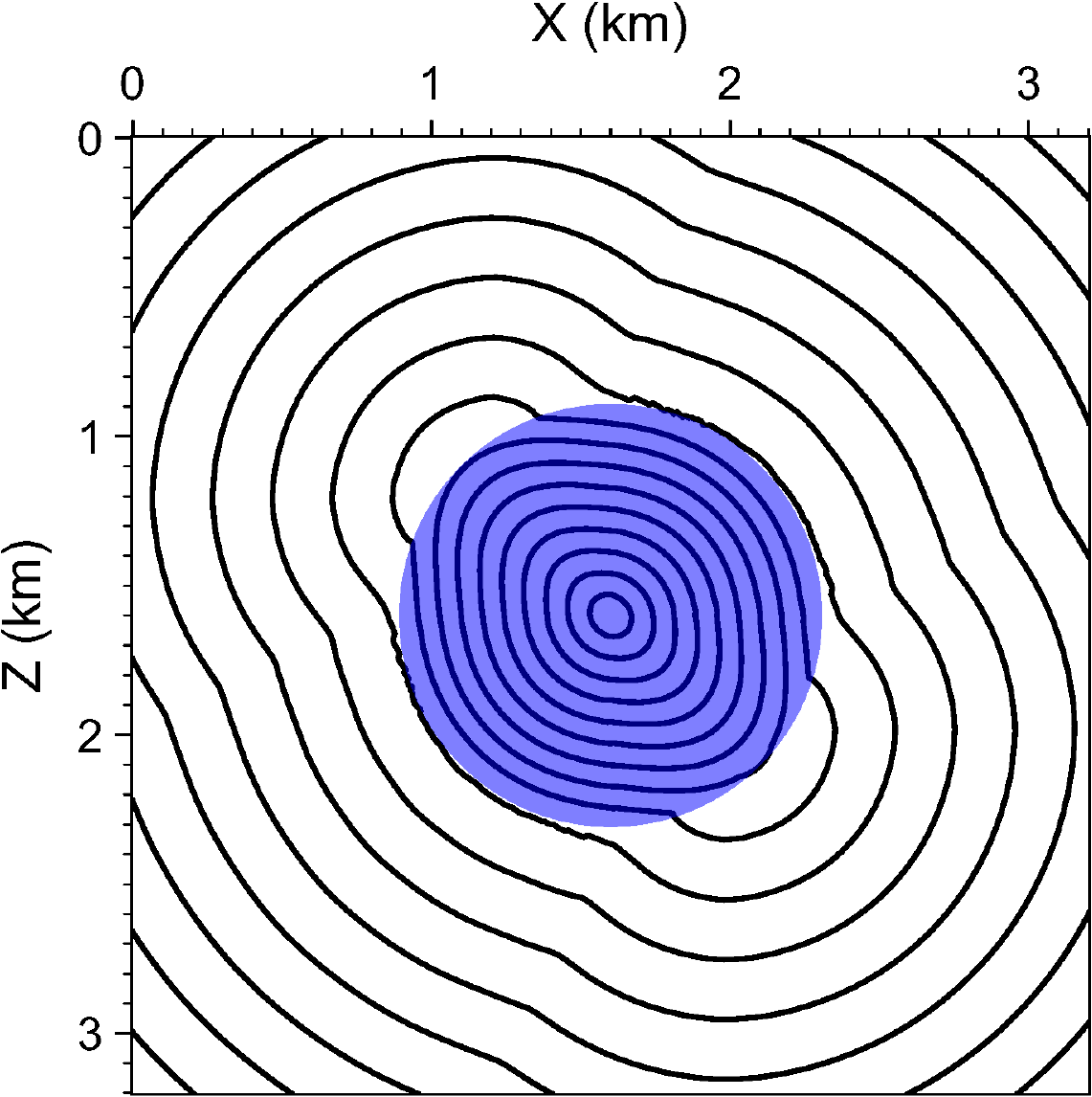}} \quad
	\subfloat[]{\includegraphics[width=0.48\textwidth]{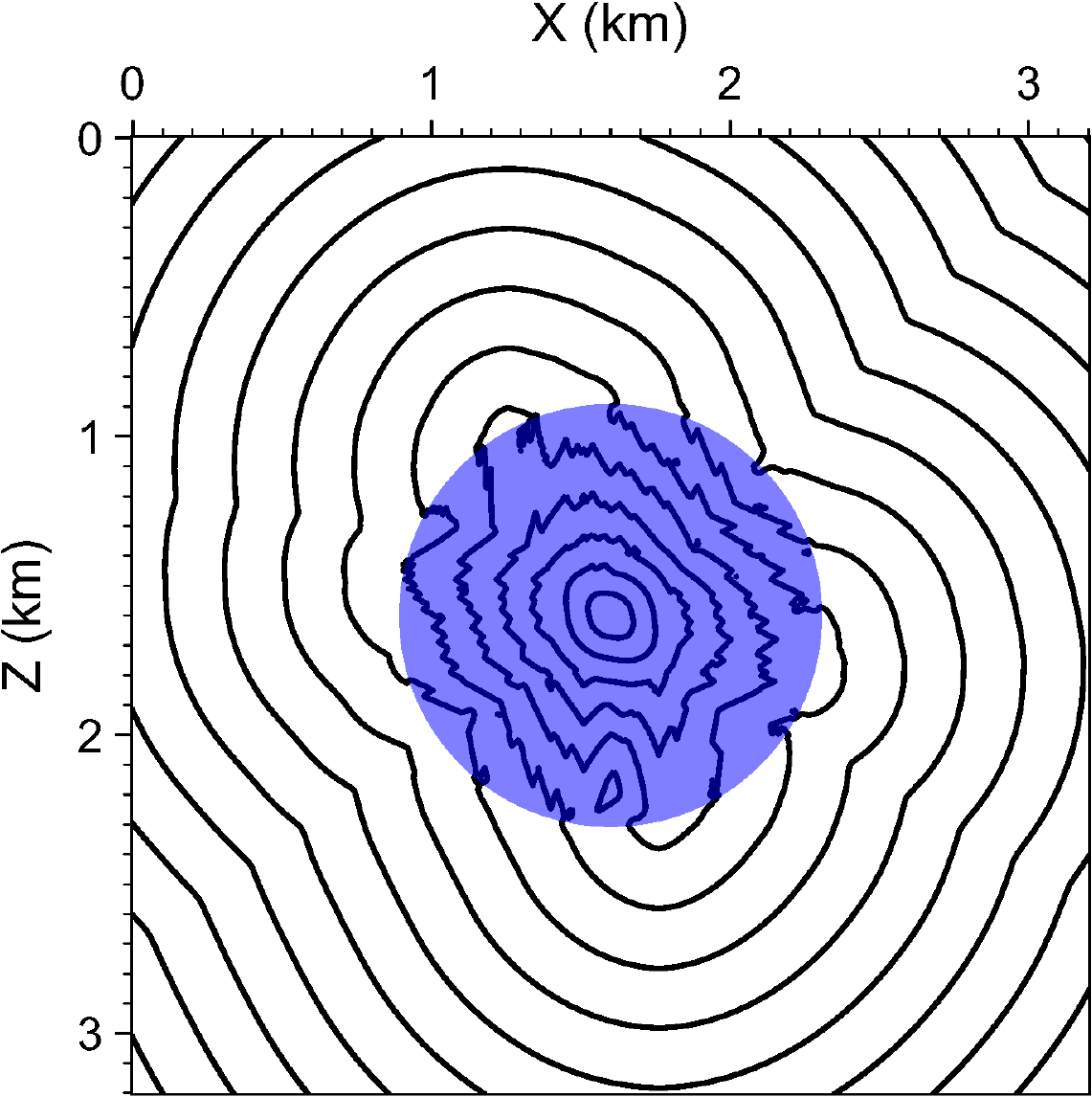}} \\
	\subfloat[]{\includegraphics[width=0.48\textwidth]{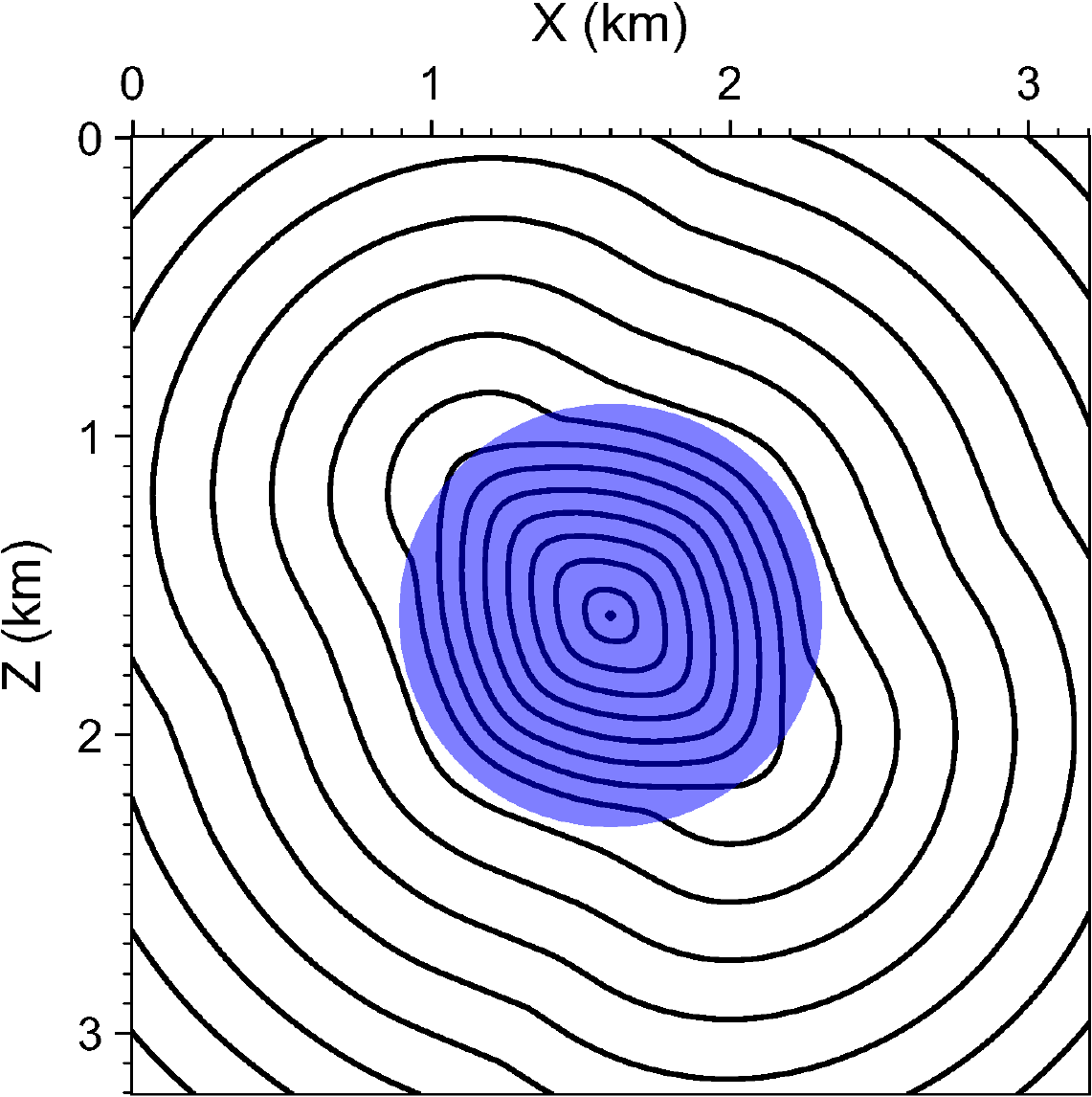}} \quad
	\subfloat[]{\includegraphics[width=0.48\textwidth]{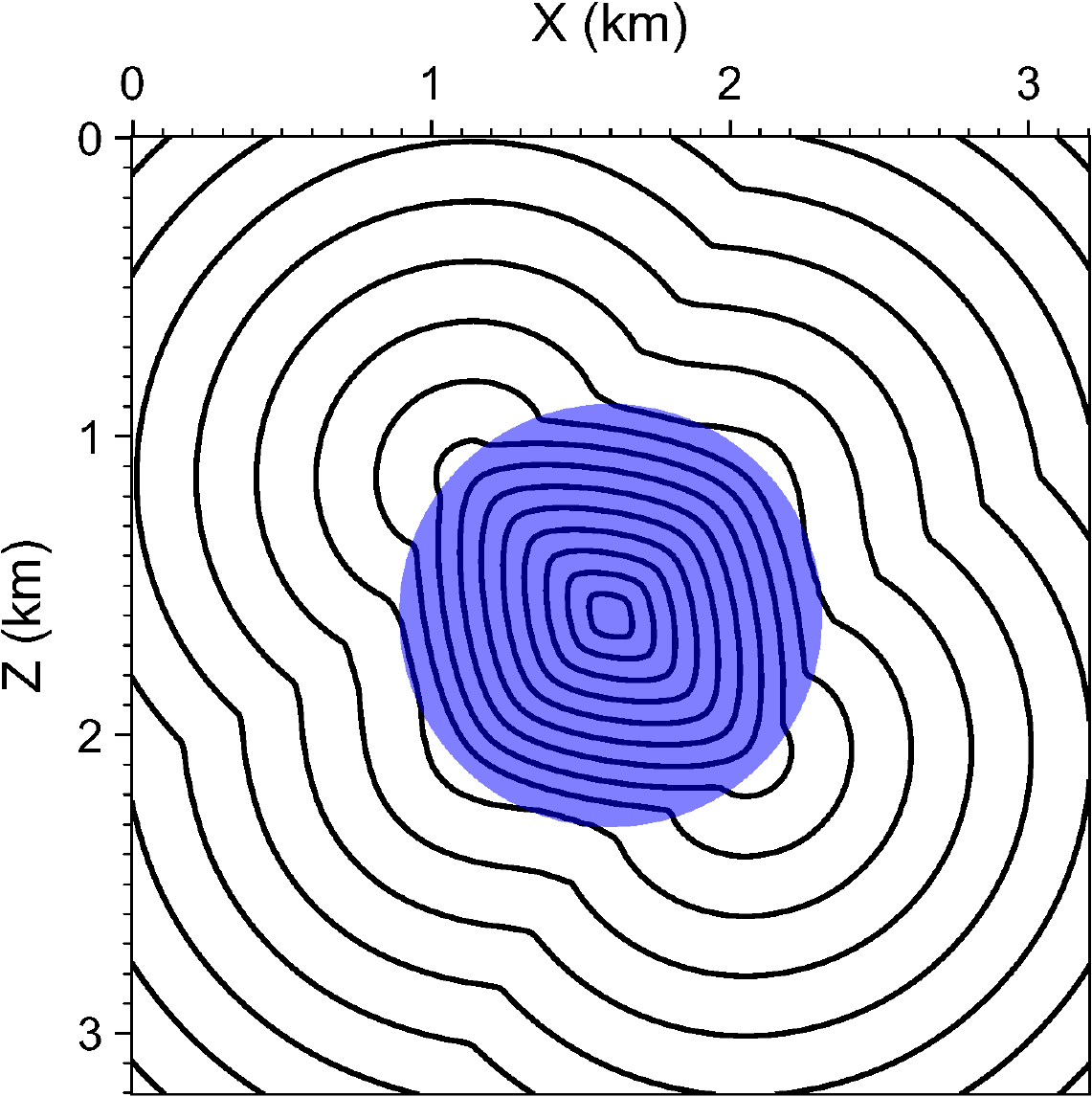}}
        \caption{Traveltime fields computed using (a) Godunov method, (b) 
        factorized Godunov method, (c) LF-3 method and (d) our hybrid 
        method. The interval between any two contours in all panels is 
        0.02~s, with contours starting from zero. Center blue-colored 
     ball region is a TTI anomaly. }
	\label{fig:tt_ball}
\end{figure}

We use the third numerical example to verify the numerical stability of 
our method in an anisotropic medium with a strong contrast as depicted   
in Figure~\ref{fig:tt_ball}. The background medium is an isotropic 
homogeneous medium with $V_p = 5,000$~m/s and $\varepsilon = \delta  
= \theta = 0$. The blue region at the center of the model indicates the 
location of a low-velocity strong TTI anisotropic anomaly with $V_p 
= 1,800$~m/s, $\varepsilon=0.3$, $\delta=-0.3$ and $\theta = \pi/4$. The 
model size is 3.2~km in both dimensions. The grid size is 10~m in both 
directions. 

Figures~\ref{fig:tt_ball}a-d show the traveltime fields computed using 
Godunov, factorized Godunov, LF-3 and our hybrid method, respectively.  
The Godunov solution exhibits weak instabilities around the boundaries of 
the TTI ball. The factorized Godunov solution in 
Figure~\ref{fig:tt_ball}b shows evident numerical instabilities. The 
computed traveltime field inside the TTI ball indicates that the 
factorized Godunov scheme becomes unstable for this TTI medium, and the 
traveltime field outside of the TTI ball is therefore mostly wrong with 
error propagating from the inside of the ball. 

Figure~\ref{fig:tt_ball}c is the LF-3 solution. The solution is more 
stable than those computed using the Godunov methods.  
Figure~\ref{fig:tt_ball}d displays our hybrid solution. Similar with the 
LF-3 solution, our hybrid method produces a stable solution with the help 
of the weighted non-oscillatory scheme. 

\begin{figure}
	\centering
	\subfloat[]{\includegraphics[width=0.48\textwidth]{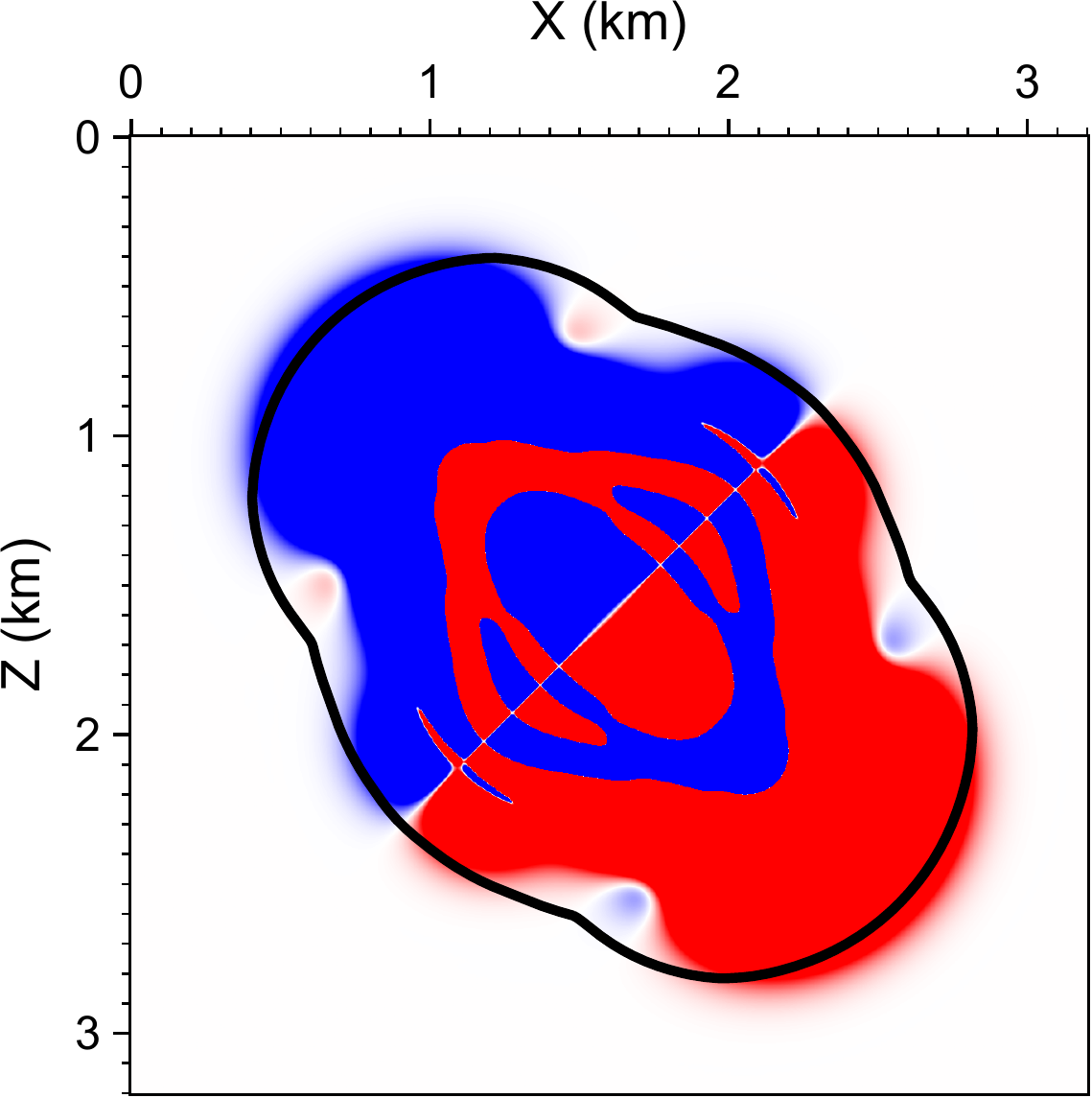}} \quad
	\subfloat[]{\includegraphics[width=0.48\textwidth]{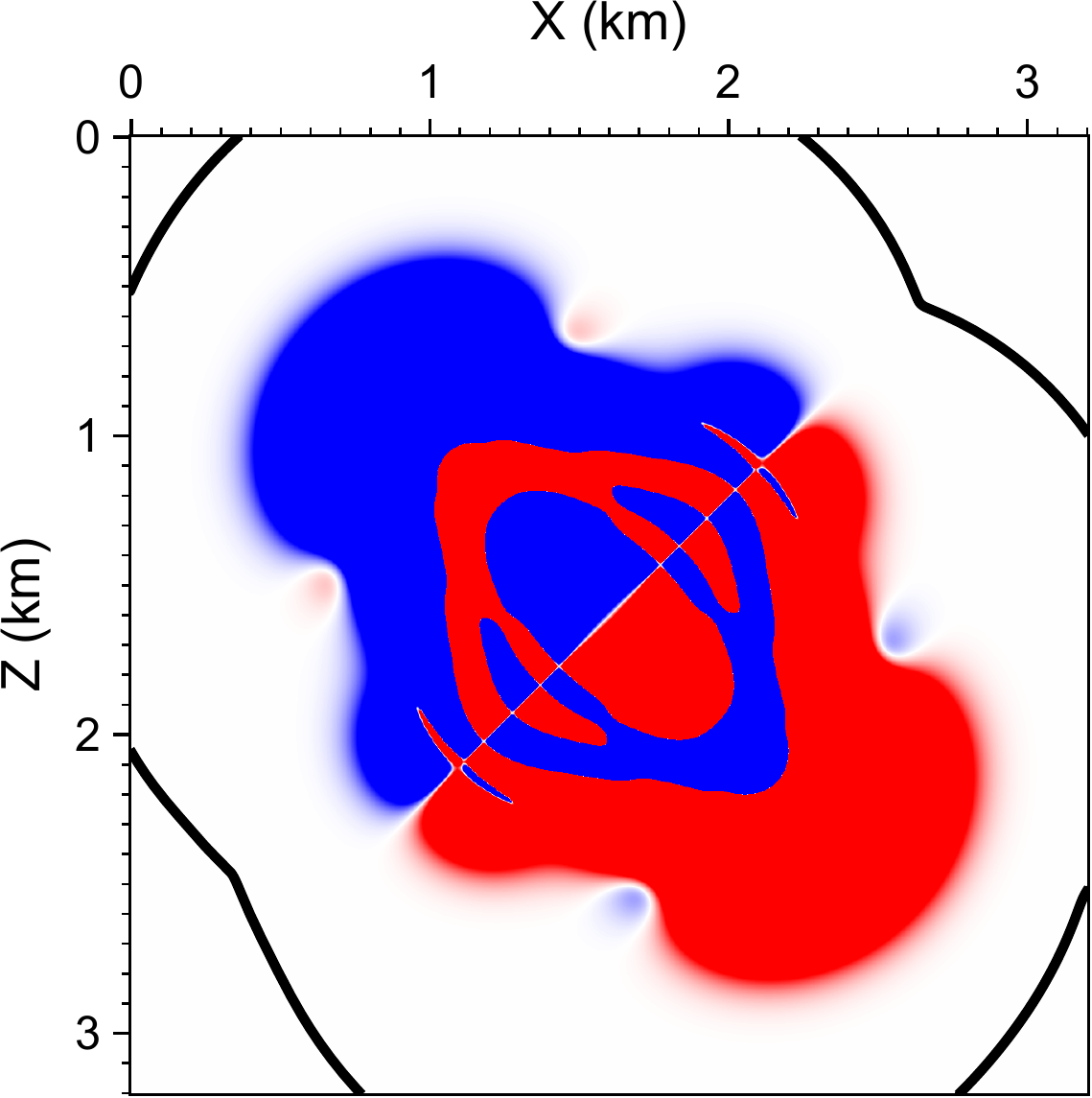}} \\
	\subfloat[]{\includegraphics[width=0.48\textwidth]{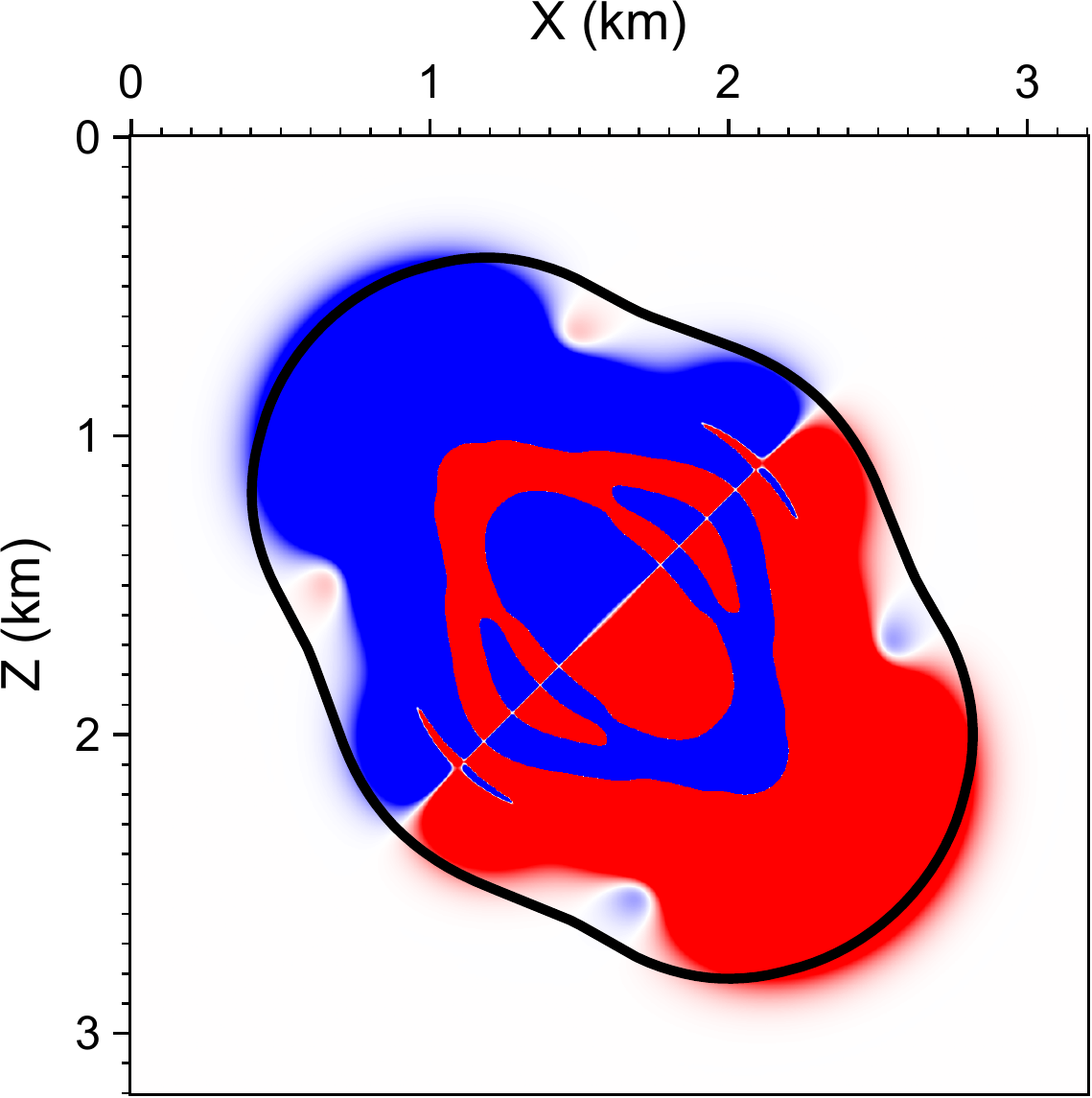}} \quad
	\subfloat[]{\includegraphics[width=0.48\textwidth]{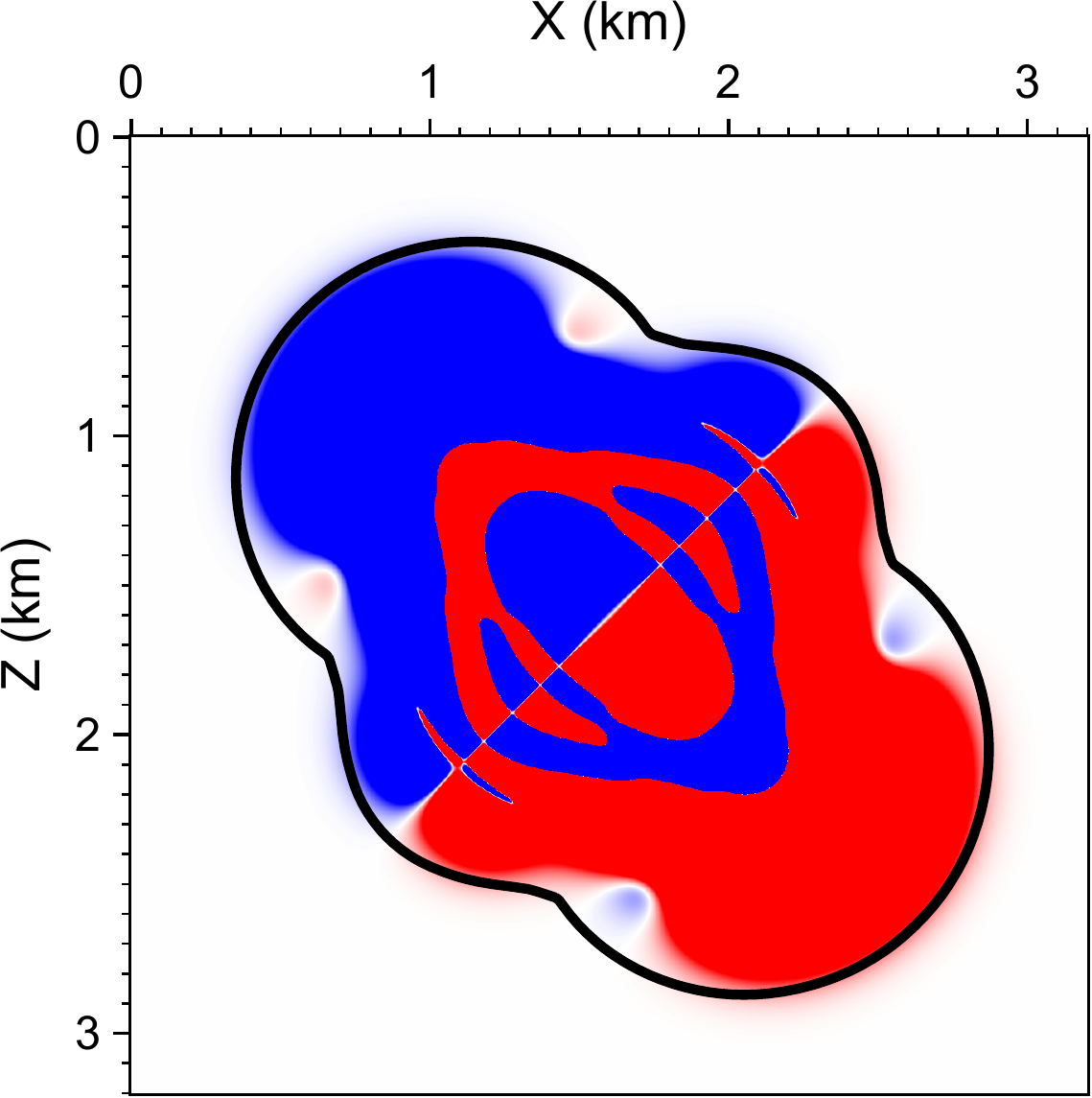}}
	\caption{Comparison between the wavefield snapshot taken at 0.55~s after source excitation and the corresponding traveltime contours (red curves) computed using (a) Godunov method, (b) factorized Godunov method, (c) LF-3 method and (d) our hybrid method. }
	\label{fig:wave_ball}
\end{figure}

It is important to compare the eikonal equation solution with the 
full-wavefield solution. We compute the full-wavefield solution using the 
fully staggered-grid finite-difference method \cite[]{Lisitsa_Vish_2010} 
with a high-order stencil, plus an optimal multi-axial perfectly matched 
layers \cite[]{Gao_Huang_2018}. Figures~\ref{fig:wave_ball}a-d show the 
full-wavefield solution at 0.55~s after the source excitation and the 
corresponding eikonal equation solutions in black curves computed using 
the Godunov method, factorized Godunov method, LF-3 method and our hybrid 
method, respectively. We find that the Godunov and the LF-3 methods are 
stable, yet are not consistent with the full-wavefield solution. The two 
solutions have an obvious delay compared with the full-wavefield solution 
wavefront. The factorized Godunov solution is completely inconsistent 
with the full-wavefield solution because it is not numerically stable.  
Only our hybrid method produces a stable and accurate solution that is 
highly consistent with the full-wavefield solution shown in 
Figure~\ref{fig:wave_ball}d. 

\subsection{Block model}

\begin{figure}
	\centering
	\includegraphics[width=\textwidth]{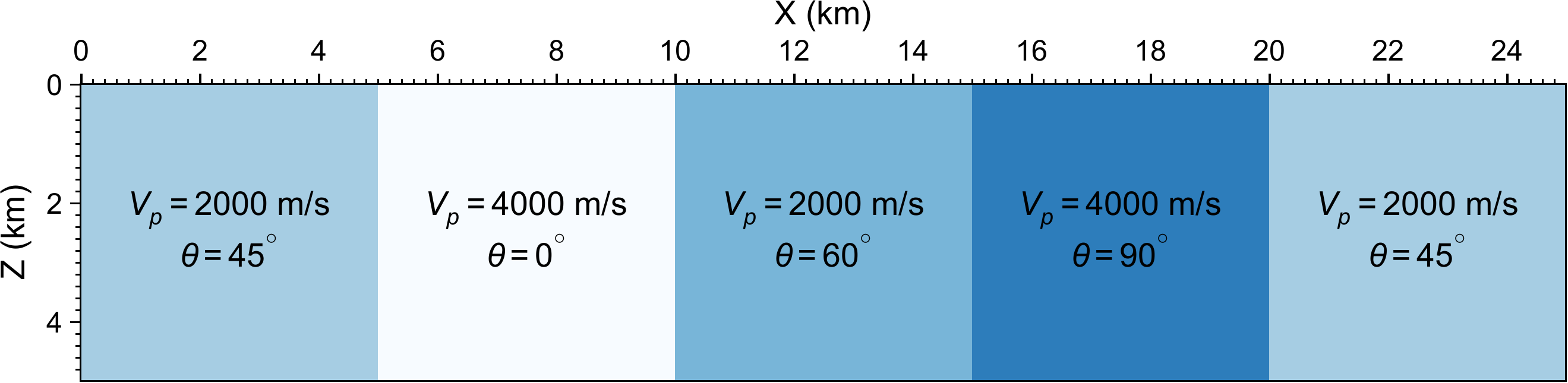}
	\caption{An TTI anisotropic model with block heterogeneities. The Thomsen parameters are $\varepsilon = 0.4$ and $\delta = -0.2$ for all the blocks. }
	\label{fig:model_block}
\end{figure}

The fourth example in Figure~\ref{fig:model_block} is a five-block 
anisotropic heterogeneous model. The model is 20~km in the X direction 
and 5~km in the Z direction, with a uniform grid sampling of 25~m in both 
directions. The model has uniform Thomsen parameters of $\varepsilon=0.4$ 
and $\delta=-0.2$, but has a strongly contrasted $V_p$ and TTI symmetry 
axis tilt angle $\theta$. The $V_p$ contrast at each interface is 
2000~m/s, and that of the tilt angle $\theta$ is at least $\pi/6$.  

\begin{figure}
	\centering
	\subfloat[]{\includegraphics[width=\textwidth]{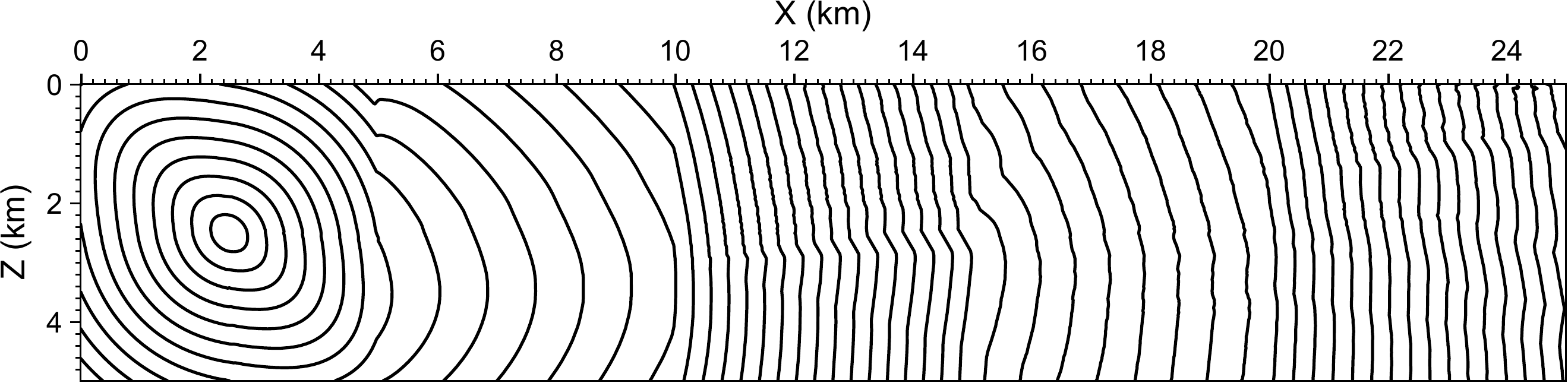}} \\
	\subfloat[]{\includegraphics[width=\textwidth]{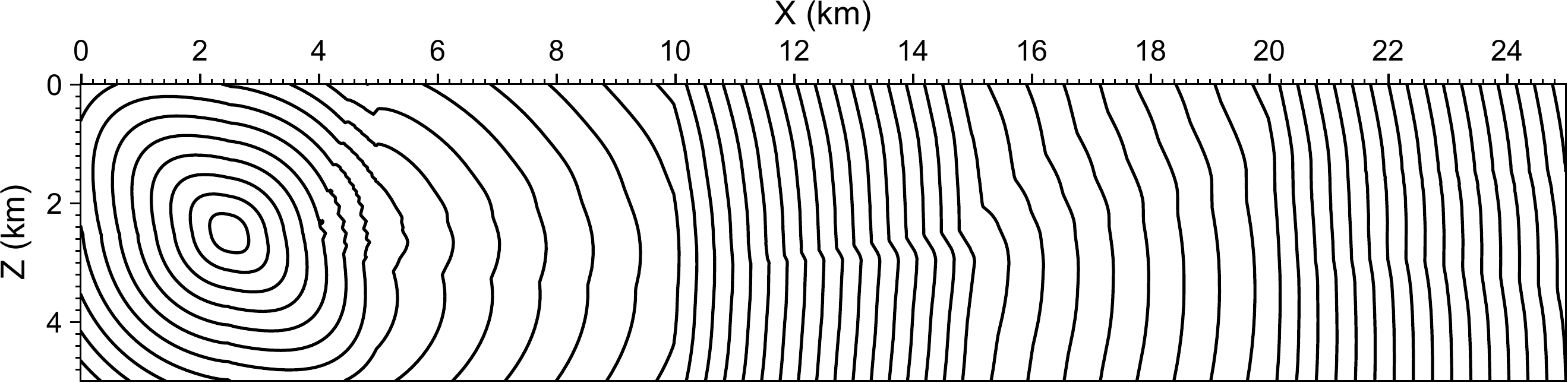}} \\
	\subfloat[]{\includegraphics[width=\textwidth]{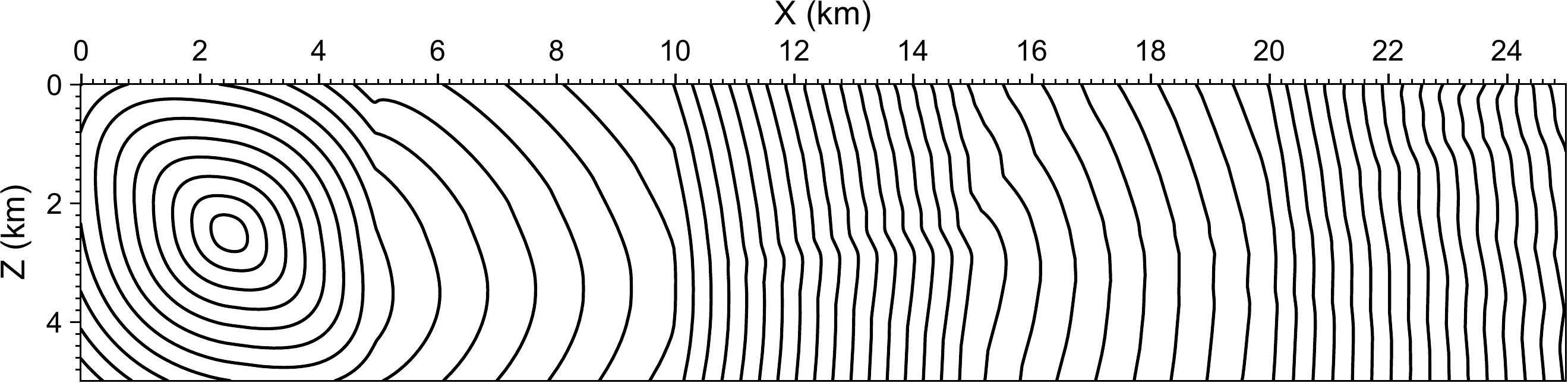}} \\
	\subfloat[]{\includegraphics[width=\textwidth]{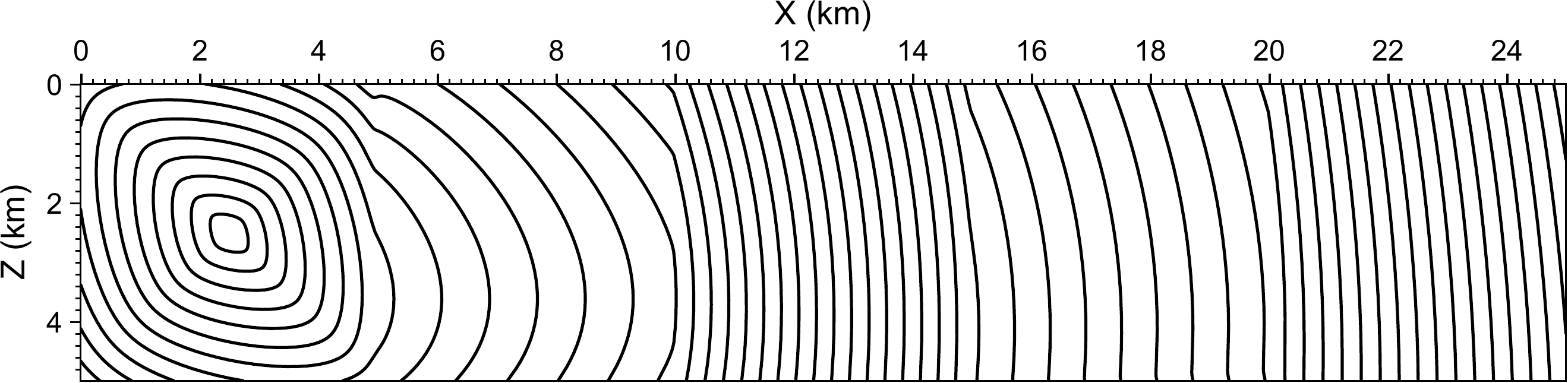}}
        \caption{Traveltime field solutions computed using (a) Godunov 
        method, (b) factorized Godunov method, (c) LF-3 method and (d) 
     our hybrid method. The contours start from 0~s and the interval 
  between any two adjacent contours is 0.15~s.}
	\label{fig:tt_block}
\end{figure}

Figures~\ref{fig:tt_block}a-d show the traveltime filed solutions 
computed using the Godunov method, factorized Godunov method, LF-3 method 
and our hybrid method, respectively. The Godunov, factorized Godunov and 
LF-3 solutions have obvious spurious modes starting from the X position 
of 10~km. This artifact is generated at the interface between the second 
and the third block and propagates to the far end of the model. In the 
factorized Godunov solution in Figure~\ref{fig:tt_block}b, we also 
observe some instabilities near the interface between the first and the 
second block. These instabilities propagates from the first interface to 
the positive direction of X, eventually deteriorating the traveltime 
field in the entire computational domain. By contrast, the solution 
computed with our hybrid method shown in Figure~\ref{fig:tt_block}d is 
the only one of the four solutions that is stable in all the five TTI 
blocks. 

\begin{figure}
	\centering
	\subfloat[]{\includegraphics[width=\textwidth]{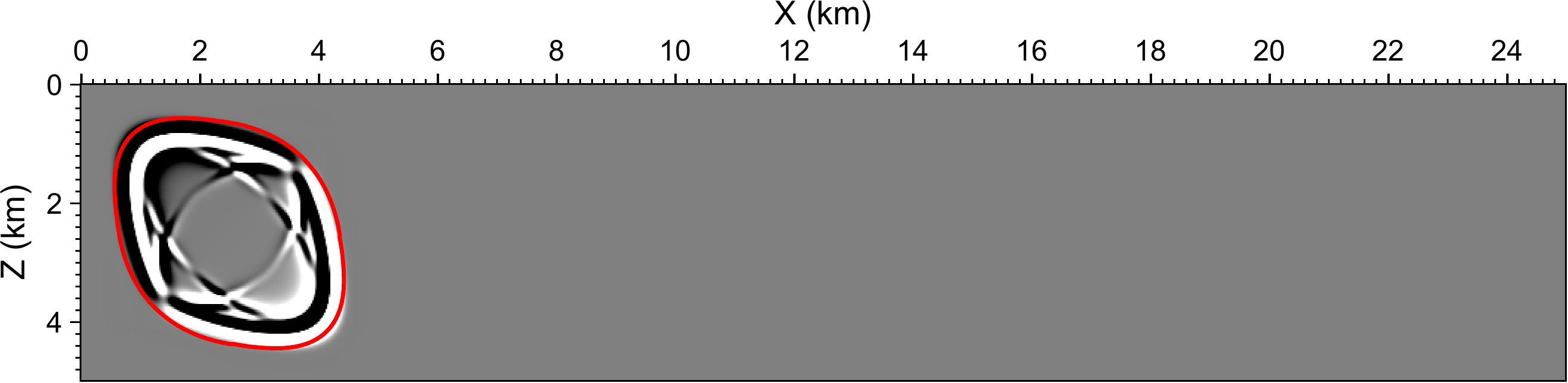}} \\
	\subfloat[]{\includegraphics[width=\textwidth]{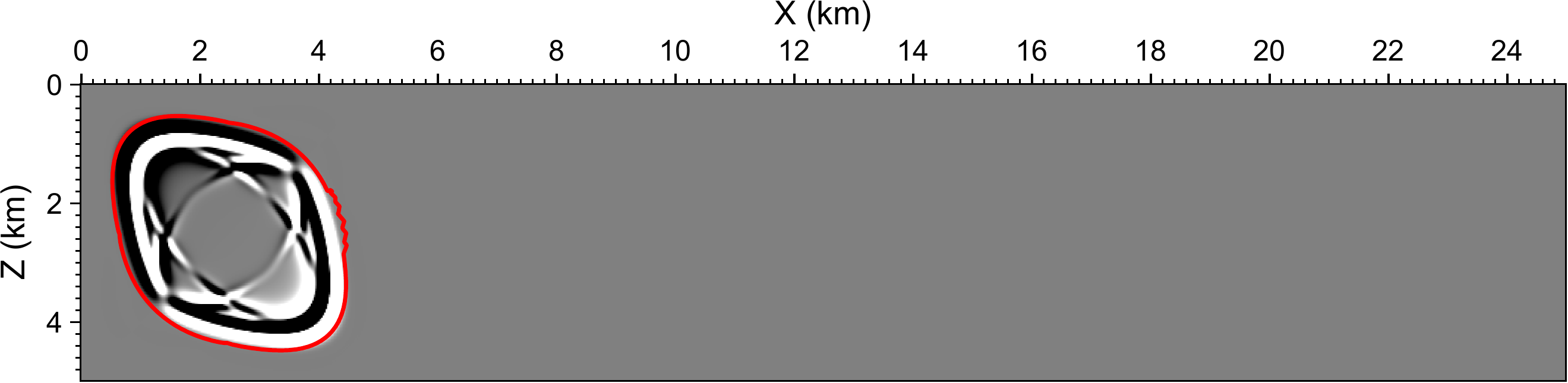}} \\
	\subfloat[]{\includegraphics[width=\textwidth]{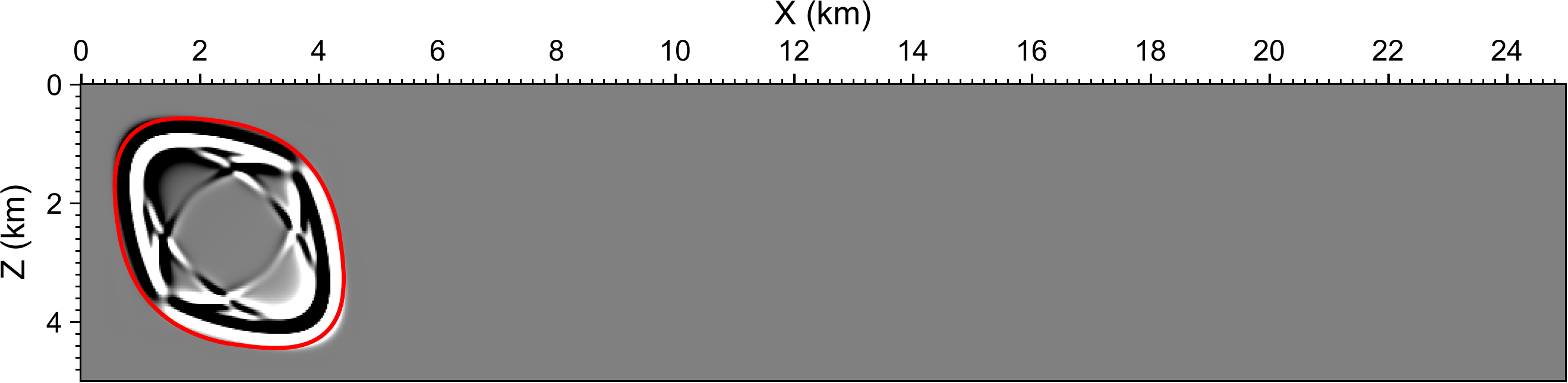}} \\
	\subfloat[]{\includegraphics[width=\textwidth]{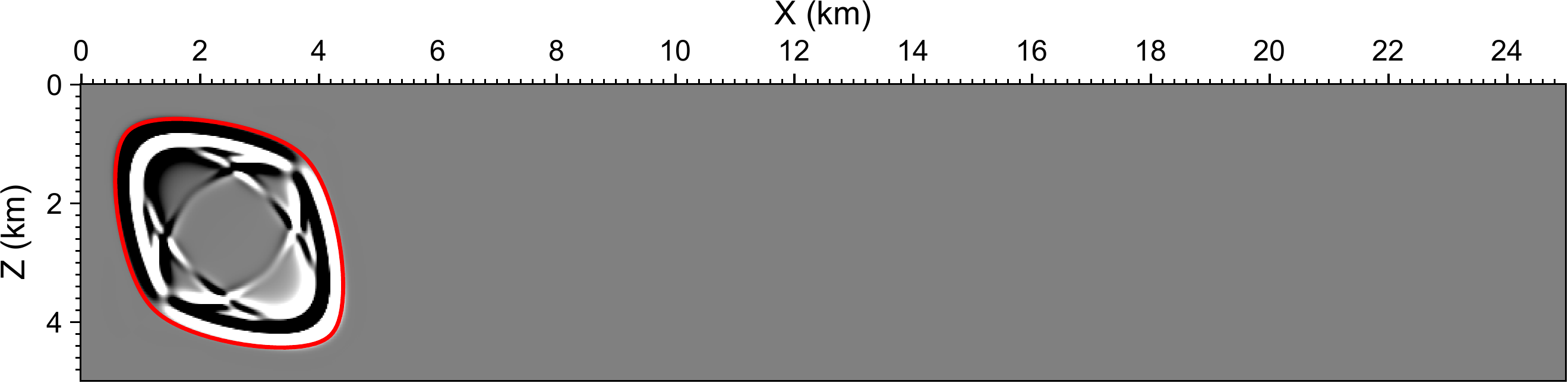}}
	\caption{Comparison between the full-wavefield snapshot at 1~s 
        and the corresponding eikonal equation solution contour (red 
     curve) computed using (a) Godunov method, (b) factorized Godunov 
  method, (c) LF-3 method and (d) our hybrid method.}
	\label{fig:wave_block_1}
\end{figure}

\begin{figure}
	\centering
	\subfloat[]{\includegraphics[width=\textwidth]{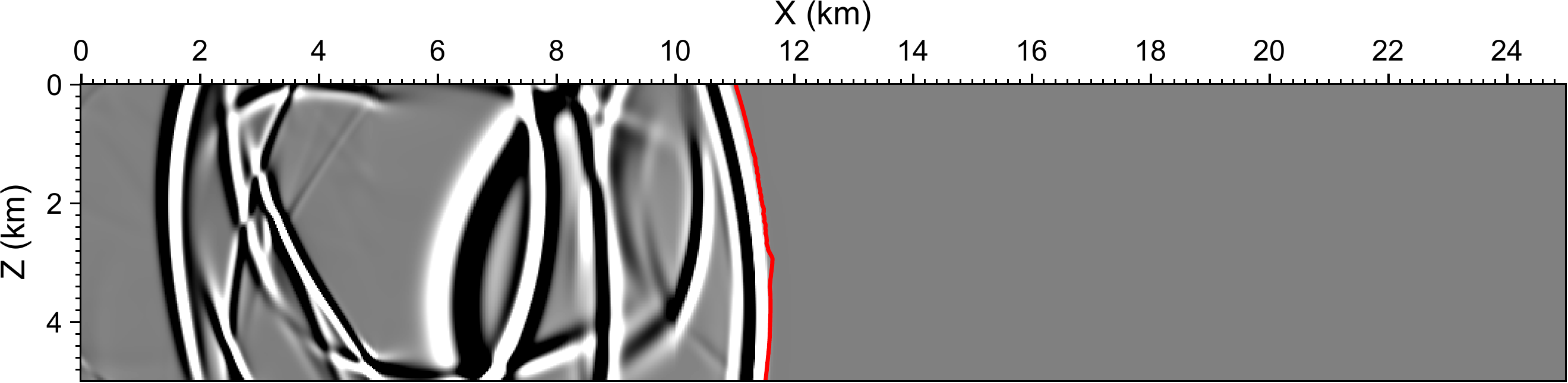}} \\
	\subfloat[]{\includegraphics[width=\textwidth]{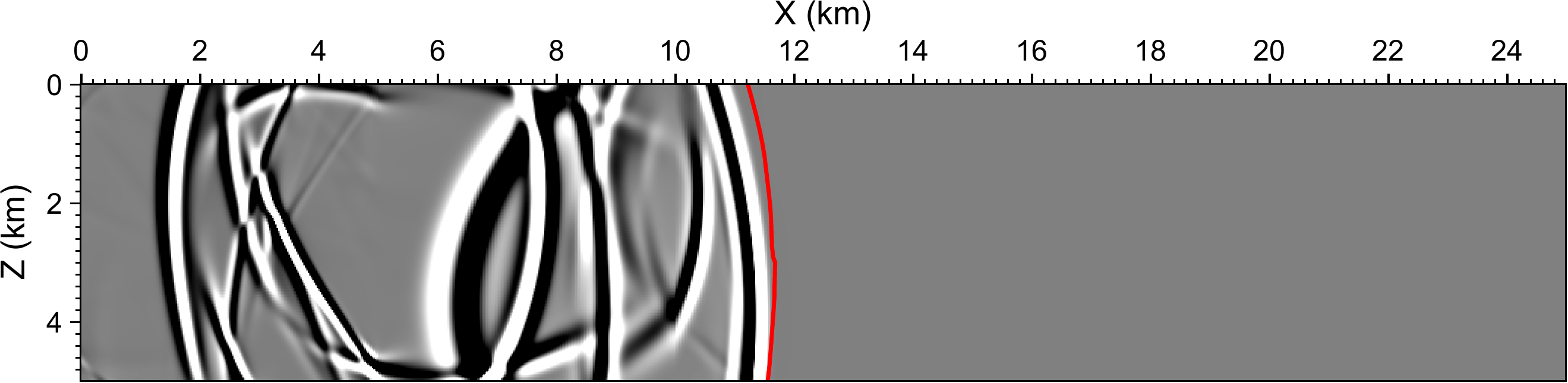}} \\
	\subfloat[]{\includegraphics[width=\textwidth]{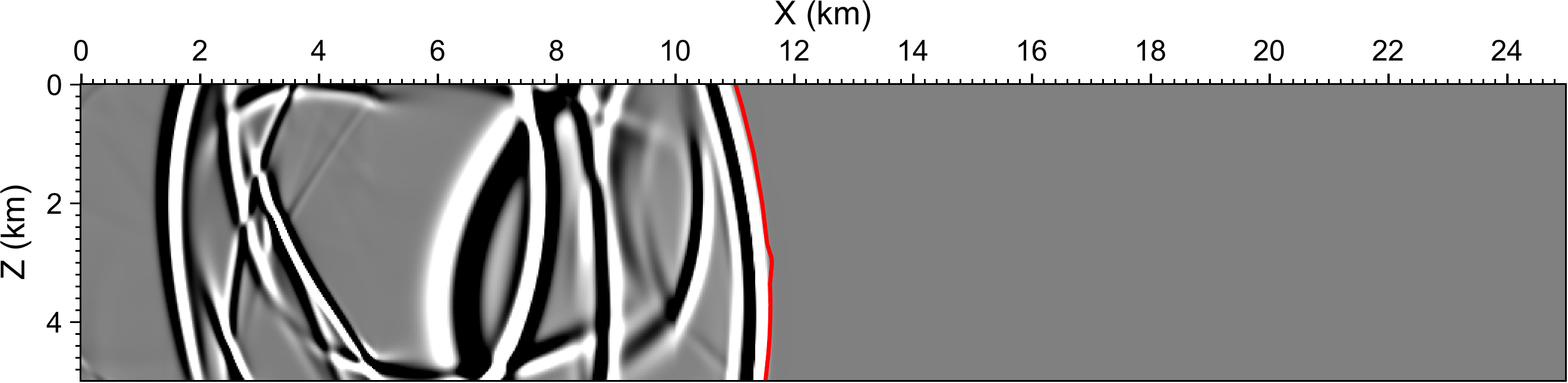}} \\
	\subfloat[]{\includegraphics[width=\textwidth]{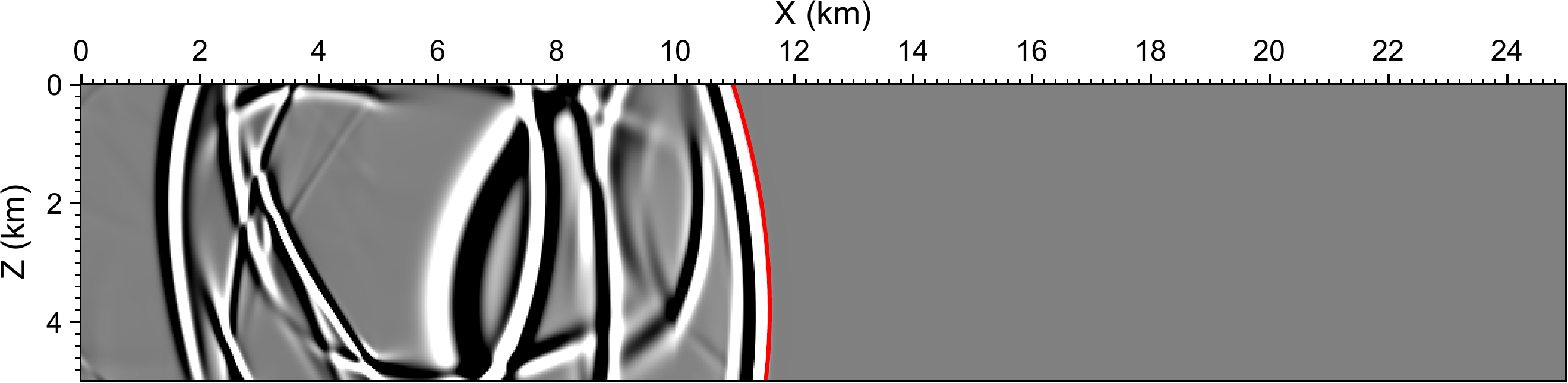}}
	\caption{Comparison between the full-wavefield snapshot at 3~s 
        and the corresponding eikonal equation solution contour (red 
     curve) computed using (a) Godunov method, (b) factorized Godunov 
  method, (c) LF-3 method and (d) our hybrid method.}
		\label{fig:wave_block_2}
\end{figure}

\begin{figure}
	\centering
	\subfloat[]{\includegraphics[width=\textwidth]{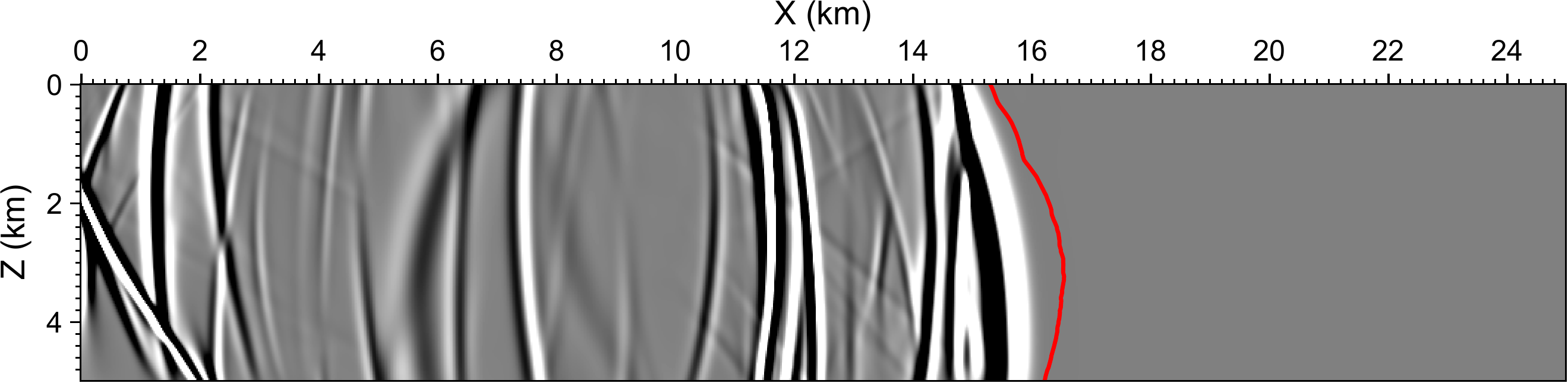}} \\
	\subfloat[]{\includegraphics[width=\textwidth]{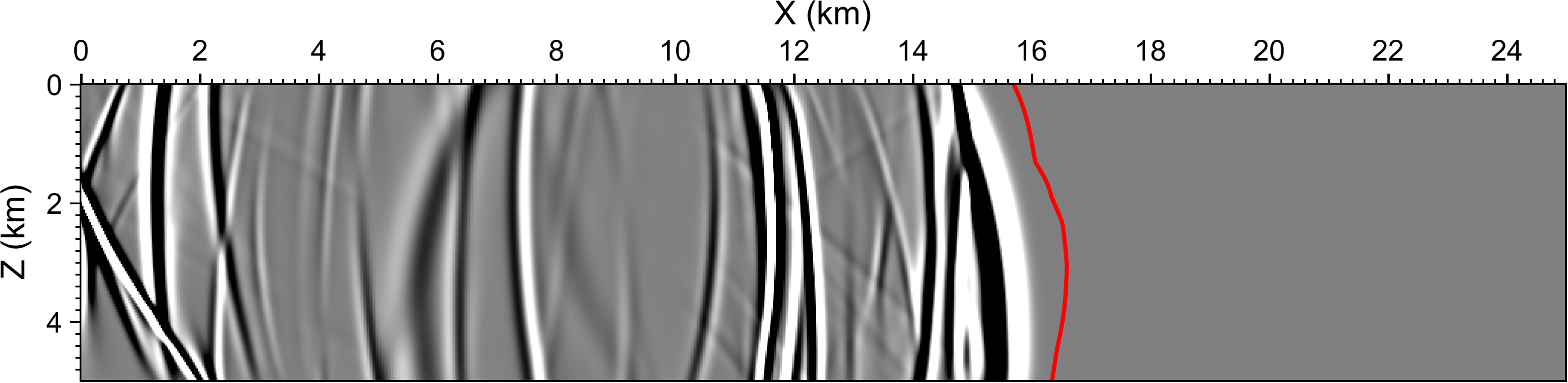}} \\
	\subfloat[]{\includegraphics[width=\textwidth]{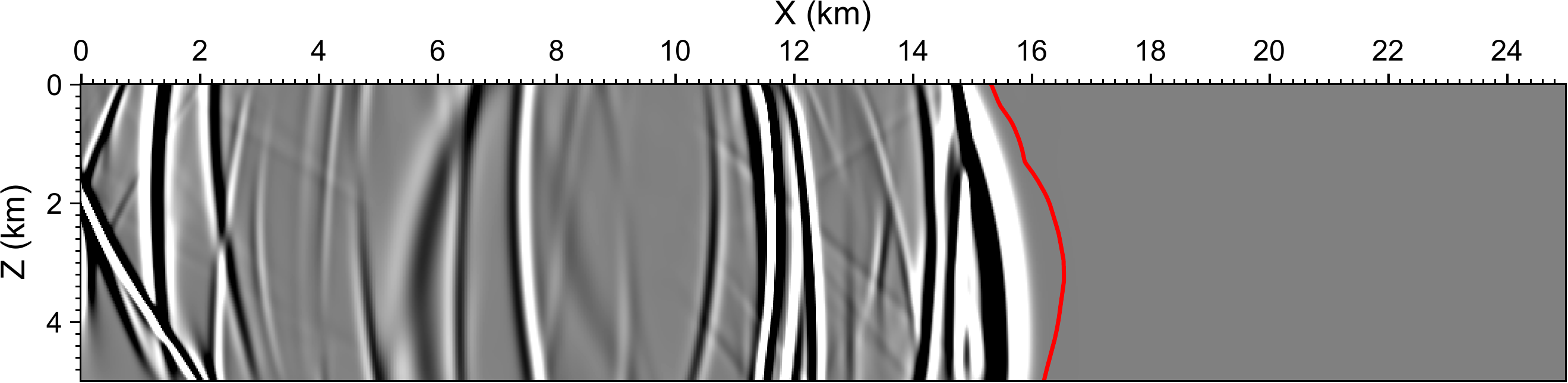}} \\
	\subfloat[]{\includegraphics[width=\textwidth]{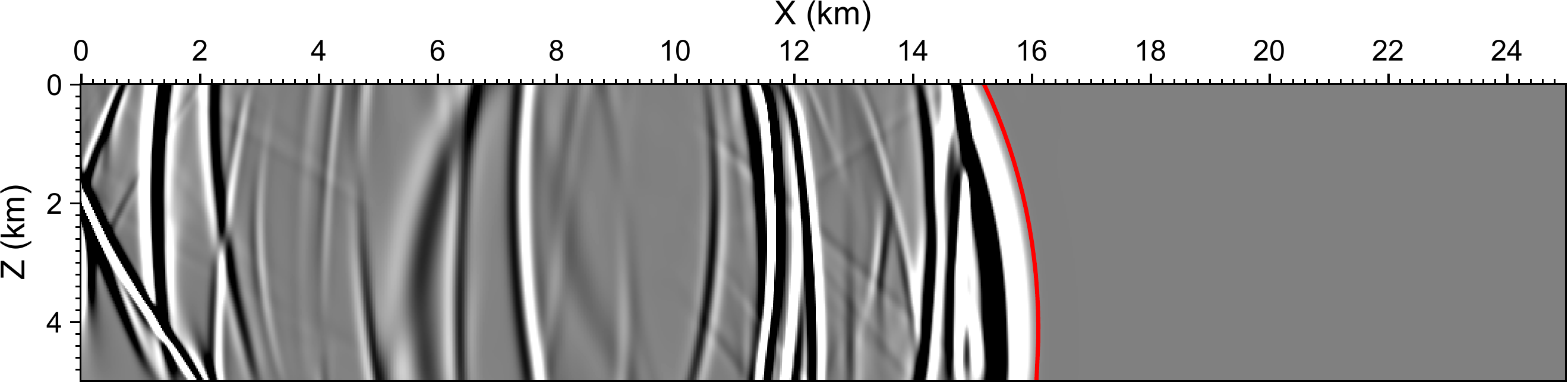}}
	\caption{Comparison between the full-wavefield snapshot at 5~s 
        and the corresponding eikonal equation solution contour (red 
     curve) computed using (a) Godunov method, (b) factorized Godunov 
  method, (c) LF-3 method and (d) our hybrid method.}
		\label{fig:wave_block_3}
\end{figure}

\begin{figure}
	\centering
	\subfloat[]{\includegraphics[width=\textwidth]{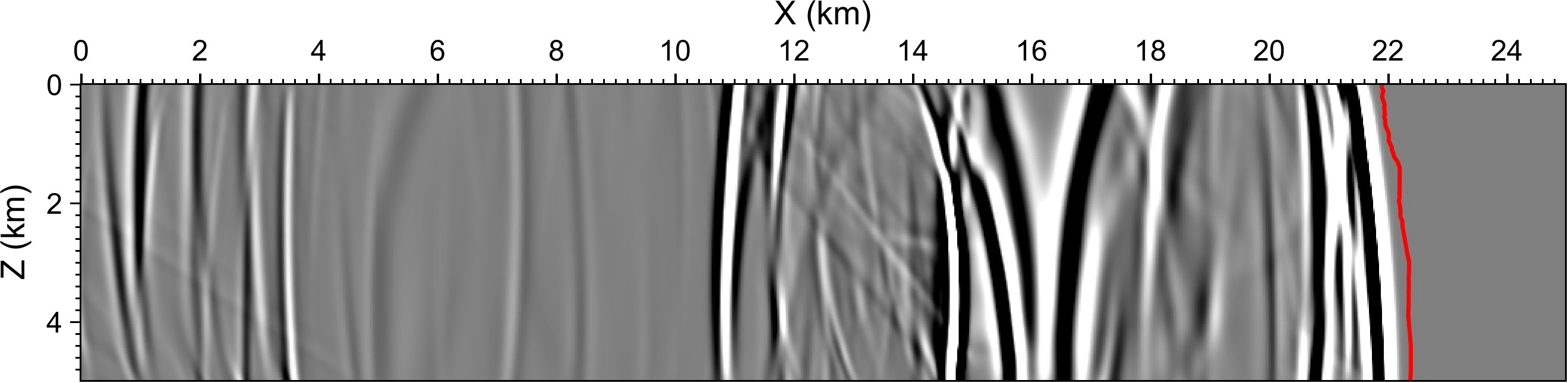}} \\
	\subfloat[]{\includegraphics[width=\textwidth]{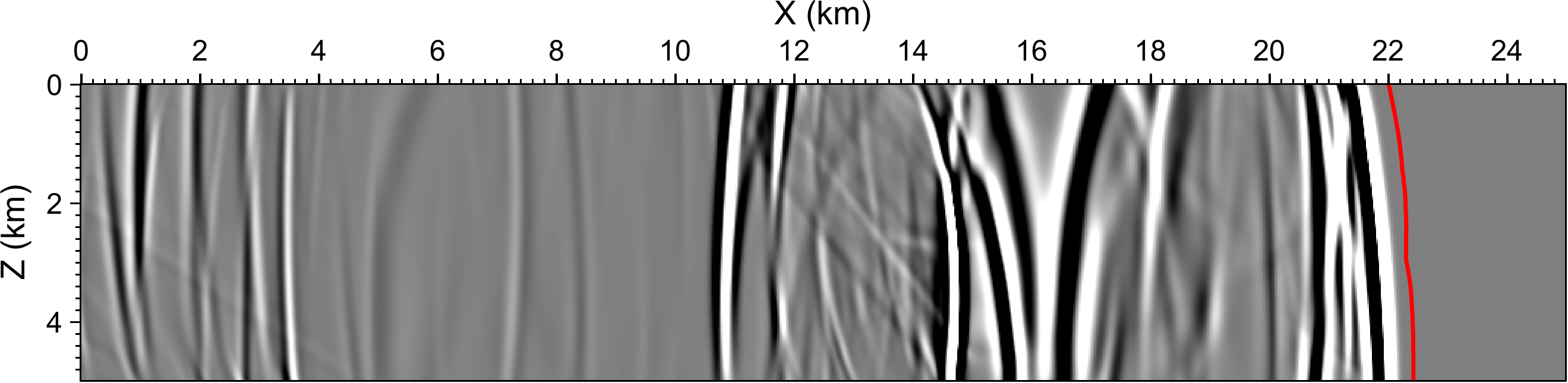}} \\
	\subfloat[]{\includegraphics[width=\textwidth]{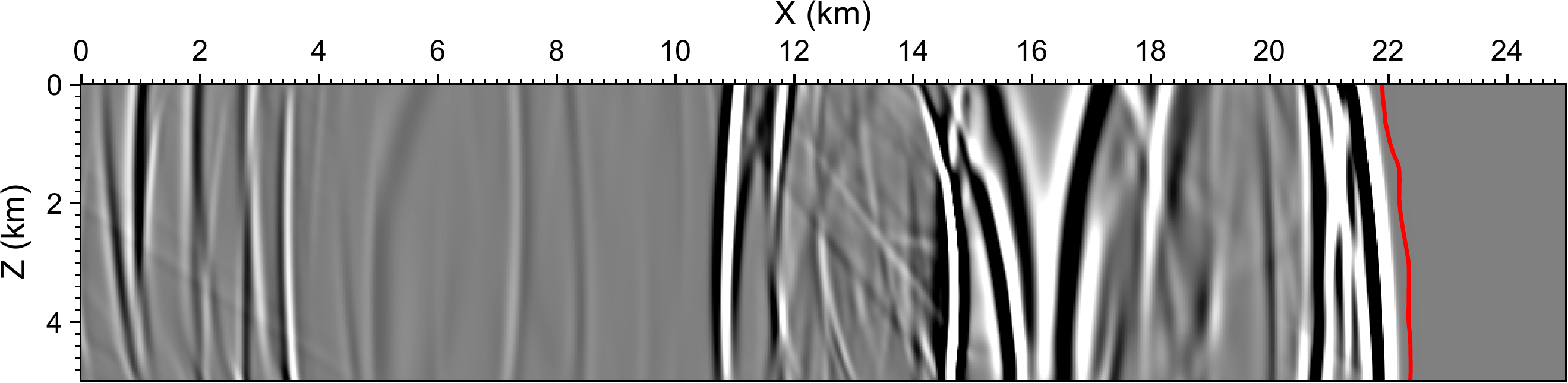}} \\
	\subfloat[]{\includegraphics[width=\textwidth]{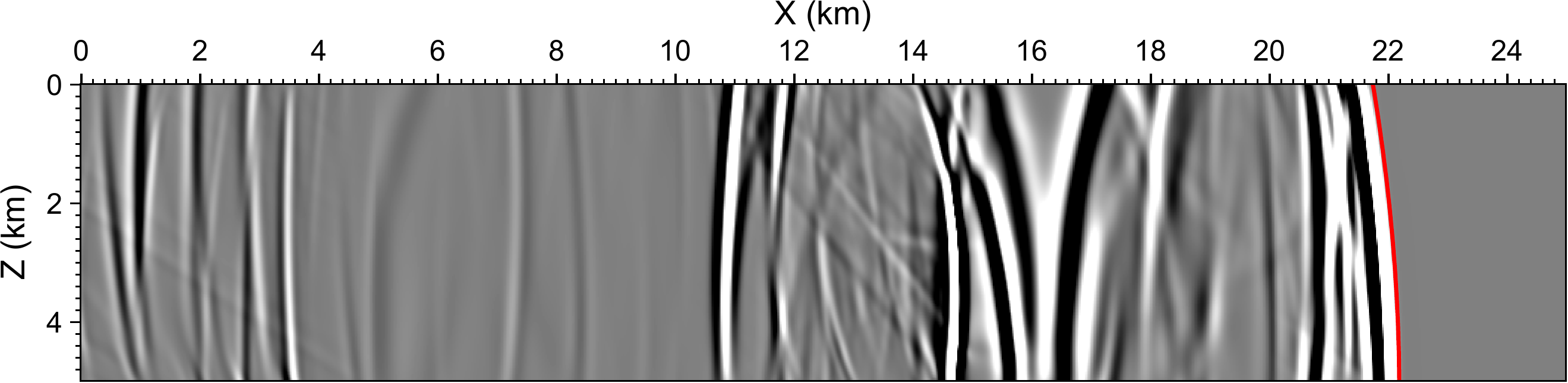}}
	\caption{Comparison between the full-wavefield snapshot at 7~s 
        and the corresponding eikonal equation solution contour (red 
     curve) computed using (a) Godunov method, (b) factorized Godunov 
  method, (c) LF-3 method and (d) our hybrid method.}
		\label{fig:wave_block_4}
\end{figure}

We further compare the accuracy of different solutions against the 
full-wavefield solution in 
Figures~\ref{fig:wave_block_1}-\ref{fig:wave_block_4}. 
Figure~\ref{fig:wave_block_1} shows the computed traveltime field 
overlying on the full-wavefield solution at 1~s after source excitation.  
Figures~\ref{fig:wave_block_1}a-d are the solutions computed using the 
Godunov method, the factorized Godunov method, the LF-3 method, and our 
hybrid method, respectively. The factorized Godunov solution contains  
some weak instabilities near the 5~km interface. The Godunov and LF-3 
solutions give a slight delay along the direction perpendicular to the 
TTI symmetry axis compared with the full-wavefield solution. In 
comparison, our hybrid method produces a solution in 
Figure~\ref{fig:wave_block_1}d that is both stable and accurate, and is 
highly consistent with the full-wavefield solution wavefront along all 
propagation directions. 

At snapshot time 2~s depicted in Figure~\ref{fig:wave_block_2}, the 
inaccuracy caused by the instability of the factorized Godunov solution 
(Figure~\ref{fig:wave_block_2}b) becomes fairly apparent, while the 
Godunov (Figure~\ref{fig:wave_block_2}a) and LF-3 solutions 
(Figure~\ref{fig:wave_block_2}c) start to show inconsistency with the 
full-wavefield solution at the Z position of approximately 3~km. This 
inconsistency is in fact the artifact in Figures~\ref{fig:tt_block}a and 
c. Only our hybrid method produces a stable and accurate solution 
(Figure~\ref{fig:wave_block_2}d) that is highly consistent with the 
full-wavefield solution.

The consistency check between the full-wavefield solution and the eikonal 
equation solution at two other time steps shown in 
Figures~\ref{fig:wave_block_3} and \ref{fig:wave_block_4} further 
verifies that our hybrid method is able to produce stable and accurate 
solutions to the anisotropic eikonal equation with strong medium property 
contrasts where conventional methods fail. 

\subsection{Salt model}

\begin{figure}
	\centering
	\subfloat[]{\includegraphics[width=0.48\textwidth]{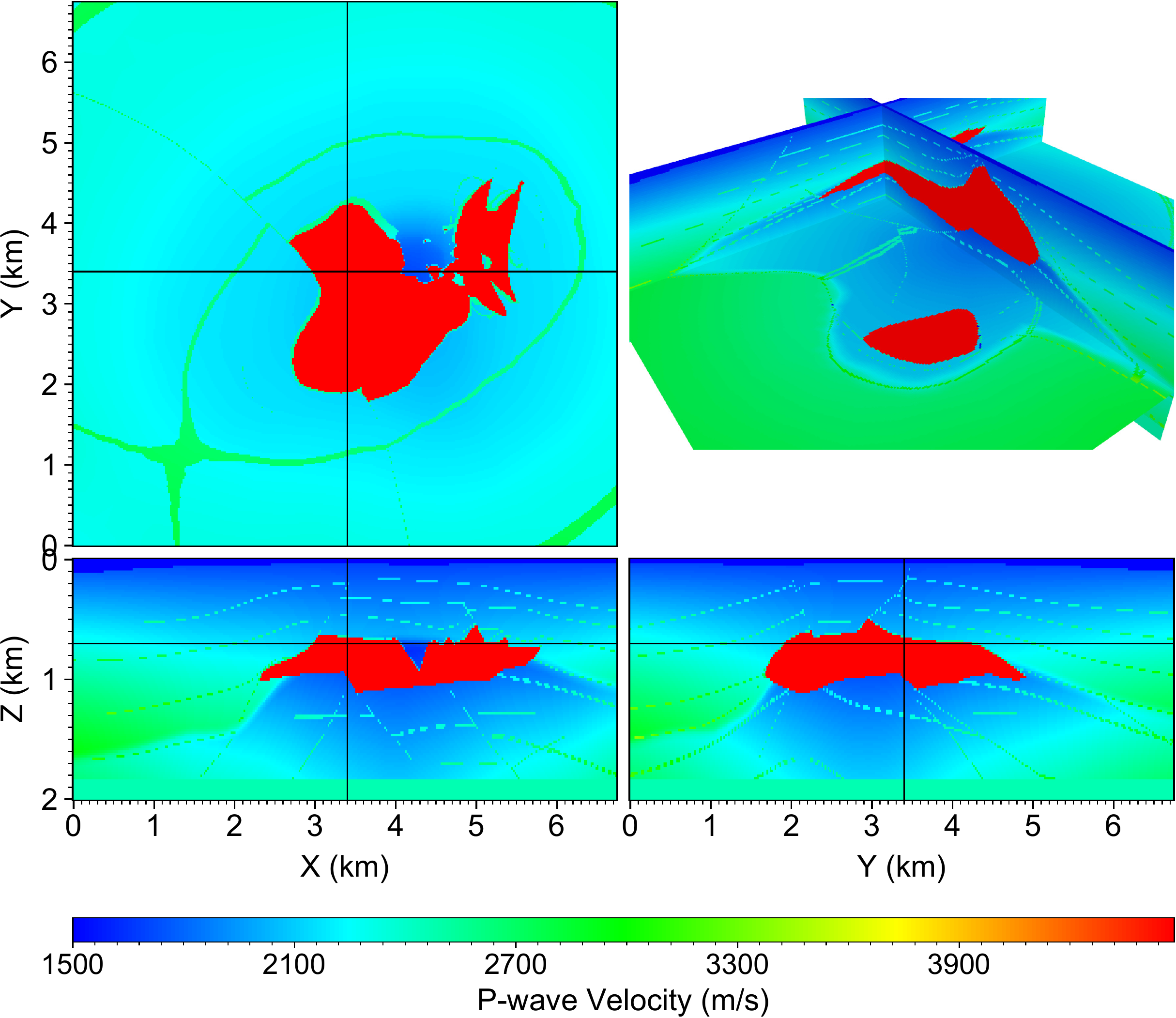}} \quad
	\subfloat[]{\includegraphics[width=0.48\textwidth]{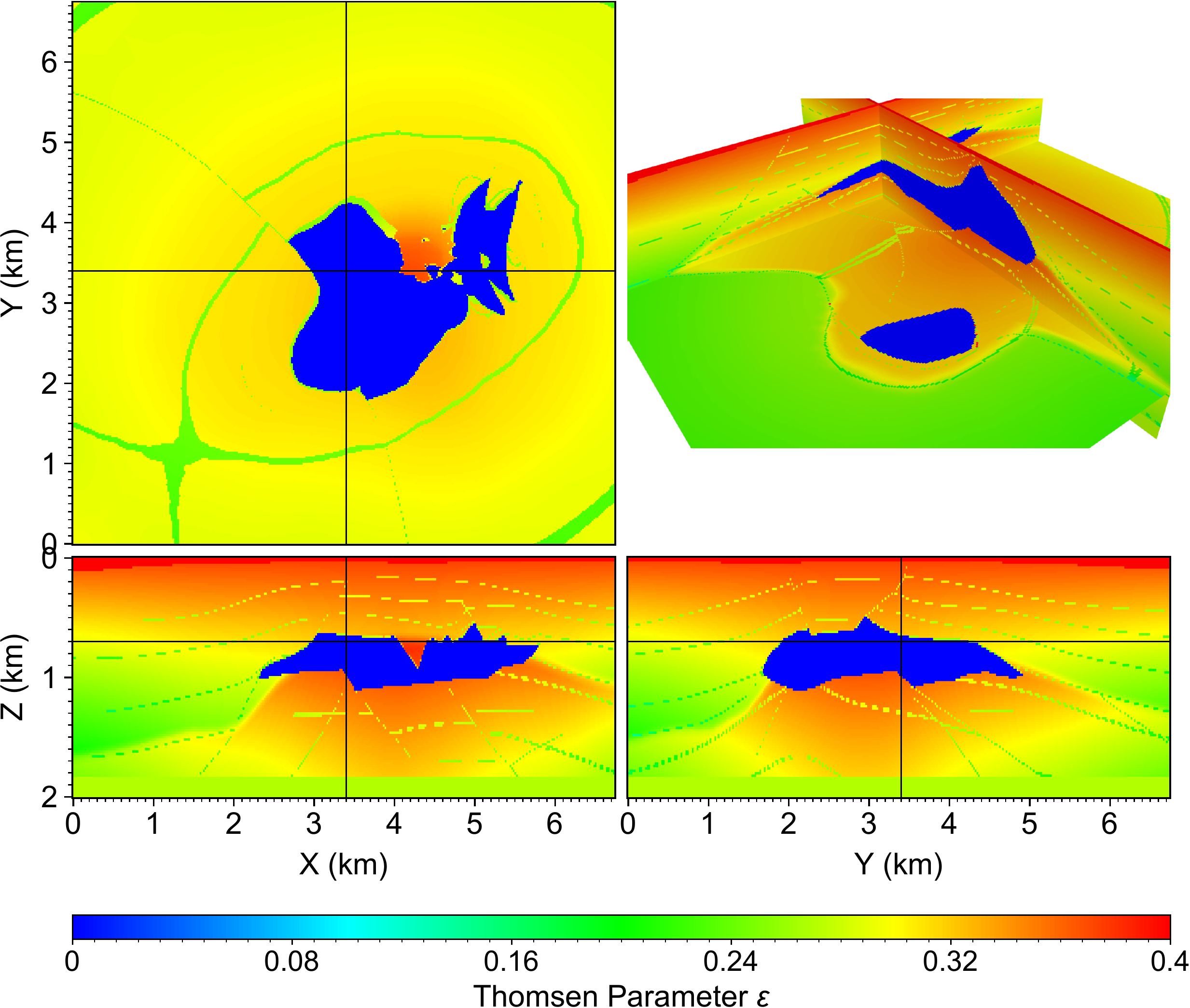}} \\
	\subfloat[]{\includegraphics[width=0.48\textwidth]{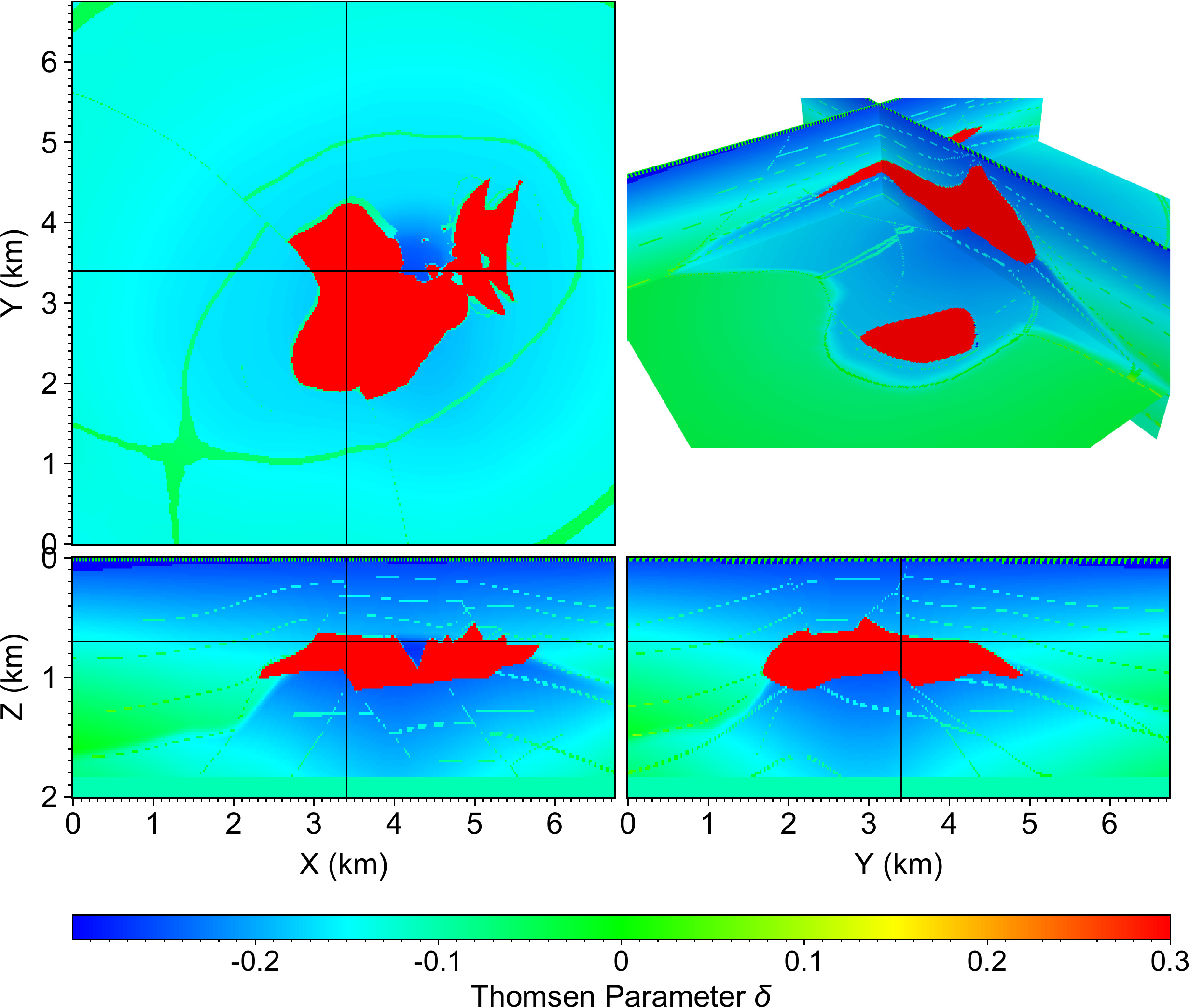}} \quad
	\subfloat[]{\includegraphics[width=0.48\textwidth]{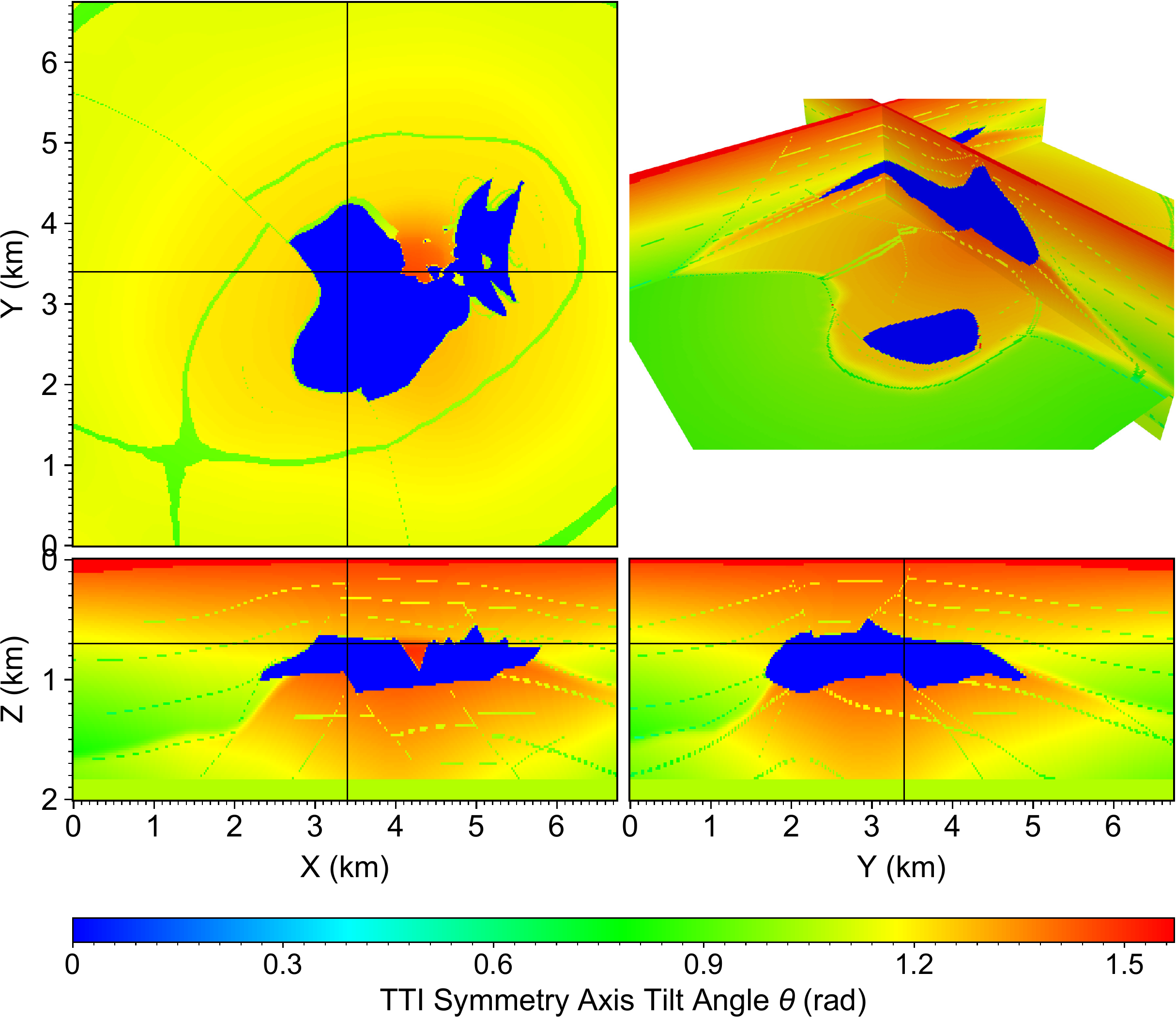}}
	\caption{The anisotropic salt model. Panels (a)-(d) show the P-wave velocity, Thomsen parameters $\varepsilon$ and $\delta$, and TTI symmetry axis tilt angle $\theta$, respectively. The TTI symmetry axis tilt angle $\phi$ has a same spatial variation pattern with $\theta$, with a value range of $[0, \pi]$. }
	\label{fig:model_salt}
\end{figure}

We verify the accuracy and stability of our hybrid method using a 3D 
anisotropic model in Figure~\ref{fig:model_salt} modified from the 
SEG/EAGE salt model.  The model dimension is 6.76~km in both the X and 
Y directions and 2~km in the Z direction. The P-wave velocity model shown 
in Figure~\ref{fig:model_salt}a has a value range from 1500~m/s to 
4500~m/s.  We create the models of Thomsen parameters $\varepsilon$ and 
$\delta$, with values varying from 0 to 0.4 and -0.3 to 0.3, 
respectively, from the original velocity model. We create the TTI 
symmetry axis tilt angle $\theta$ and $\phi$ models with a value range 
from 0 to $90^{\circ}$ and 0 to $180^{\circ}$, respectively. The $\phi$ 
model has the same spatial pattern as the $\theta$ model shown in 
Figure~\ref{fig:model_salt}d. 

\begin{figure}
	\centering
	\subfloat[]{\includegraphics[width=0.8\textwidth]{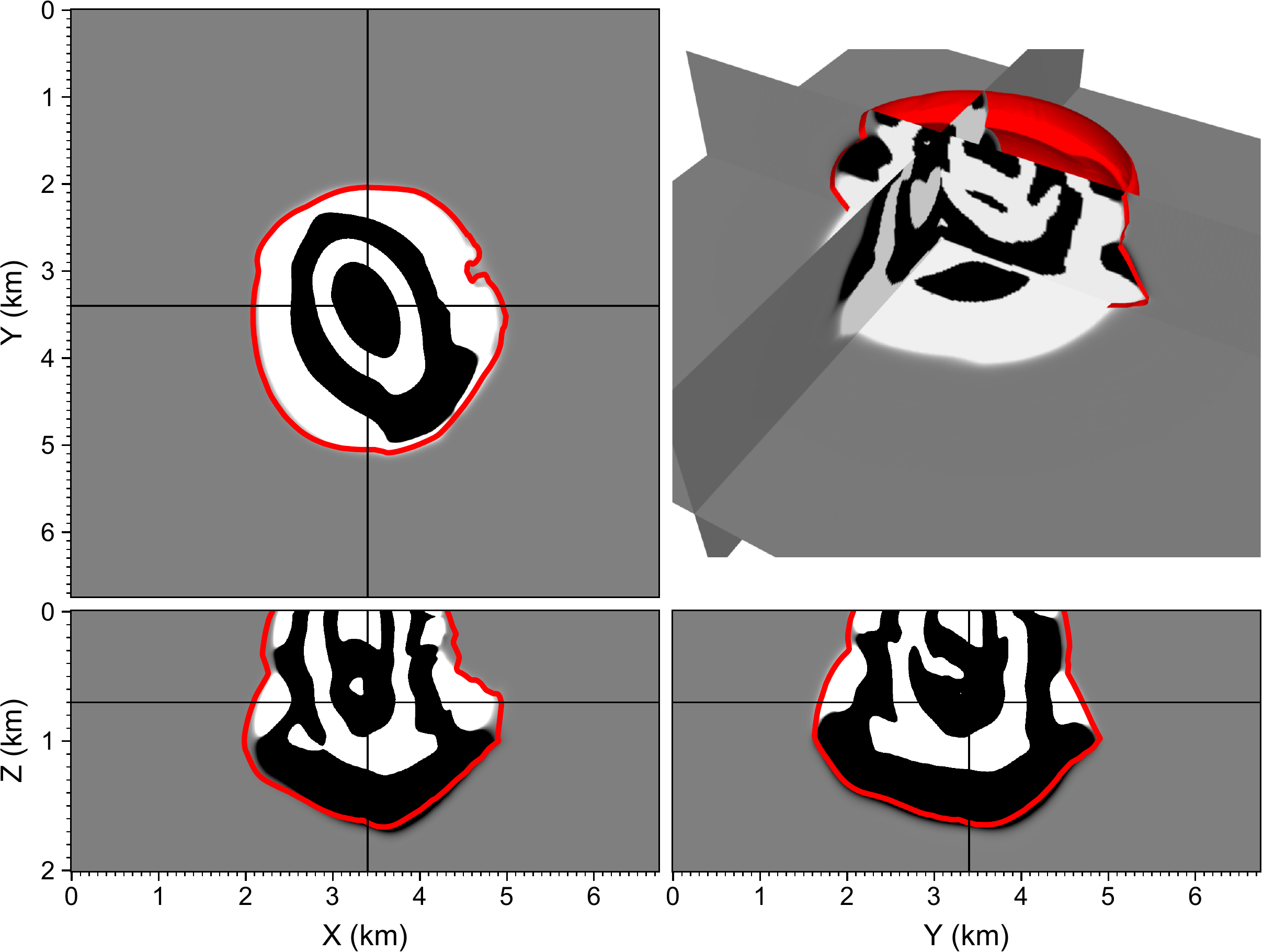}} \\
	\subfloat[]{\includegraphics[width=0.8\textwidth]{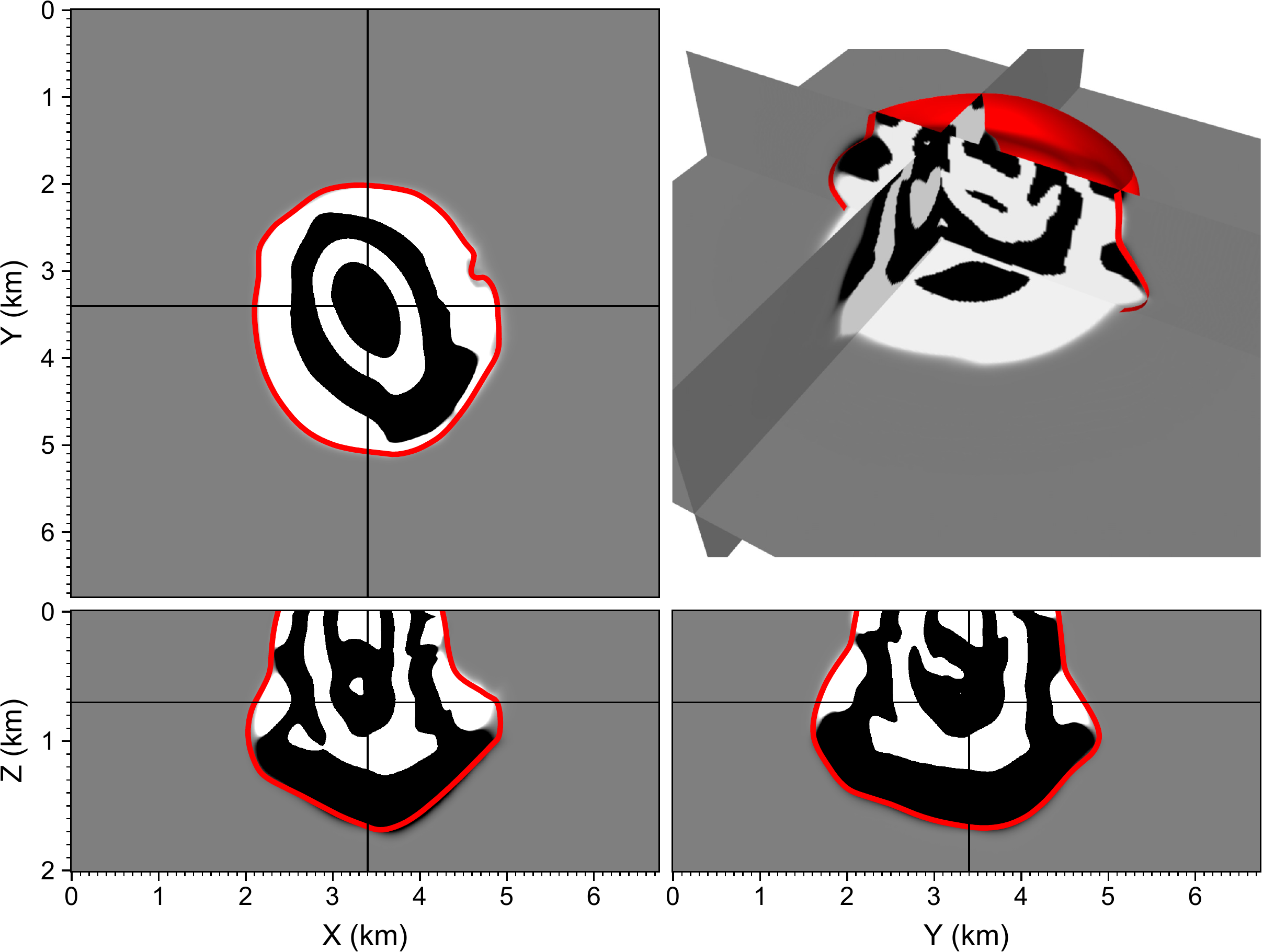}}
	\caption{Comparison between the full-wavefield snapshot at 0.2~s and the corresponding traveltime contour (red curves and isosurface) computed using (a) Godunov method and (b) our hybrid method.}
	\label{fig:wave_salt_1}
\end{figure}

Figures~\ref{fig:wave_salt_1}a and b compares the full-wavefield solution 
at 0.2~s with the Godunov and our hybrid method solutions, respectively.  
There exist obvious inconsistency between the Godunov solution and the 
full-wavefield solution in the X-Z slice of Figure~\ref{fig:wave_salt_1}a 
at a depth of approximately 0.3~km. The Godunov solution is faster than 
the wavefront of the full-wavefield solution around this depth.  At 
a depth of approximately 1.7~km, the Godunov solution is slower than the 
wavefront of the full-wavefield solution. By contrast, our hybrid method 
solution in Figure~\ref{fig:wave_salt_1}b shows good consistency with the 
full-wavefield solution in both shallow and deep regions. 

\begin{figure}
	\centering
	\subfloat[]{\includegraphics[width=0.8\textwidth]{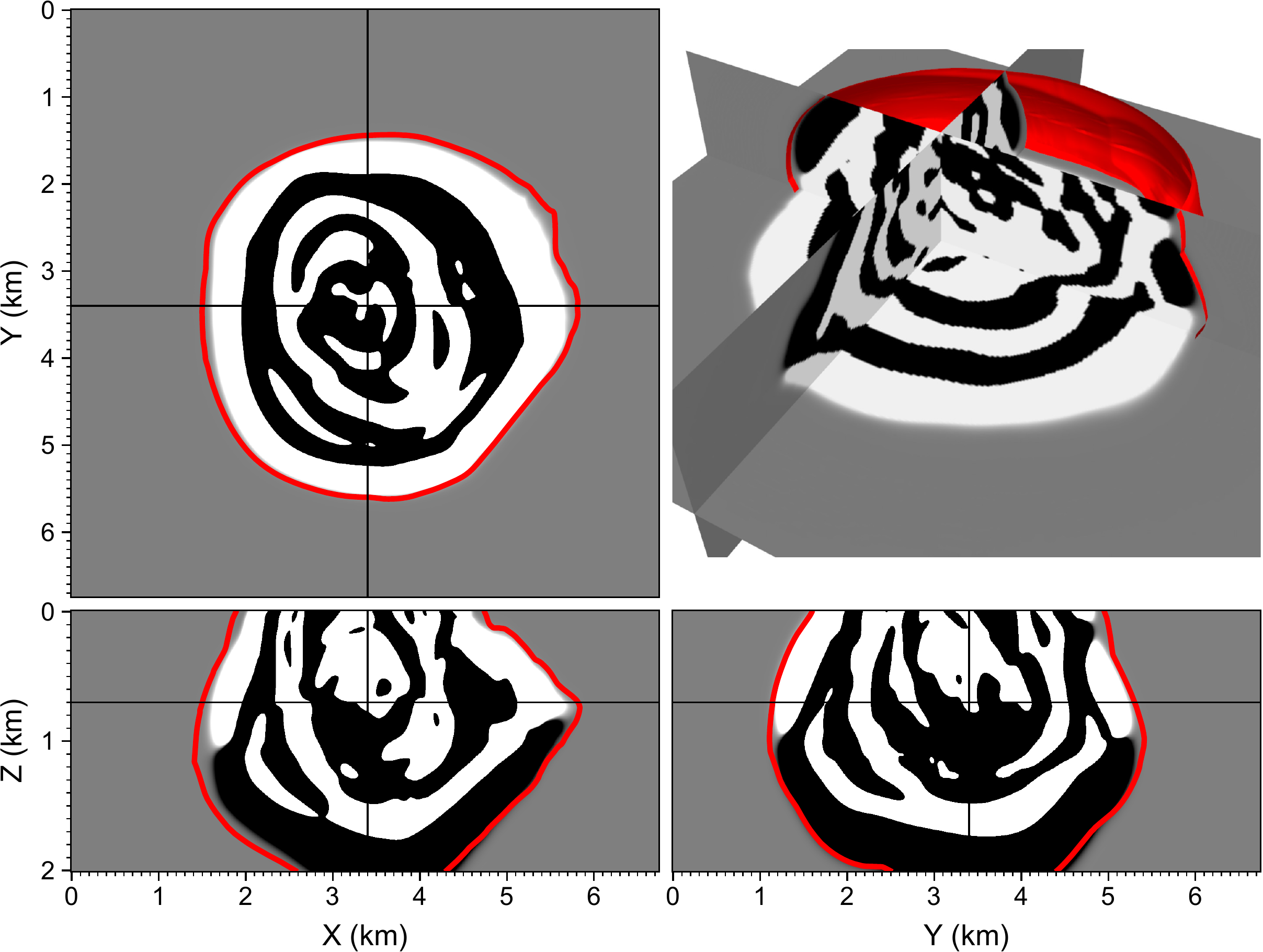}} \\
\subfloat[]{\includegraphics[width=0.8\textwidth]{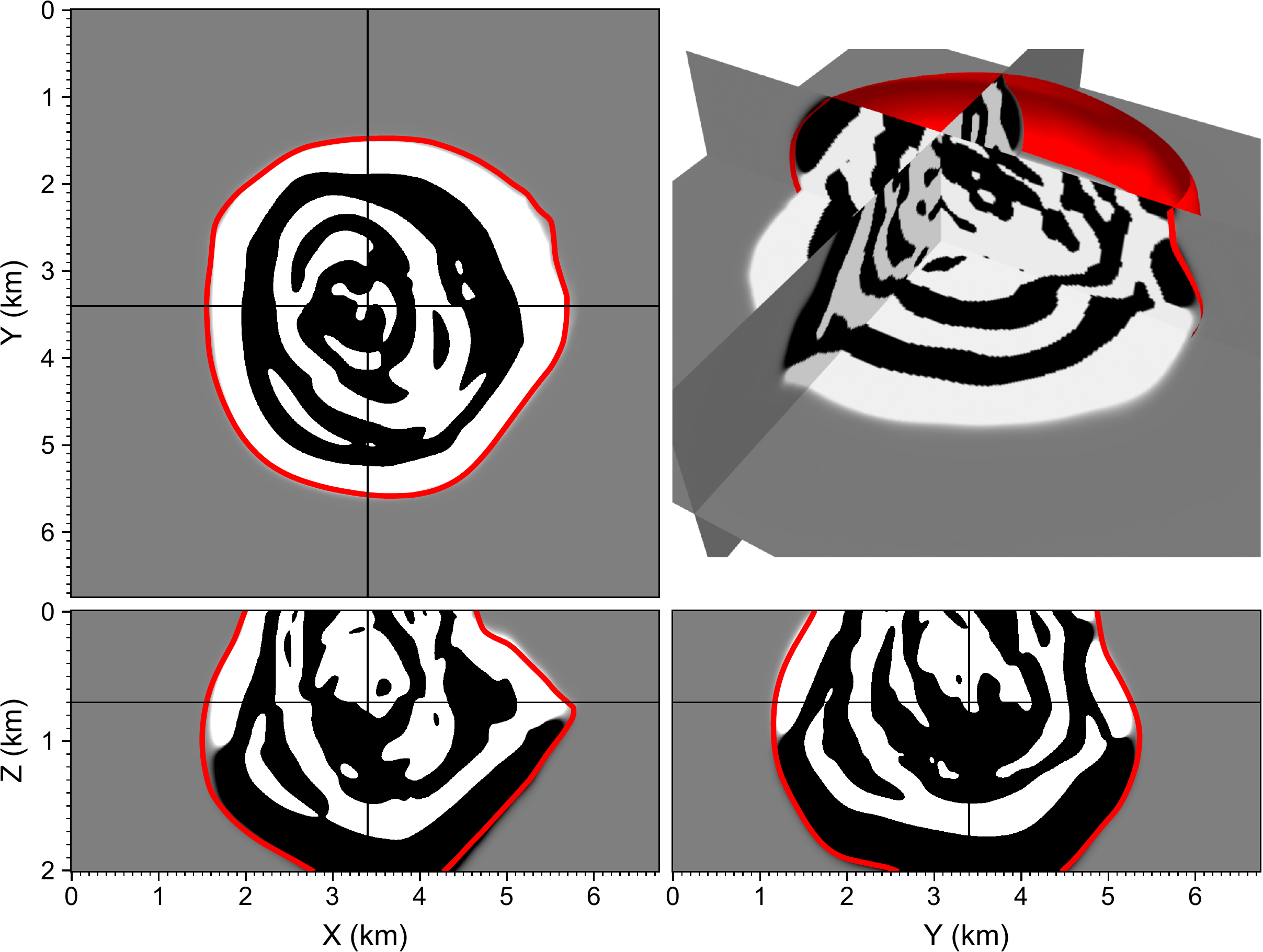}}
	\caption{Comparison between the full-wavefield snapshot at 0.3~s and the corresponding traveltime contour (red curves and isosurface) computed using (a) Godunov method and (b) our hybrid method.}
		\label{fig:wave_salt_2}
\end{figure}

The full-wavefield and Godunov solution consistency check in 
Figure~\ref{fig:wave_salt_2}a for the snapshot of 0.3~s shows that the 
Godunov solution is faster than the wavefront of the full-wavefield 
solution at the depth around 1.1~km in the X-Z slice. There exist obvious 
inconsistency between the two solutions in the X-Y slice in 
Figure~\ref{fig:wave_salt_2}a. By contrast, our hybrid method produces 
a solution in Figure~\ref{fig:wave_salt_2}b that is consistent with the 
full-wavefield solution in all three slices.

\begin{figure}
	\centering
	\subfloat[]{\includegraphics[width=0.8\textwidth]{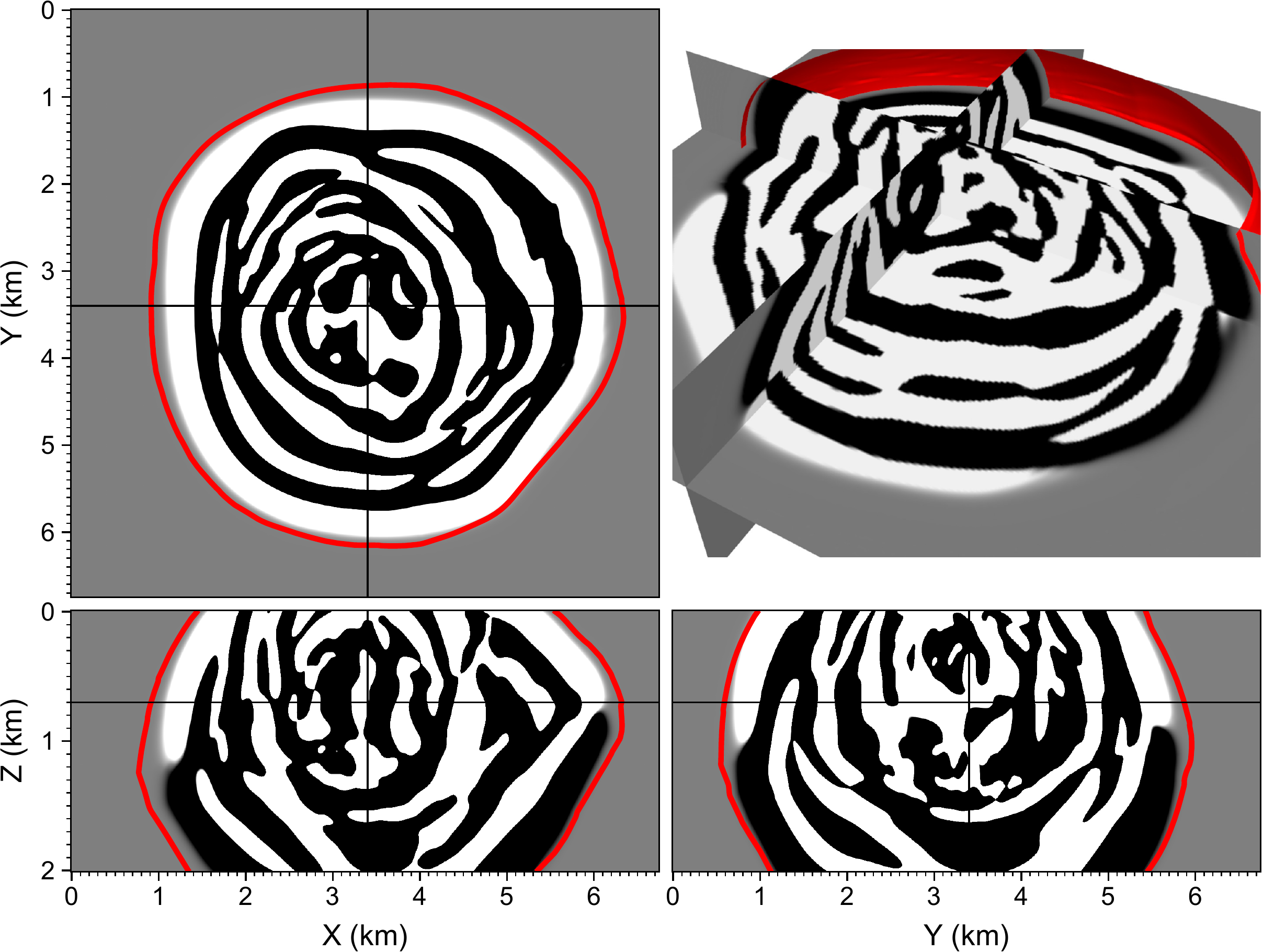}} \\
\subfloat[]{\includegraphics[width=0.8\textwidth]{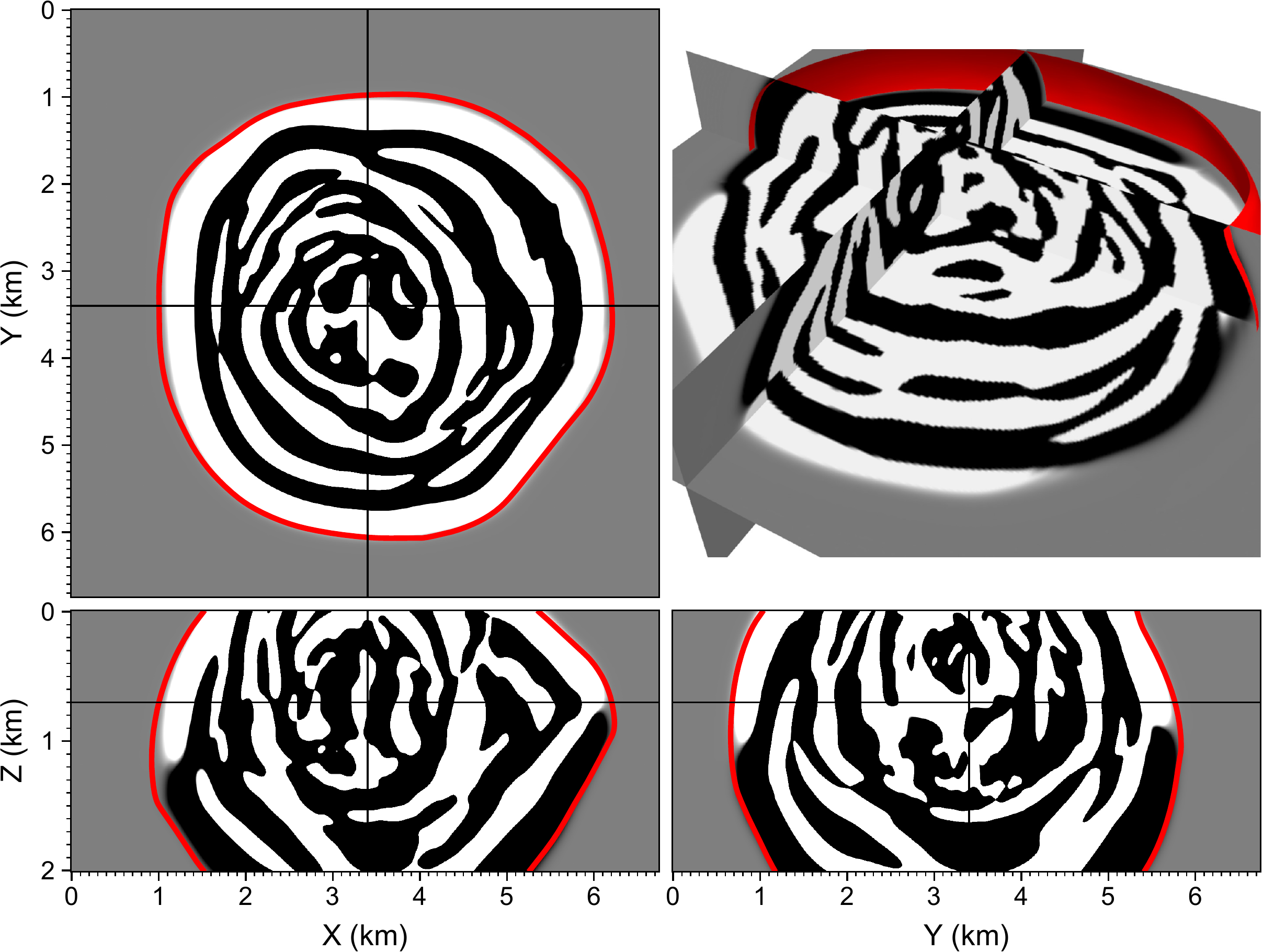}}
	\caption{Comparison between the full-wavefield snapshot at 0.4~s and the corresponding traveltime contour (red curves and isosurface) computed using (a) Godunov method and (b) our hybrid method.}
		\label{fig:wave_salt_3}
\end{figure}

Figure~\ref{fig:wave_salt_3} depicts a full-wavefield snapshot at 0.4~s 
superimposed with the corresponding traveltime contours obtained using 
the Godunov method and our hybrid method. The results further verify that 
that our hybrid method is stable and accurate for 3D heterogeneous 
anisotropic media with strong contrasts. 
 
\section{Conclusions}

We have developed a hybrid eikonal solver for computing first-arrival 
traveltime in 2D and 3D anisotropic media. The numerical scheme of our 
hybrid eikonal solver consists of three steps: the Godunov fast 
locking-sweeping step, the base traveltime computation step in 
anelliptically anisotropic media, and the third-order Lax-Friedrichs fast 
sweeping step. There are three advantages in our hybrid eikonal solver 
compared with existing eikonal solvers for anisotropic media.  (1) The 
solver avoids the source singularity issue by multiplicative traveltime 
factorization and requires no specification of near-source traveltime 
values. (2) It is high-order accurate in space. (3) It can produces 
stable and accurate solution in models with strong anelliptically 
anisotropy, strong medium property contrasts, and complex structures. We 
have used five numerical examples, including four 2D examples and one 3D 
example, to verify the stability and high-order accuracy of our hybrid 
eikonal solver. The results show that our new method is advantageous in 
terms of stability and accuracy compared with conventional approaches. 
Future work aims at extending our method to address media with more 
complex anisotropies such as orthorhombic and rotated orthorhombic 
anisotropies. 

\section{Acknowledgments}

This work was supported by the U.S.\ Department of Energy (DOE) 
Geothermal Technologies Office through the Los Alamos National Laboratory 
(LANL).  LANL is operated by Triad National Security, LLC, for the U.S.\ 
DOE National Nuclear Security Administration (NNSA) under Contract No.\  
89233218CNA000001.  This research used resources provided by the LANL 
Institutional Computing Program supported by the U.S.\ DOE NNSA under 
Contract No.~89233218CNA000001. 

\bibliographystyle{seg}
\bibliography{refs}

\section*{Appendix A: Hybrid eikonal solver for 3D TTI media}

In 3D, our hybrid eikonal solver is based on the following eikonal 
equation in TTI media:
\begin{equation}
v_x^2 \left(\frac{\partial t}{\partial \hat{x}}\right)^2 + v_y^2 \left(\frac{\partial t}{\partial \hat{y}}\right)^2 + v_z^2 \left(\frac{\partial t}{\partial \hat{z}}\right)^2 \left\{1-2(\varepsilon - \delta) v_z^2 \left[\left(\frac{\partial t}{\partial \hat{x}}\right)^2 + \left(\frac{\partial t}{\partial \hat{y}}\right)^2\right] \right\} = 1, \label{eq:aniso_eikonal_3d}
\end{equation}
where $v_x (\mathbf{x}) = v_y(\mathbf{x}) = V_p(\mathbf{x}) \sqrt{1+2\varepsilon(\mathbf{x})}$ is the qP-wave velocity along the $x$- and $y$-axes, $v_z(\mathbf{x}) = V_p(\mathbf{x})$ is the qP-wave velocity along the $z$-axis (i.e., the anisotropy symmetry axis), $\varepsilon = \varepsilon(\mathbf{x})$ and $\delta = \delta(\mathbf{x})$ are Thomsen parameters describing a VTI medium's anisotropy properties. 

Equation~\eqref{eq:aniso_eikonal_3d} is written in the rotated coordinates, and the spatial derivatives are combinations of the spatial derivatives in the original coordinates:
\begin{align}
\frac{\partial t}{\partial \hat{x}} &= \alpha_x \frac{\partial t}{\partial x} + \beta_x \frac{\partial t}{\partial y} + \gamma_x \frac{\partial t}{\partial z}, \\
\frac{\partial t}{\partial \hat{y}} &= \alpha_y \frac{\partial t}{\partial x} + \beta_y \frac{\partial t}{\partial y} + \gamma_y \frac{\partial t}{\partial z}, \\
\frac{\partial t}{\partial \hat{z}} &= \alpha_z \frac{\partial t}{\partial x} + \beta_z \frac{\partial t}{\partial y} + \gamma_z \frac{\partial t}{\partial z},
\end{align}
with the coordinate transformation matrix $\mathbf{R}$ written as
\begin{equation}
\mathbf{R} = \begin{bmatrix}
\alpha_x & \beta_x & \gamma_x \\
\alpha_y & \beta_y & \gamma_y \\
\alpha_z & \beta_z & \gamma_z 
\end{bmatrix} = \begin{bmatrix}
\cos\theta \cos\phi & \cos\theta \sin\phi & \sin\theta \\
-\sin\phi & \cos\phi & 0 \\
-\sin\theta \cos\phi & -\sin\theta \sin\phi & \cos\theta
\end{bmatrix},
\end{equation}
where $\theta = \theta(\mathbf{x})$ is the tilt angle of a VTI medium's symmetry axis (i.e., the counterclockwise rotation angle of the symmetry axis w.r.t.\ the $y$-axis), $\phi = \phi(\mathbf{x})$ is the azimuth angle of a VTI medium's symmetry axis (i.e., the counterclockwise rotation angle of the symmetry axis w.r.t.\ the $z$-axis). 

We then solve the factorized eikonal equation using the Lax-Friedrichs scheme based on a third-order WENO discretization. To facilitate our description, we define
\begin{align}
F_x(\tau, \tau_x, \tau_y, \tau_z) &= \alpha_x \left(t_0 \tau_x + t_{0x} \tau \right) +\beta_x \left(t_0 \tau_y + t_{0y} \tau \right) + \gamma_x \left(t_0 \tau_z + t_{0z} \tau\right), \\
F_y(\tau, \tau_x, \tau_y, \tau_z) &= \alpha_y \left(t_0 \tau_x + t_{0x} \tau \right) +\beta_y \left(t_0 \tau_y + t_{0y} \tau \right) + \gamma_y \left(t_0 \tau_z + t_{0z} \tau\right), \\
F_z(\tau, \tau_x, \tau_y, \tau_z)& = \alpha_z \left(t_0 \tau_x + t_{0x} \tau \right) +\beta_z \left(t_0 \tau_y + t_{0y} \tau \right) + \gamma_z \left(t_0 \tau_z + t_{0z} \tau\right). 
\end{align}
Then the Hamiltonian for equation~\eqref{eq:aniso_eikonal_2d} can be 
written as
\begin{align}
H(\tau, \tau_x, \tau_y, \tau_z) &= F_x^2 (\tau, \tau_x, \tau_y, \tau_z) + F_y^2 (\tau, \tau_x, \tau_y, \tau_z)+ F_z^2(\tau, \tau_x, \tau_y, \tau_z) \nonumber \\
&- \xi F_x^2(\tau, \tau_x, \tau_y, \tau_z) F_z^2(t, \tau_x, \tau_y,\tau_z) - \xi F_y^2(\tau, \tau_x, \tau_y, \tau_z) F_z^2(t, \tau_x, \tau_y,\tau_z), 
\end{align}
which leads to
\begin{align}
\frac{\partial H}{\partial \tau} & = 2\left[(1-\xi F_z^2) \frac{\partial F_x}{\partial \tau} F_x + (1-\xi F_y^2) \frac{\partial F_y}{\partial \tau} F_y+ (1-\xi F_x^2 - \xi F_y^2) \frac{\partial F_z}{\partial \tau} F_z\right] \nonumber \\
& = 2\left[(1-\xi F_z^2) (\alpha_x t_{0x} + \beta_x t_{0y} + \gamma_x t_{0z}) F_x + (1-\xi F_z^2) (\alpha_y t_{0x} + \beta_y t_{0y} + \gamma_y t_{0z}) F_y \right. \nonumber \\
& \quad \left. + (1-\xi F_x^2 - \xi F_y^2) (\alpha_z t_{0x} + \beta_y t_{0y} + \gamma_z t_{0z}) F_z\right], \\
\frac{\partial H}{\partial \tau_x} & = 2\left[(1-\xi F_z^2) \frac{\partial F_x}{\partial \tau_x} F_x +(1-\xi F_z^2) \frac{\partial F_y}{\partial \tau_x} F_y+ (1-\xi F_x^2 - \xi F_y^2) \frac{\partial F_z}{\partial \tau_x} F_z\right] \nonumber \\
& = 2t_0\left[(1-\xi F_z^2) \alpha_x F_x + (1-\xi F_z^2) \alpha_y F_y+ (1-\xi F_x^2 - \xi F_y^2) \alpha_z F_z\right], \\
\frac{\partial H}{\partial \tau_y} & = 2\left[(1-\xi F_z^2) \frac{\partial F_x}{\partial \tau_y} F_x +(1-\xi F_z^2) \frac{\partial F_y}{\partial \tau_y} F_y+ (1-\xi F_x^2 - \xi F_y^2) \frac{\partial F_z}{\partial \tau_y} F_z\right] \nonumber \\
& = 2t_0\left[(1-\xi F_z^2) \beta_x F_x + (1-\xi F_z^2) \beta_y F_y+ (1-\xi F_x^2 - \xi F_y^2) \beta_z F_z\right], \\
\frac{\partial H}{\partial \tau_z} & = 2\left[(1-\xi F_z^2) \frac{\partial F_x}{\partial \tau_z} F_x +(1-\xi F_z^2) \frac{\partial F_y}{\partial \tau_z} F_y+ (1-\xi F_x^2 - \xi F_y^2) \frac{\partial F_z}{\partial \tau_z} F_z\right] \nonumber \\
& = 2t_0\left[(1-\xi F_z^2) \gamma_x F_x + (1-\xi F_z^2) \gamma_y F_y+ (1-\xi F_x^2 - \xi F_y^2) \gamma_z F_z\right].
\end{align}

We then compute the following artificial viscosities along the three spatial axes:
\begin{align}
\omega_x &= \max\limits_{\Omega} \left(\left|\frac{\partial H}{\partial \tau}\right| + \left|\frac{\partial H}{\partial \tau_x}\right|\right), \\
\omega_y &= \max\limits_{\Omega} \left(\left|\frac{\partial H}{\partial \tau}\right| + \left|\frac{\partial H}{\partial \tau_y}\right|\right), \\
\omega_z &= \max\limits_{\Omega} \left(\left|\frac{\partial H}{\partial \tau}\right| + \left|\frac{\partial H}{\partial \tau_z}\right|\right), 
\end{align}
where $\Omega$ is the entire computational domain. 

This results in the following Lax-Friedrichs update scheme for the 
traveltime field $\tau$:
\begin{align}
\tau_{i,j,k}^{(m+1)} = \dfrac{1 - H\left(\tau_{i,j,k}^{(m)}, \tau_{0x; i,j,k}^{(m), \dagger}, \tau_{0y; i,j,k}^{(m), \dagger}, \tau_{0z; i,j,k}^{(m),\dagger}\right) + \omega_x \tau_{0x; i,j,k}^{(m), *} +\omega_y \tau_{0y; i,j,k}^{(m), *}+ \omega_z \tau_{0z; i,j,k}^{(m), *}}{\omega_x/\Delta x + \omega_y/\Delta y +  \omega_z /\Delta z} + \tau_{i,j,k}^{(m)}, 
\end{align}
where $\Delta x$, $\Delta y$ and $\Delta z$ are the regular grid sample intervals along the $x$-, $y$- and $z$-axis, respectively. 

The fast sweepings in 3D consist of the following sweepings:
\begin{align}
& i = 1, \cdots, N_x, \quad j=1, \cdots, N_y, \quad k = 1, \cdots, N_z, \\
& i = N_x, \cdots, 1, \quad j=1, \cdots, N_y, \quad k = 1, \cdots, N_z, \\
& i = 1, \cdots, N_x, \quad j=N_y, \cdots, 1, \quad k = 1, \cdots, N_z, \\
& i = N_x, \cdots, 1, \quad j=N_y, \cdots, 1, \quad k = 1, \cdots, N_z, \\
& i = 1, \cdots, N_x, \quad j=1, \cdots, N_y, \quad k = N_z, \cdots, 1, \\
& i = N_x, \cdots, 1, \quad j=1, \cdots, N_y, \quad k = N_z, \cdots, 1, \\
& i = 1, \cdots, N_x, \quad j=N_y, \cdots, 1, \quad k = N_z, \cdots, 1, \\
& i = N_x, \cdots, 1, \quad j=N_y, \cdots, 1, \quad k = N_z, \cdots, 1. 
\end{align}
where, again, the sweeping order is not important as along as all the listed sweepings are implemented. 

\end{document}